\documentclass[longauth]{aa}  

\usepackage{natbib}
\usepackage{graphicx}
\usepackage{txfonts}
\usepackage{hyperref}
\usepackage{lscape}
\usepackage{xcolor}
\usepackage{ulem}
\usepackage{color, soul} 

\def\ms{\hbox{\,m\,s$^{-1}$}}         
\def\cms{\hbox{\,cm\,s$^{-1}$}}       
\def\kms{\hbox{\,km\,s$^{-1}$}}       

\def\rearth{\hbox{R$_\oplus$}}

\begin{document} 
  \title{ESPRESSO@VLT}
  \subtitle{On-sky performance and first results\thanks{Based on Guaranteed Time Observations collected at the European Southern Observatory under ESO program(s) 1102.C-0744, 1102.C-0958 and 1104.C-0350 by the ESPRESSO Consortium}
  } 
 
  \author{F. Pepe\inst{1}
  \and S.\,Cristiani\inst{2,16}
  \and R.\,Rebolo\inst{3,18,19}
  \and N.\,C.\,Santos\inst{4,12}
  \and H.\,Dekker \inst{5}
   \and A.\,Cabral \inst{6,11}
  \and P.\,Di\,Marcantonio \inst{2}
  \and P.\,Figueira \inst{4,15}
  \and G.\,Lo Curto \inst{15}
  \and C.\,Lovis \inst{1}
% ...
  }

\institute{D\'epartement d'astronomie, Universit\'e de Genève, ch. des Maillettes 51, CH-1290 Versoix, Switzerland\\
\email{francesco.pepe@unige.ch}
\and INAF -- Osservatorio Astronomico di Trieste, Via Tiepolo 11, I-34143 Trieste, Italy
\and Instituto de Astrofisica de Canarias,  Via Lactea, E-38200 La Laguna, Tenerife, Spain
\and Instituto de Astrof\'isica e Ci\^encias do Espa\c{c}o, Universidade do Porto, CAUP, Rua das Estrelas, 4150-762 Porto, Portugal
\and ESO, European Southern Observatory, Karl-Schwarzschild-Stra\ss{}e 2, 85748 Garching, Germany 
\and Instituto de Astrofísica e Ciências do Espaço, Universidade de Lisboa, Edifício C8, 1749-016 Lisboa, Portugal
\and INAF -- Osservatorio Astronomico di Brera, Via Bianchi 46, I-23807 Merate, Italy
\and Centro de Astrobiolog\'ia (CSIC-INTA), Carretera de Ajalvir km 4, E-28850 Torrej\'on de Ardoz, Madrid, Spain
\and Physics Institute of University of Bern, Gesellschaftsstrasse\,6, CH-3012 Bern, Switzerland
\and INAF -- Osservatorio Astrofisico di Torino, Via Osservatorio 20, I-10025 Pino Torinese, Italy
\and Departamento de Física da Faculdade de Ciências da Univeridade de Lisboa, Edifício C8, 1749-016 Lisboa, Portugal
\and Departamento de F\'isica e Astronomia, Faculdade de Ci\^encias, Universidade do Porto, Rua do Campo Alegre, 4169-007 Porto, Portugal     
\and INAF -- Osservatorio Astronomico di Palermo, Piazza del Parlamento 1, 90134 Palermo, Italy
\and NRCC-HIA, 5071 West Saanich Road Building VIC-10, Victoria, British Columbia V9E 2E, Canada
\and ESO, European Southern Observatory, Alonso de Cordova 3107, Vitacura, Santiago
\and Institute for Fundamental Physics of the Universe, IFPU, Via Beirut 2, 34151 Grignano, Trieste, Italy
\and Centro de Astrof\'isica da Universidade do Porto, Rua das Estrelas, 4150-762 Porto, Portugal
\and Universidad de La Laguna, Departamento de Astrof\'isica, E- 38206 La Laguna, Tenerife, Spain
\and Consejo Superior de Investigaciones Cient\'ificas, E-28006 Madrid, Spain
% ...
}
    
  \date{Received April 17, 2020; accepted September 29, 2020}

  \abstract
   {ESPRESSO is the new high-resolution spectrograph of ESO's Very-Large Telescope (VLT). It was designed for ultra-high radial-velocity precision and extreme spectral fidelity with the aim of performing exoplanet research and fundamental astrophysical experiments with unprecedented precision and accuracy. It is able to observe with any of the four Unit Telescopes (UT) of the VLT at a spectral resolving power of 140\,000 or 190\,000 over the 378.2 to 788.7\,nm wavelength range, or with all UTs together, turning the VLT into a 16-m diameter equivalent telescope in terms of collecting area, while still providing a resolving power of 70\,000.}
   {We provide a general description of the ESPRESSO instrument, report on the actual on-sky performance, and present our Guaranteed-Time Observation (GTO) program with its first results.}
   {ESPRESSO was installed on the Paranal Observatory in fall 2017. Commissioning (on-sky testing) was conducted between December 2017 and September 2018. The instrument saw its official start of operations on October 1st, 2018, but improvements to the instrument and re-commissioning runs were conducted until July 2019.}
   {The measured overall optical throughput of ESPRESSO at 550\,nm and a seeing of 0.65\arcsec exceeds the 10\% mark under nominal astro-climatic conditions. We demonstrate a radial-velocity precision of \textit{better} than 25\cms during one night and 50\cms over several months. These values being limited by photon noise and stellar jitter show that the performanceis compatible with an instrumental precision of 10\cms. No difference has been measured across the UTs neither in throughput nor RV precision.}
   {The combination of the large collecting telescope area with the efficiency and the exquisite spectral fidelity of ESPRESSO opens a new parameter space in RV measurements, the study of planetary atmospheres, fundamental constants, stellar characterisation and many other fields.}


   \maketitle
%
\section{Introduction}
High-resolution spectroscopy enables key physical insight in the study of stars, galaxies, and interstellar and intergalactic medium. High precision and high dispersion has emerged in recent years as a crucial element in spectroscopy of faint objects, which called in turn for the use of larger telescopes to increase the photon-collecting area. Furthermore, many science cases require repeatable observations over a long temporal  baseline. For instance, the HARPS spectrograph at the ESO 3.6-m telescope established itself as a pioneering instrument for radial-velocity (RV) measurements \citep{mayor:2003} because of its ultra-high precision, but also thanks to its operational efficiency and continuous availability on a medium-size telescope.

The search for and characterization of terrestrial planets in the habitable zone of their parent stars is one of the most exciting science topics of the next decades and one of the main science drivers for the new generation of space missions and extremely-large telescopes. These goals rely on the definition of a suitable target list that has been cleaned from false positives and is filled with good targets for follow-up and characterization. There is a gap between past observations and surveys with small- and medium-size telescopes ($<4$\,m diameter), and the target-focused observations on future large facilities (ELTs), which can and must be filled by the optimum use of new instruments on currently accessible 10-m class telescopes.

Another aspect that must be considered is the fast evolution of the field from exoplanet search to exoplanet characterization. A sudden change was induced by space-based transit searches such as \textit{CoRoT} and \textit{Kepler}. Especially the latter one started delivering thousands of transiting candidates that had to be confirmed, and more importantly, for which we were lacking mass measurements. This situation emphasized the synergy between radial velocities and transit observations, and thus the need for ground-based radial-velocity follow-up in view of better understanding the internal structure and composition of exoplanets. Unfortunately, most of the transiting candidates delivered by \textit{Kepler} are too faint for RV follow-up with small and medium-size telescopes and ``even fainter'' when aiming at observing planetary atmospheres by the means of ground-based transit spectroscopy.

These facts were recognized by the ESO-ESA working group report on extrasolar planets in 2005 who emphasized the need of HARPS-like instruments on large telescopes \citep{Perryman2005}. Simultaneously, the idea of a HARPS-like instrument for the VLT was directly derived from the CODEX project \citep{pasquini:2005,molaro2006,molaro2009}, a phase-A study for the development of an ultra high-precision spectrograph for the European ELT, and presented by \citet{pasquini:2009} in 2007 at the conference \textit{Science with the VLT in the ELT Era}. Following this idea, ESO's Scientific-Technical Committee (STC) recommended in October 2007 the development of second-generation VLT instruments, and this proposal was endorsed by the ESO Council in December of the same year. Among others, a high-resolution, ultra-stable spectrograph for the VLT combined-Coud\'e focus was suggested. Following these recommendations, in March 2008 ESO issued a call for proposals to carry out the Phase-A study for such an instrument driven by three main science cases: \textit{1) Detection and characterization of rocky planets in the habitable zone}, \textit{2) Study of the variability of fundamental constants}, and \textit{3) Analysis of the chemical composition of stars in nearby galaxies}. The proposal submitted by the ESPRESSO Consortium\footnote{The ESPRESSO Consortium was composed of: Observatoire Astronomique de l'Universit\'e de Gen\`eve (project head, Switzerland); Centro de Astrof\'isica da Universidade do Porto (Portugal), Faculdade de Ciencias da Universidade de Lisboa (Portugal); INAF-Osservatorio Astronomico di Brera (Italy); INAF-Osservatorio Astronomico di Trieste (Italy); Instituto de Astrof\'isica de Canarias (Spain);	Physikalisches Institut der Universit\"at Bern (Switzerland); ESO participated to the ESPRESSO project as Associated Partner} was accepted by ESO and selected to carry out the construction of this spectrograph in collaboration with ESO as a partner. In exchange for funding and building the instrument, the ESPRESSO Consortium was awarded 273 nights of Guaranteed-Time Observation (GTO), in order to pursue part of the science cases described above.

The present paper aims at providing a general description of the ESPRESSO instrument and its performance as delivered to the community by summer 2019. Chapter\,\ref{sec:hist} summarizes the project history. In Chapter\,\ref{sec:inst} we present a first-level description of the instrument and its subsystems to provide the reader with a general overview without entering technical details; instead, references to technical papers will be given. Also, we shall describe the current operation scheme of ESPRESSO at the Paranal Observatory. Chapter\,\ref{sec:perf} focuses on the \textit{on-sky} performance derived from the commissioning phases and the first periods of operation. In Chapter\,\ref{sec:obs}, we describe our GTO program and give an overview of suitable science cases of ESPRESSO. Examples of first results obtained during the initial periods of observation are presented and references to the corresponding papers given.

\section{The ESPRESSO project}
\label{sec:hist}
The project kick-off was held in February 2011 and the design phase ended with the Final Design Review in May 2013. The procurement of components and manufacturing of subsystems took roughly three years, being considerably longer than originally planned. Integration started in early 2016 with the first subsystem, the large vacuum vessel, in the integration hall of the Geneva Observatory, overlapping with the procurement phase of other subsystems in the various Consortium-Partner institutes. Assembly, integration and verification (AIV) was completed with the Provisional Acceptance Europe (PAE) in June 2017. In parallel to the procurement and AIV phase of the spectrograph in Europe, the unexploited Combined-Coudé Laboratory (CCL) of the VLT was cleared and equipped with an air-conditioned environment to accept the ESPRESSO spectrograph. A considerable effort was devoted by the Consortium and ESO to equip \textit{all} the Unit Telescopes (UT) of the VLT with the so-called Coudé-Train (CT), an ensemble of optics, mechanics and control units to route the light from each UT through the ``incoherent'' Coudé-path and the tunnels to the CCL, where it can be either combined to use ESPRESSO in the 4-UT configuration, or individually selected to feed the spectrograph in its 1-UT configuration. By PAE, both the CCL (2015-2016) and the CT (October 2016) had been commissioned and were ready to welcome the spectrograph, even though some UT branches of the CT would have to be commissioned and re-aligned later on, during the spectrograph's commissioning. The spectrograph was shipped from Geneva to Paranal in summer 2017.

During fall 2017, the on-site reassembly in the CCL started, followed by the integration with the CT. The CT had been commissioned to verify all interfaces, functionalities and performance, especially with respect to telescope selection, optical path configuration, telescope pre-setting, pointing, acquisition, tracking and guiding. ESPRESSO required on-sky configuration and testing on each of the UTs individually in the 1-UT configuration and on all four UTs together for 4-UT configuration. The commissioning plan and strategy was adapted to scheduling requirements and the commissioning runs were performed on the dates reported in Table\,\ref{tab:comm_runs}.

\begin{table*}
\caption{List of Commissioning runs and technical missions on ESPRESSO}             
\label{tab:comm_runs}      
\centering          
\begin{tabular}{l l l l}    
\hline\hline       
Commissioning run	& Start date &	End date	& Tested modes and telescopes \\
\hline                    
\textbf{First light} &		\textbf{27/11/2017} &	 & \\
Commissioning 1A &  	27/11/2017 &	06/12/2017 &	HR and UHR modes, all UTs\\
Commissioning 1B  & 	26/01/2018 &	07/02/2018 &	All modes and all configurations\\
Commissioning 2A  &	25/02/2018 &	07/03/2018 &	All modes, all UTs\\
\textit{Technical activities}  &		07/03/2018 &	21/03/2018 & All subsystems\\
Commissioning 2B &	27/04/2018 &	05/05/2018 &	HR and UHR mode, all UTs\\
\textit{Technical activities} &		13/09/2018 &	21/09/2018 & Vacuum and cryogenics optimization\\
\textbf{Official start of Operations}	&	\textbf{01/10/2018} &	 & \\
\textit{Technical activities}  &	03/10/2018 &	23/03/2018 & Optical efficiency tests and re-alignment\\
Commissioning 2C -- July 2018&		03/07/2018 &	09/07/2018 &	All modes and all configurations\\
4-UT Commissioning -- December 2018&	30/11/2018 &	01/12/2018 &	MR mode, all UTs\\
\textit{Technical activities}  &	08/06/2019 &	27/06/2019 & Fiber upgrade\\
Recommissioning -- July 2019 &		27/06/2019 &	08/07/2019 &	All modes and all configurations\\
\hline\hline                  
\end{tabular}
\end{table*}

\textit{First light} of ESPRESSO in \textit{1-UT configuration} took place on November 27, 2017, the date on which the official ESPRESSO commissioning started (ESO Release eso1739). This commissioning run was followed by others in January, February, April and July 2018. During these runs, all modes of ESPRESSO (including the 4-UT configuration) were tested 'on sky' and characterized. \textit{First light in 4-UT configuration} took place during the January 2018 run in presence of ESO's Director General (ESO Release eso1806). The simultaneous use of the 4 UTs implies a great complexity of the control system. Despite this, all processes -- communication with the telescopes, configuration of the instrument in this new mode, slewing and pointing of the telescopes, acquisition of the target on all the telescopes, guiding and exposing -- were carried out without failure and produced, at the first attempt, the first visible high-resolution spectrum of an astronomical target obtained with the collecting area of a 16-m diameter telescope-equivalent.

Minor improvements on hardware and software were performed in parallel to the commissioning phase, but the instrument demonstrated excellent operational performance and reliability from the beginning. While radial-velocity and stability requirements were essentially met, optical throughput turned out to suffer an unexpected deficiency. For this reason, the spectrograph was re-opened and individual elements were optically tested in September 2018. It was found that the main culprit was the fiber-link, but it was neither possible nor suitable to perform corrective actions on the spot because throughput and RV precision in the 1-UT configuration were satisfactory and the instrument was about to start operations. Two additional nights of 4-UT commissioning period took place in December 2018 in order to perform general functional and preliminary performance tests. 

The official start of operations in 1-UT configuration took place on October 1st, 2018, the date on which ESPRESSO was offered by ESO for both open-time observations to the community and Guaranteed-Time Observations (GTO) to the ESPRESSO Consortium. During the first semester of operations, and in view of the start of operations in the 4-UT configuration by April 2019, the spare fiber link was returned to Europe, repaired and upgraded. Laboratory tests at the Osservatorio Astronomico di Brera in Merate immediately showed that an improvement in throughput was within reach. After returning the spare fiber link to Paranal, the component was installed and tested on the instrument in June 2019. This major intervention required partial re-commissioning of the 1-UT configuration and completion of the 4-UT configuration commissioning. Both were conducted successfully, despite bad weather conditions, from July 3 to 8, 2019. By July 9, 2019, the instrument can be considered to be in its final configuration and fully operational. No changes have been done after this date apart from developing and implementing a new instrument mode, the HR42 mode, being a standard-resolution mode with heavier binning and slow read-out speed, in order to push the read-out noise limited magnitude range to fainter magnitudes. While the practical implementation of this mode in the instrument software turned out to be straight forward, the adaptation of the data-reduction software demanded quite some effort due to the different spectral format.


\section{The ESPRESSO instrument}
\label{sec:inst}

\subsection{Technical description}
\subsubsection{Overview} ESPRESSO is a fiber-fed, cross-dispersed, high-resolution echelle spectrograph located in the Combined-Coud\'e Laboratory (CCL) at the incoherent focus of the VLT, where it can be fed by the light of either one UT (1-UT configuration) or of all the four UTs (4-UT configuration). In each of its configurations two fibers illuminate the spectrograph: 
One fiber carries the light from the science target, and the second one either the light from the sky background (7\arcsec~away) or the light from a reference source for simultaneous drift measurements. A detailed description of the instrument and all its subsystems is given in \citet{pepe:2014}, \citet{megevand:2014}, \citet{gonzalez:2018} and references therein. The control software and electronics are described more specifically in \citet{calderone:2018}, \citet{baldini:2016} and \citet{calderone:2016}. An overview of the scientific software and the description of its flow and its products is presented in Section\,\ref{sec:dfs}. The operational concept of ESPRESSO and its observing modes are sketched in Section\,\ref{sec:operations}

\subsubsection{Light path from the telescope to the CCL} 
\begin{figure}[h]
   \centering
   \includegraphics[width=\hsize]{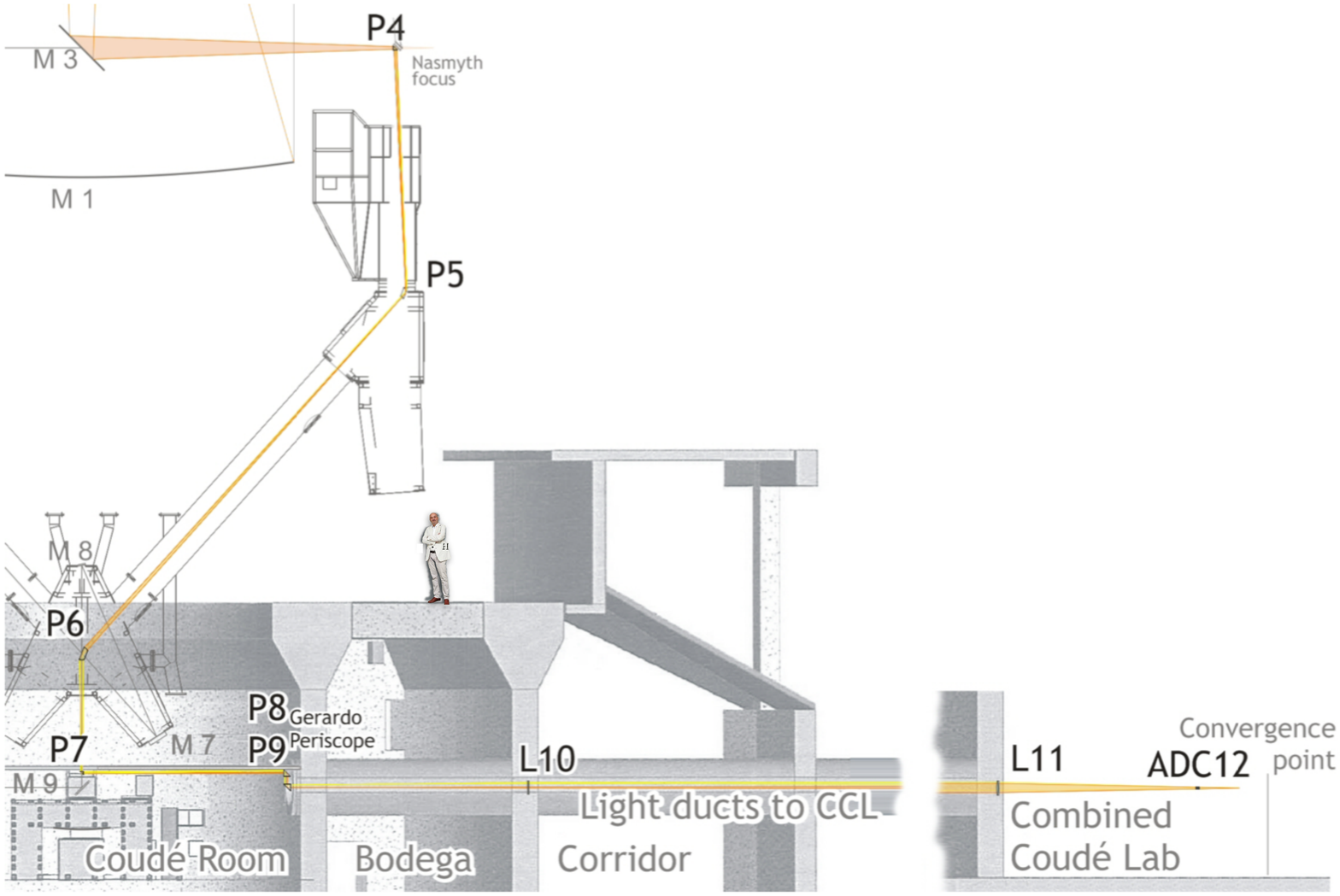}
   \caption{Optical light path of the Coud\'e Train from the Nasmyth platform to the CCL. P4 to P7 are folding prisms that route the beam from the Nasmyth platform to the telescope's Coud\'e room. The mirrors R8 and R9 form a periscope that redirects the beam into the tunnel. At the entrance and the exit of the tunnel respectively, the two lenses L10 and L11 re-condition the light beam in order to enter the Front-End Unit inside the CCL with a specified F-number (fixed for all telescopes). ADC12, two counter-rotating double prisms, eventually corrects for atmospheric dispersion across the wide spectral range of the spectrograph. Figure from \citet{pepe:2014} and ESPRESSO-Project documentation.}
   \label{fig:coude-train}
    \end{figure}

The telescope light is brought to the CCL via the so-called Coud\'e Train (CT, Figure\,\ref{fig:coude-train}). The CT is an all-optical system of prisms, mirrors and lenses that guide the light from the Nasmyth focus through the elevation axis, the telescope structure and the azimuth axis to the Coud\'e room (basement) of the specific telescope. There, the beam is bent and launched through the ``incoherent'' tunnels to the CCL, where the light passes through an Atmospheric Dispersion Compensator (ADC). A detailed description of the quite complex CT can be found in \citet{cabral:2019,cabral:2014,cabral:2013}. Its installation and alignment \citep{avila:2016} represented a considerable share of the overall project effort, but it was necessary to transform ESPRESSO to the first (and only) instrument able to make use of the incoherent focus of the VLT, converting it to the largest collecting-area optical telescope presently available.

\subsubsection{Front-end unit and calibration system} 
In the CCL, the Front-End (FE) shapes and collects the light beams from one to four telescopes. Its design is described in \citet{pepe:2014} and \citet{riva:2014-2}. The FE is composed of 4 units, one per telescope, that combine several functions: 1) A telescope-pupil stabilisation by means of a pupil camera \citep{calderone:2016} and a tip-tilt mirror. 2) A field-stabilization by means of a second tip-tilt mirror and a guiding camera \citep{calderone:2016}. 3) A focusing mechanism to optimize light injection into the optical fiber. 4) A calibration-light injection system that selects whether the spectrograph fibers ``see'' the sky or the calibration light, and allows to inject the light of any calibration source into any of the two fibers. Each FE was integrated and aligned in Merate, Italy, 'blindly', i.e. using only very precise mechanical references, and installed in Paranal adjusting the FE as a whole with respect to the incoming light beam \citep{pariani:2016,aliverti:2016}.

One important role of the FE is to perform field stabilisation (or secondary guiding) that corrects for tip-tilt motion of the star not corrected by the primary stabilisation of the telescope or introduced along the optical path of the CT at frequencies $<10$\,Hz. The images acquired by the field-stabilisation camera are analysed by a control loop to determine the difference in position of the stellar image with regard to the one of the fiber entrance, compute the error and apply the correction to the tip-tilt mirror. By this system both slit efficiency is optimized and RV-effects due to off-centered light injection are avoided \citep{pepe:2002}.

The light from various calibration sources can be injected through the FE in any of the fibers in order to perform order-definition and spectral flat-fielding (white light), wavelength calibration (absolute spectral reference) and drift measurement (stable spectral source) in simultaneous-reference mode \citep{mayor:2003,pepe:2002}. The white light consists of a laser-driven light source (LDLS) EQ-99X from Energetiq. Wavelength calibration is obtained by combining ThAr-lamp spectra with a white-light illuminated Fabry-P\'erot. The former provides accuracy, the latter delivers a high number of uniform spectral emission lines across the whole spectrum that ensure ``local'' wavelength precision (see Section\,\ref{sec:drs} and \citet{cersullo:2019} and references therein). All these sources are coupled by optical fibers to two calibration units \citep{megevand:2014}, one for each fiber channel of the spectrograph. The calibration units have the functionality of attenuating the light of the calibration source and re-direct it through another optical fiber to the corresponding FE Unit.

A Laser-Frequency Comb (LFC) is also available and should have replaced both the ThAr lamp and the Fabry-P\'erot \citep{pasquini:2018,frank:2018,huke:2018}, but, due to lack of reliability and limited spectral coverage, it is currently not integrated in the operational scheme. At the time of writing, it is still unclear when the LFC will become operational. It must be noted that, given the currently limited wavelength range, the LFC will not be used \textit{alone} for complete wavelength calibration. This situation might compromize the ability of ESPRESSO of guaranteeing RV repeatability at the 10\cms-level over years, since the wavelength calibration will have to rely on hollow-cathode lamps.

A toggling system (rotational plate) in the FE carries six fiber heads. Four  fiber heads belong to the medium-resolution mode (MR) and can be positioned simultaneously at the output of the respective FE-unit of each telescope in order to operate ESPRESSO in the 4-UT configuration. Alternatively, in the 1-UT configuration, either the High-Resolution (HR) or the Ultra-High Resolution (UHR) head can be positioned by rotating the toggling system at the output of any of the four FE units.

\begin{figure}[h]
   \centering
   \includegraphics[width=\hsize]{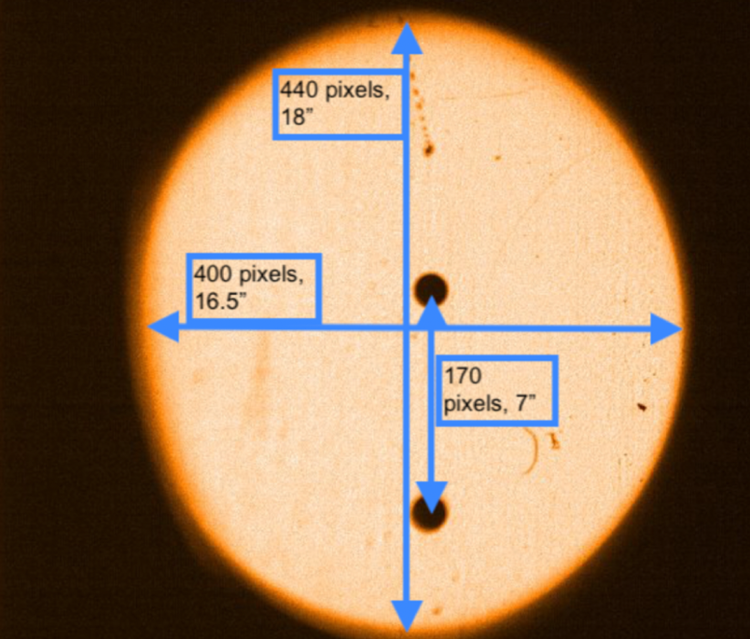}
   \caption{Image of the pierced focal mirror, field, as seen by the field-stabilisation or secondary-guiding camera. The physical direction is rotated by $90^\circ$.}%
   \label{fig:field}
    \end{figure}

Each fiber head carries a pierced mirror \citep{gracia:2018-1} located in the focal plane of the FE Unit. It reflects the light from the telescope to the field and pupil cameras. Figure\,\ref{fig:field} shows an image of the pierced focal mirror, the ``field'', as seen by the field-stabilisation or secondary-guiding camera. The size of the field is about 17\arcsec~ in diameter, the distance between fiber A and fiber B is of 7\arcsec~ and the scale-plate is 0.041\arcsec~ per ($2\times2$-binned) pixel. 

\subsubsection{Fiber feed} 
At the back of each of the two pinholes in the mirror, relay optics re-image the light entering the pinholes onto two octagonal fibers, the target fiber and the reference/sky fiber for simplicity called hereafter fiber A and B, respectively. Each fiber individually carries the light to the double scrambler, which is mounted on the vacuum vessel. The main functionality of the double scrambler is to convert the far-field of the incoming optical fiber into the near field and vice-versa, while injecting the beam into a second section of an identical fiber. This system ensures, in combination with the use of octagonal fibers, a homogeneous and stable illumination of the spectrograph, which is fundamental for repeatable RV measurements \citep{chazelas:2012,cosentino:2012,chazelas:2010}. The double scrambler integrates also a vacuum window that provides the optical vacuum feed-through of the scientific light. The exposure shutters, one for each mode, are installed on the air-side of the feed-through within the double scrambler mechanics. In the 4-UT configuration the double scrambler has furthermore the task of combining the light of the four fibers of each channel A and B from the four telescopes into one single square-shaped fiber for each channel \citep{gracia:2018-2}. All the fibers from the scrambler are then routed to the focal plane of the spectrograph to form a single entrance `slit' (see the right-hand side of Figure\,\ref{fig:spectral-format}). The two fibers A and B of each mode are aligned vertically, along the cross-dispersion direction, while all A and all B fibers are arranged side-by-side horizontally along the main-dispersion direction. The three separate exposure shutters for the three modes ensure that only one mode at the time feeds the spectrograph. The fiber-link design has been based on the concepts originally developped for HARPS and HARPS-N \citep{cosentino:2012,locurto:2015} and was adapted to ESPRESSO with its various observing and instrument modes \citep{megevand:2014}.

\subsubsection{The spectrograph} 
The light of the fibers entering the spectrograph is dispersed and the spectra imaged through two different camera arms, a red and a blue one, onto two distinct detectors. First, an anamorphic pupil slicing unit (APSU, \citet{riva:2014-1} and \citet{oggioni:2016}) shapes the beam in order to reduce the size of the spectrograph optics; without the APSU, the size of the Echelle grating would have to be $60\times240$\,cm to achieve the desired spectral resolution. Instead, the APSU slices and superimposes the so produced half-pupils in the main dispersion direction, while in the other direction it anamorphically compresses the beam by a factor 3. The collimator, an off-axis parabola used in double pass, collimates the beam to a rectangular shape of $20\times30$\,cm that matches the $20\times120$\,cm physical surface of the $76^\circ$ (R4) Echelle mosaic \citep{lizon:2018}. A dichroic beamsplitter placed before the intermediate spectrum sends the light to the blue and the red channels. For each channel a specific (spherical) transfer collimator re-images the white pupil of the Echelle grating onto the cross-dispersers, a Volume-Phase Holographic (VPH) grating \citep{arns:2016}. Two camera lenses eventually separately focus the blue and the red portion of cross-dispersed echelle spectra onto a large CCD detector each.

Compared to a classical UVES-like white pupil design, ESPRESSO uses several design innovations: 1) By adjusting the power of the red field lens and the blue field mirror at the intermediate focal planes (white spectrum), the field curvature is compensated, thus avoiding the need for a cylindrical field lens in front of the detector. 2) The transfer collimator is a tilted sphere with a focal length much shorter than the main collimator; this significantly reduces the beam size on the cross-dispersers (asymmetric white pupil). 3) By adjusting the tilt angle of the spherical mirrors and deliberately operating the VPH crossdisperser in a non-collimated beam, aberrations introduced by the main collimator are compensated, except for residual spherical aberration that are corrected for in the camera lenses. 4) A prism cemented to the VPH lifts the residual 1:1.5 anamorphism and returns a square beam. Furthermore, it compresses by the same factor 1.5 the spectral orders in cross-dispersion direction, reducing spatial sampling and needed CCD space. 5) The camera lenses consist of only three lenses of inexpensive glass each, the last being used as the cryostat window. In each camera, the residual longitudinal chromatism is compensated by titling the CCD chip.

\begin{figure}[h]
   \centering
   \includegraphics[width=\hsize]{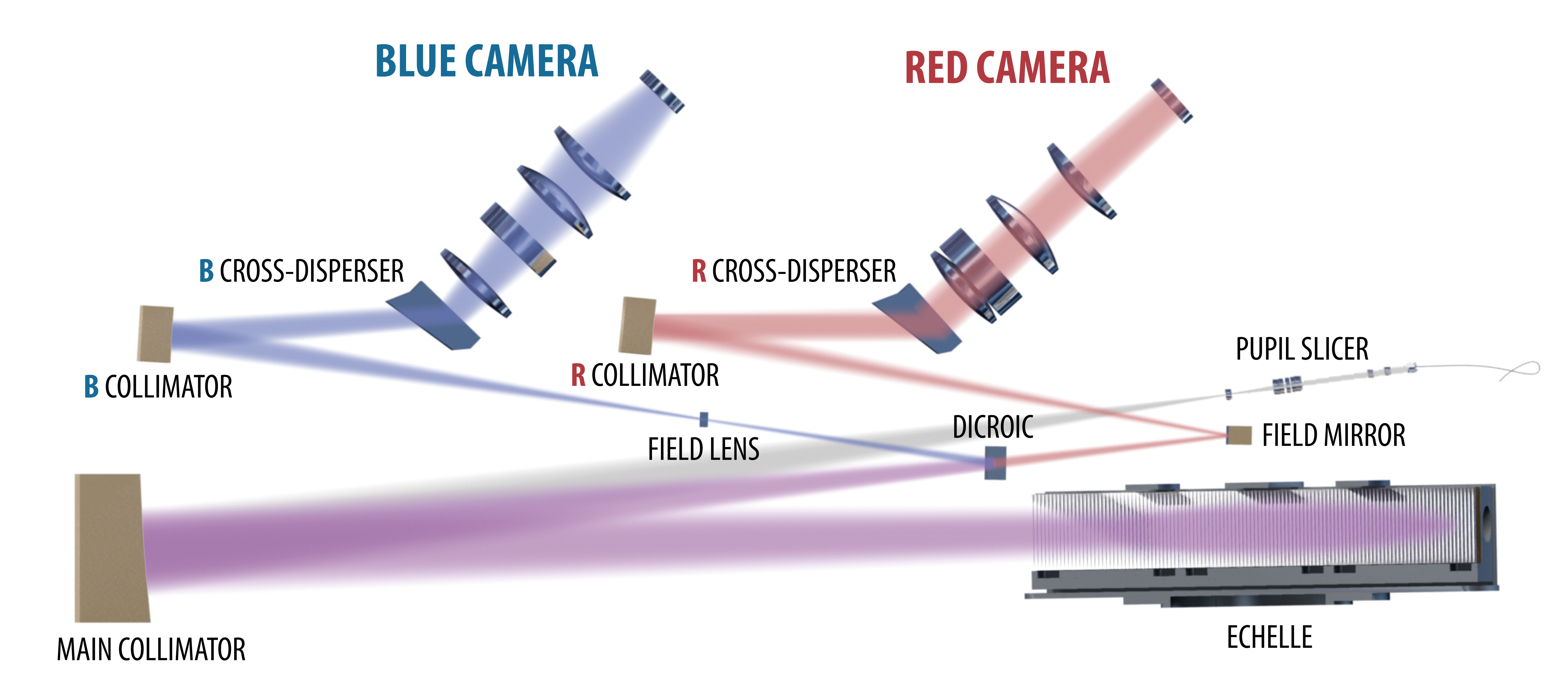}
   \caption{Schematic view of the optical design and layout of the spectrograph and its components. Figure from \citet{gonzalez:2018}, courtesy of Gonzalez-Hernandez.}
   \label{fig:spectrograph}
    \end{figure}

Two individual E2V CCDs of 9K$\times$9K and 10\,$\mu$m-square pixels record the blue and the red spectra separately. The CCDs are different in thickness and were optimized in sensitivity separately for the blue and the red channels. The two chips are cooled and stabilized in temperature to $160\pm 0.01$\,K by two individual Continuous-Flow Cryostats \citep{lizon:2016} providing the detectors with separate high vacuum within the large ESPRESSO detector vessel. The CCDs are controlled and read-out by ESO's New-Generation Controllers (NGC) that provide selectable read-out speed, gain and binning. Special attention was paid to equalize the power consumption of the large CCD detectors during the different phases of idling, integrating and reading also through dummy clocking and dithering (Iwert et al., internal review documentation), such to avoid ``thermal breathing'' of the pixel grid, as seen in the HARPS detector mosaic. A schematic view of the spectrograph and its optical design is shown in Figure\,\ref{fig:spectrograph}.

\subsubsection{Spectral format} 
Given the use of a pupil slicer, the spectrograph delivers two simultaneous spectra of the light from the two fibers in a wavelength range between 378.2 and 788.7\,nm and at a resolving power reaching $R>$190\,000 in the Ultra-High-Resolution mode (UHR) and $R \sim$ 140\,000 in the High-Resolution mode (HR), the two instrument modes available in the 1-UT configuration. In the 4-UT configuration, also called Medium-Resolution mode (MR) that combines the light collected by all the UTs, the same wavelength range is covered, but the increased Etendue of the telescope comes at the price of a larger input fiber and thus a reduced resolving power of $R \sim$70\,000. This value is nevertheless remarkable and exceptional, considering that ESPRESSO is the first optical spectrograph working with a 16-m diameter equivalent telescope.

\begin{figure}[h]
   \centering
   \includegraphics[width=\hsize]{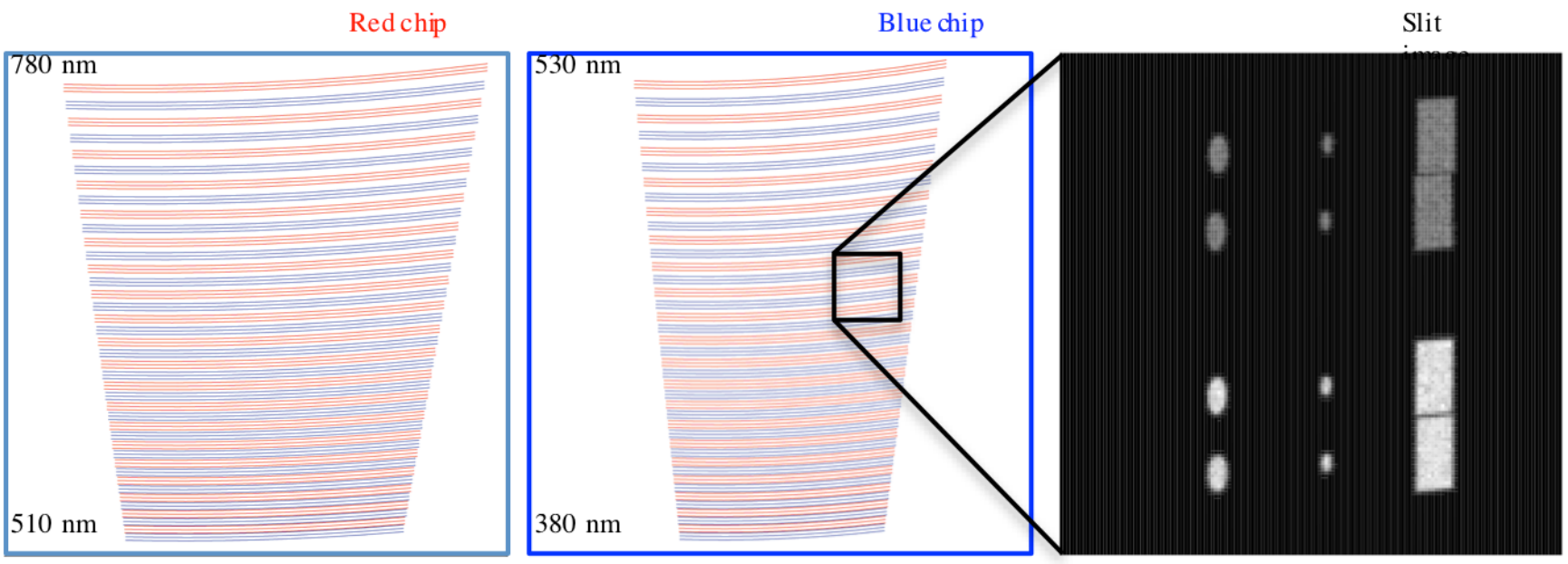}
   \caption{Left: Spectral format (left) of the spectrum on the two detectors (90$\times$90\,mm) showing the traces of the various orders over one FSR. To better distinguish neighboring orders, even and odd order numbers have been drawn in blue and red, respectively. Each order is represented by three traces indicating the center of the order in cross-dispersion direction (vertical), as well as the upper and lower boundary of the full field produced by the pseudo-slit. Right: Zoom on the virtual 'pseudo-slit' showing the images of the various fibers projected on the CCDs as if they were simultaneously illuminated by a monochromatic light source. The image covers on the detector a square of approximately 120\,pixels (1.2\,mm) in size. From left to right the images of the HR, UHR and MR fibers are shown. For each mode the lower two images arise from the two slices of fiber A (target) and the upper two from fiber B (reference). In the actual application, only one mode is illuminated at the time, leading to four images (2\,fiber images $\times$ 2\,slices) aligned along the vertical direction (cross-dispersion). The total height of the image determines the width of the echelle order and thus the required cross-dispersion to avoid overlap of neighboring orders in the blue end of the spectral range. Figure from \citet{pepe:2014}.}
   \label{fig:spectral-format}
    \end{figure}

Figure\,\ref{fig:spectral-format} shows the spectral formats of the blue and red arms, respectively. The traces represent the individual echelle orders over one free-spectral range (FSR). The right-hand panels show a zoom on the 'pseudo-slit' (combination of all fibers entering the spectrograph) as being re-imaged on the CCDs. A tabulated extract of wavelengths and spectral bin values is given in Table\,\ref{tab:format}. The complete and detailed description of the spectral format can be obtained through the official Exposure Time Calculator (ETC)\footnote{\url{www.eso.org/observing/etc/bin/gen/form?INS.NAME=ESPRESSO+INS.MODE=spectro}}.

\begin{table*}
\caption{ESPRESSO's spectral format. For some representative diffraction order of the echelle spectrograph we provide central wavelength (CW) in the middle of the order, start and end wavelength of the FSR, as well as the start and end wavelength of the total (extracted) spectral range (TSR). Also, the spectral size of a physical detector pixel (resolution bin) located in the center of the order is given.}             
\label{tab:format}      
\centering                      
    \begin{tabular}{l l l l l l l l l}
    \hline\hline
        PDO \# & CW & FSR & FSR start & FSR end & Order start & Order End & TSR & Res. bin) \\
         & [nm] & [nm] & [nm] & [nm] & [nm] & [nm] & [nm] & [nm] \\
         \hline
        \multicolumn{9}{c}{red end of red arm}\\
        78 & 784.45 & 10.06 & 779.45 & 789.51 & 778.98 & 790.64 & 11.66 & 0.00129 \\
        79 & 774.52 & 9.80 & 769.65 & 779.45 & 769.11 & 780.65 & 11.54 & 0.00128 \\ 
         \multicolumn{9}{c}{...}\\
         87 & 703.30 & 8.08 & 699.28 & 707.37 & 698.34 & 708.89 & 10.56 & 0.00116 \\ 
        88 & 695.31 & 7.90 & 691.38 & 699.28 & 690.40 & 700.84 & 10.45 & 0.00115 \\
         \multicolumn{9}{c}{...}\\
        101 & 605.81 & 6.00 & 602.83 & 608.83 & 601.50 & 610.66 & 9.16 & 0.00100 \\
        102 & 599.87 & 5.88 & 596.95 & 602.83 & 595.60 & 604.67 & 9.07 & 0.00099 \\
         \multicolumn{9}{c}{...}\\
         111 & 551.24 & 4.97 & 548.76 & 553.73 & 547.30 & 555.65 & 8.36 & 0.00091 \\ 
        112 & 546.31 & 4.88 & 543.89 & 548.76 & 542.41 & 550.69 & 8.29 & 0.00090 \\ 
         \multicolumn{9}{c}{...}\\
         116 & 527.48 & 4.55 & 525.21 & 529.76 & 523.70 & 531.71 & 8.01 & 0.00087 \\ 
        117 & 522.97 & 4.47 & 520.74 & 525.21 & 519.22 & 527.16 & 7.94 & 0.00086 \\
         \multicolumn{9}{c}{blue end of red arm}\\
         \hline
        \multicolumn{9}{c}{red end of blue arm}\\
        117 & 522.97 & 4.47 & 520.74 & 525.21 & 519.13 & 527.03 & 7.89 & 0.00086 \\ 
        118 & 518.54 & 4.39 & 516.35 & 520.74 & 514.72 & 522.57 & 7.84 & 0.00086 \\
         \multicolumn{9}{c}{...}\\
        135 & 453.24 & 3.36 & 451.57 & 454.92 & 449.83 & 456.82 & 6.99 & 0.00075 \\ 
        136 & 449.91 & 3.31 & 448.26 & 451.57 & 446.52 & 453.46 & 6.94 & 0.00074 \\
         \multicolumn{9}{c}{...}\\
        152 & 402.55 & 2.65 & 401.23 & 403.88 & 399.48 & 405.75 & 6.26 & 0.00067 \\ 
        153 & 399.92 & 2.61 & 398.61 & 401.23 & 396.87 & 403.10 & 6.22 & 0.00066 \\
         \multicolumn{9}{c}{...}\\
        160 & 382.42 & 2.39 & 381.23 & 383.62 & 379.50 & 385.47 & 5.96 & 0.00063 \\
        161 & 380.04 & 2.36 & 378.87 & 381.23 & 377.15 & 383.07 & 5.93 & 0.00063 \\
         \multicolumn{9}{c}{blue end of blue arm}\\
\hline\hline
\multicolumn{9}{l}{PDO Physical Diffraction Order; CW Central Wavelength; [F/T]SR [Free/Total]-Spectral Range} \\
\end{tabular}
\end{table*}

\subsubsection{Exposure meter} 
In order to track the flux entering the spectrograph in real time, and thus monitor the instantaneous efficiency, the part of the spectrograph beam that would fall outside the clear aperture of the spectrograph is collected within the APSU by a flat mirror with a central hole (pupil stop for the spectrograph). The light beam is focused by a lens onto the tip of two other fibers, one per channel, that transport the light to the outside of the spectrograph and inject it in the so-called exposure meter. The exposure meter produces a low-resolution spectrum and records the light entering the spectrograph in a blue, green and red channels for the two spectrograph-fiber channels independently. The exposure meter not only allows us to monitor the efficiency of the instrument but also to record the flux of the astronomical target such to compute precisely the flux-weighted mean time of exposure ($MTE$), that is in turn used to determine the precise value of the Earth's barycentric velocity along the line of sight for which the recorded spectrum must be corrected for \citep{landoni:2014}.

\subsubsection{Ensuring instrumental stability} 
The spectrograph is enclosed into a vacuum vessel that is in turn located inside a thermally controlled room. This environment offers stable conditions in terms of temperature and air pressure, minimizing its impact on the stability of the instrument \citep{alvarez:2018}. The whole spectrograph mechanics is designed and manufactured for best-possible metrological stability \citep{tenegi:2016,santana:2014}. Materials are matched as much as possible for the thermal expansions coefficients, the setup is symmetric relative to a vertical plane (and thus gravity), no alignment mechanisms are available (alignment only by manufacturing and shimming), and no movable components exist inside the vacuum vessel; in other words, nothing can move on the entire light path after the optical fibers that feed the spectrograph. As for the FE, a fully mechanical alignment technique was developed and adopted during integration in Europe and Chile \citep{pariani:2018,genoni:2016}.

ESPRESSO is designed for an extreme radial-velocity precision of 10\cms, which requires a perfect control of the metrological stability of the spectrograph. This extreme stability, which corresponds to a physical shift of the spectrum in the focal plane of the spectrograph by a few nanometers, cannot be achieved passively and the physical parameters ruling it must be permanently controlled, or at least monitored. ESPRESSO applies therefore the 'simultaneous-reference' technique \citep{mayor:2003,pepe:2002}, where the second fiber B can be fed with a spectral reference instead of the sky light, and is used to track possible residual instrumental drifts down to the requested RV precision level. The simultaneous reference is only needed when the (photonic) radial-velocity error achieved on the scientific target is expected to be comparable to or lower than the typical  instrumental drifts of the order of $~$\ms. Consequently, the simultaneous-reference technique is applied for bright objects and when high radial-velocity precision is aimed at. For faint targets, the second fiber is fed with sky light, such that sky background can be measured and subtracted.

It must be noted however that the aimed (long-term) RV-repeatability of 10\cms eventually relies on the concept of repeatable wavelength calibration, and thus on the availability of an absolute spectral reference of wide spectral range. This spectral reference was supposed to be provided by the LFC, which however was still not operational by the time of writing this paper. For these reasons, the instrument was adapted to the combined use of a ThAr lamp and the Fabry-Pérot etalon, providing accuracy and precision, respectively, to a satisfactory level. 


   

\subsection{Observing with ESPRESSO}
\label{sec:operations}

\subsubsection{Operations}   
The operational flexibility of being able to feed light of any UT into ESPRESSO is used to maximize the overall scientific output of both the instrument and the observatory. In spite of the large number of subsystems and optical components involved, ESPRESSO observations can be triggered and started in any UT in less than ten minutes. The operators can thus optimize telescope time by selecting the UT where the observing pressure is the lowest at a given moment. The information of the used telescope, along with several environmental parameters, are stored in the night reports and in the FITS headers of the collected data.

\begin{figure}[h]
   \centering
   \includegraphics[width=\hsize]{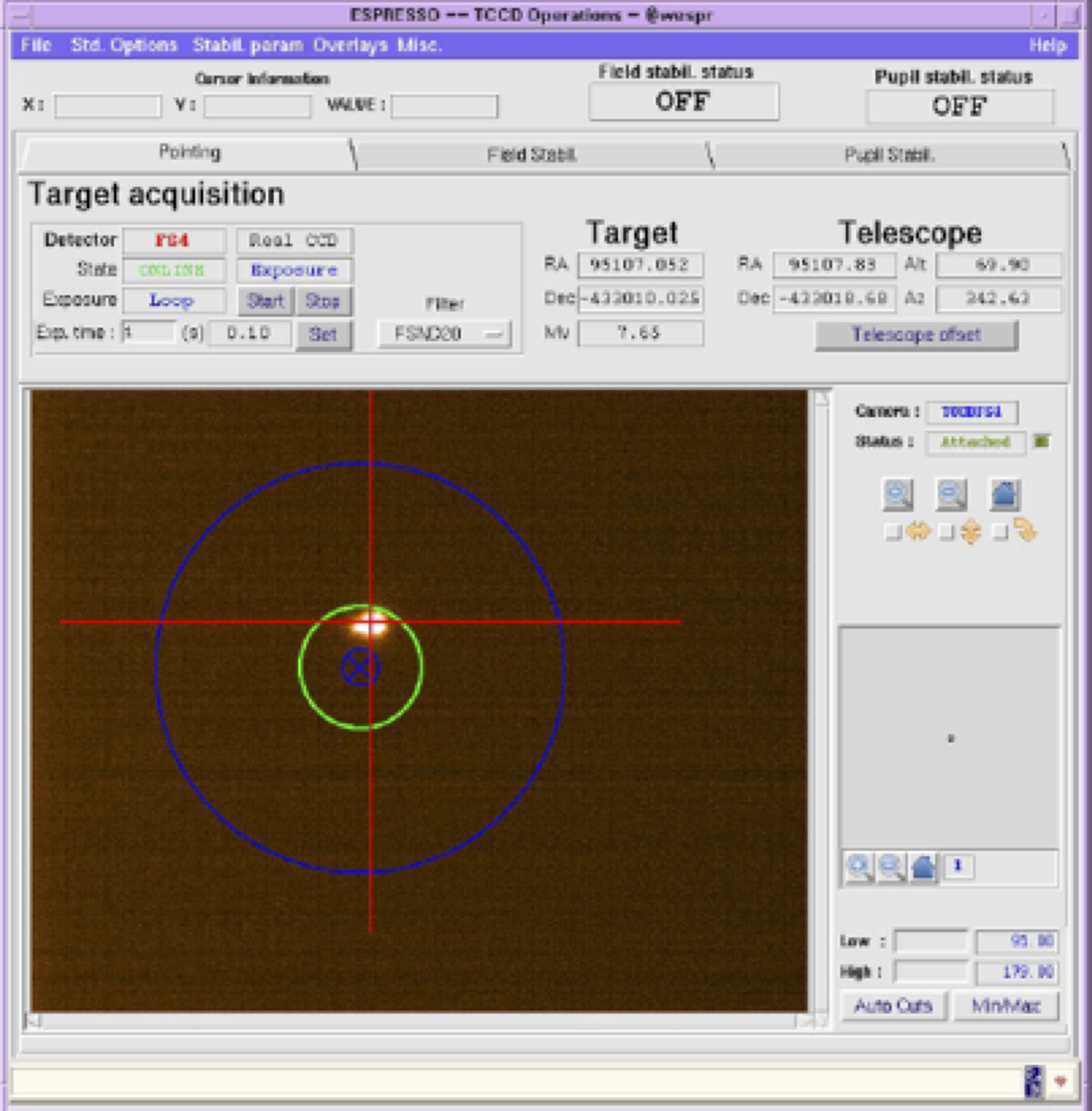}
   \includegraphics[width=\hsize]{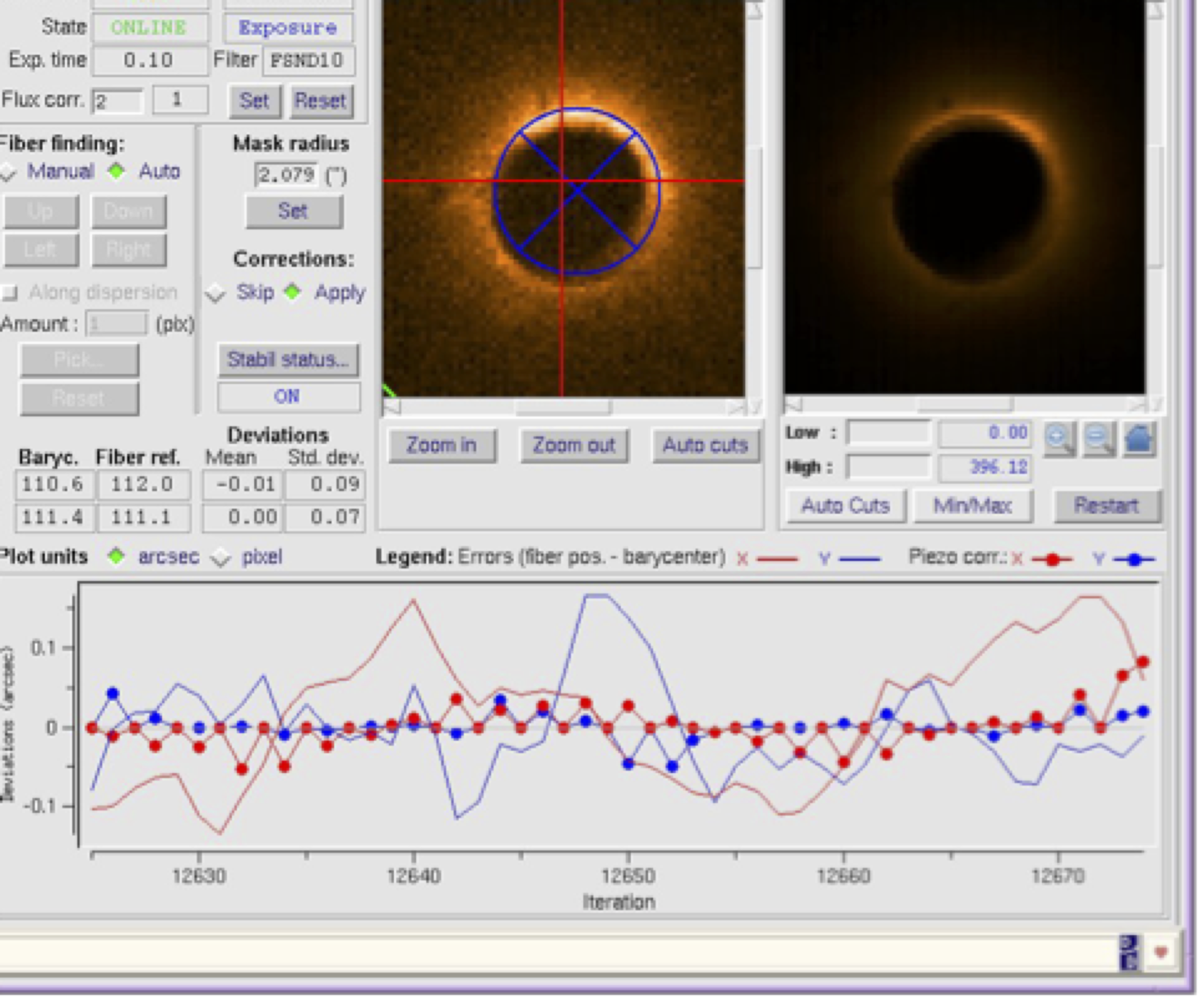}
   \caption{Acquisition field (top) and secondary-guiding panel in the 1-UT configuration during field stabilisation (target centered on the fiber).}%
   \label{fig:acquisition}
    \end{figure}

An observation sequence consists of an {\it acquisition} and an {\it observation} part. The {\it acquisition} defines the observing mode and configures the hardware accordingly. The control software sets up the configuration of the chosen instrument mode, i.e. selection of the fiber head, exposure type, binning pattern, read-out speed and simultaneous reference. Meanwhile, the telescope is moved into position, the target acquired and positioned over the fiber (see the acquisition panel in Figure\,\ref{fig:acquisition}, top, and the field image in Figure\,\ref{fig:field}). At the end of the acquisition, and during the full length of the observation, a secondary field-stabilization loop is activated, keeping the target well centered in the fiber by moving the tip-tilt mirror \citep{landoni:2016}. We refer to Figure\,\ref{fig:acquisition}, bottom, for a view of the secondary-guiding panel. The maximum correction frequency is of 10\,Hz but the actual speed depends on the magnitude of the target. The secondary guiding system also provides pupil stabilisation. However, this correction in not performed during observations given the excellent stability of the pupil. The {\it observation} consists of one or more exposures (integrations) on the same target, which will end by the read-out and the storage of the raw individual frames. The observation is assisted by the Exposure Meter that measures the flux entering the fiber in real time and computes the flux-weighted $MTE$.

\begin{figure}[h]
   \centering
   \includegraphics[width=\hsize]{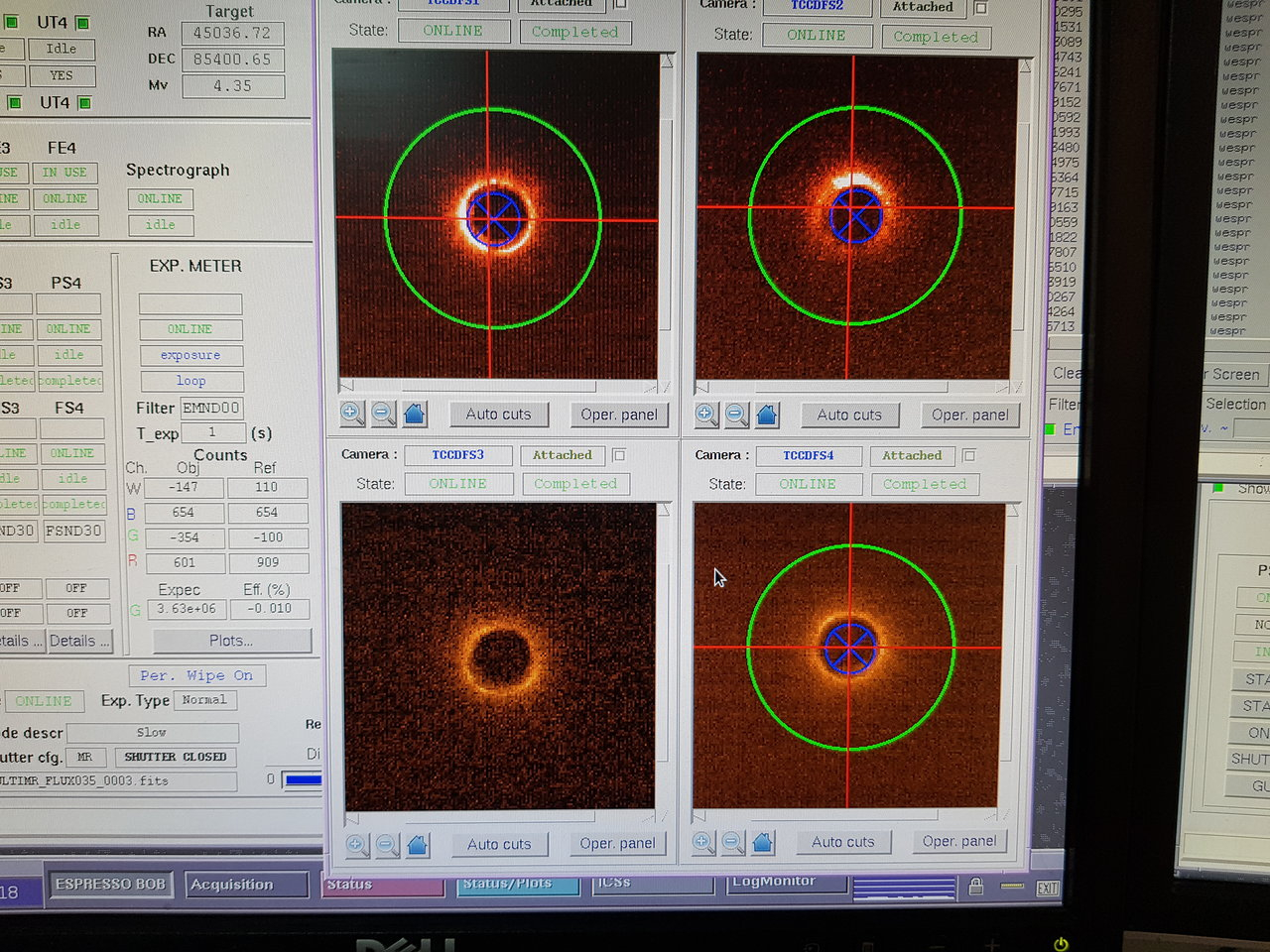}
   \caption{Screenshot of the field stabilisation panel in 4-UT configuration}%
   \label{fig:4utstabilisation}
    \end{figure}

Operation in 4-UT configuration is very similar to that in 1-UT configuration. While in 1-UT configuration the observing software dispatches instructions to the selected UT and ESPRESSO, in the 4-UT configuration it sends instructions to the 4 UTs plus the instrument simultaneously. Presetting and target acquisition is done on all the four UTs \textit{simultaneously}. After this step, one needs to center the target of every UT on the specific fiber located inside the corresponding FE unit. This step is performed \textit{sequentially}, moving the star's position on the focal plane of the fiber by offsetting the telescope. Once the target is centered and the secondary guiding activated and stable for every telescope, the exposure is started. During the exposure, all the four FE units, including the respective ADCs and secondary guiding, are active and independent. A screenshot of the field-stabilisation panel in the 4-UT configuration is shown in Figure\,\ref{fig:4utstabilisation}.

\begin{figure}[h]
   \centering
   \includegraphics[width=\hsize]{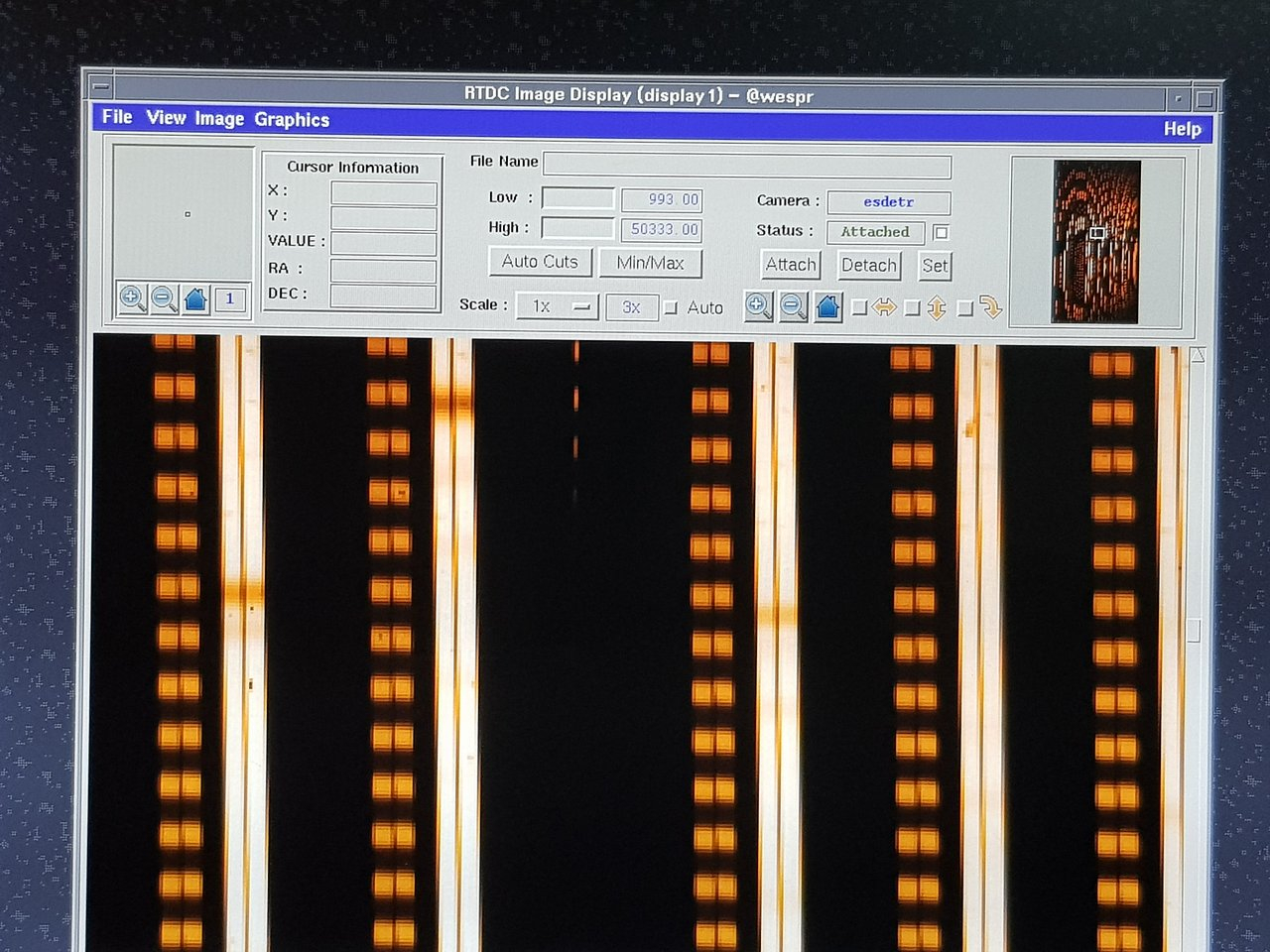}
   \caption{Real-time display of a raw spectrum. A small portion of a star + Fabry-P\'erot spectrum acquired in 4-UT configuration is shown. Main dispersion is vertical. Each order consists of two spectral traces (slices) of the Fabry-P\'erot with its characteristics equidistant spectral lines, and two traces of the stellar spectrum plotting essentially the continuum with a few absorption features. This figure is best understood when compared to Figure\,\ref{fig:spectral-format} when tilted by 90$^\circ$. }
   \label{fig:raw_spectrum}
    \end{figure}

Within seconds after the end of the exposure, the raw data are acquired and shown by ESO's real-time display tool RTD on the instrument machine (Figure\,\ref{fig:raw_spectrum}). The raw data are available for download through the ESO archive\footnote{\url{www.eso.org/archive/}}. The ESPRESSO pipeline installed in Paranal reduces automatically the acquired data to provide a quick-look and a general check of the basic properties of the data and operational status of the instrument. For updated information on how to use ESPRESSO and latest news on the instrument, the interested reader is referred to the \textit{ESPRESSO User Manual}\footnote{\url{www.eso.org/sci/facilities/paranal/instruments/espresso/doc.html}} and dedicated \textit{News} pages\footnote{\url{www.eso.org/sci/facilities/paranal/instruments/espresso/news.html}} on ESO's website.

\subsubsection{Optimizing observations}
The main instrument characteristics in the various observing configurations and instrument modes are summarized in Table\,\ref{tab:modes}. An ESPRESSO user can choose among two observing modes: the 1-UT configuration (one single telescope, any of the UT) or the 4-UT configuration\footnote{Combinations with two or three telescopes (degraded mode) are technically possible but not formally offered.} (all UTs simultaneously). In the 1-UT configuration four different instrument modes exist. In HR mode the light is collected using (octagonal) fibers of projected sky diameter of 1\arcsec. This mode is intended for spectroscopy and RV monitoring. Three different detector read-out modes offer optimized solutions depending on the target magnitude. In case of bright targets (HR11, 1$\times$1 binning), the detector will be saturated quickly and the exposure time has therefore to be short. Overheads are reduced by a faster read-out and the dynamic range is maximized by avoiding binning. Given the high signal level expected for a bright star, the higher read-out noise (RON) is acceptable. For faint or very-faint objects, instead, the HR21 (2$\times$1 binning) or HR42 (4$\times$2 binning) should be used. The difference between these two modes is that in the HR42 mode the total RON per spectral bin is further reduced and the limiting magnitude, at which the noise starts to be dominated by the detector RON, improved. This gain comes at the cost of a slightly reduced resolving power, since the resulting pixel sampling gets close to the Nyquist limit. This is especially the case on the blue side of the longest (reddest) echelle orders, where the anamorphic de-magnification reduces the size of the projected slit/fiber and thus the sampling. At the very blue end of the reddest echelle order, the sampling of the HR42 mode (worst case) is reduced by 30\%, i.e. to about 1.7 pixels. The UHR11 mode, finally, is the one delivering the highest resolving power, but the required use of a 0.5\arcsec diameter fiber leads to reduced spectral sampling and higher slit losses. For this latter reason, this mode is particularly suitable for ultra-high resolution observations on relatively bright targets.

It must be noted that the pixel sampling given in Table\,\ref{tab:modes} for the individual modes refers to the value expected at the center of an extracted echelle order. This value remains constant \textit{across} all the orders and over both detectors, so does also the resolution bin (pixel) size expressed in \ms, which results from a typical feature of cross-dispersed echelle spectrographs. Another very elegant characteristic of echelle spectrographs, when used in Littrow condition, is that the resolution bin size (again expressed in \ms or $\Delta\lambda/\lambda$) remains essentially constant \textit{along} the echelle order, since the anamorphic effect on the projected slit/fiber size and the variation of linear dispersion compensate almost perfectly. Finally, our design choice of oversizing the detector in order to extract at least 20\% more than the FSR of the longest (reddest) orders (see Figure\,\ref{fig:spectral-format}) leads to the third elegant feature of echelle spectrographs: By summing up the flux of a given wavelength recorded at the edges of two subsequent echelle orders one recovers almost the same number of photo-electrons as for wavelengths located close to the blaze (center) of the echelle order. This means that after extracting and merging the orders, the detection efficiency obtained at a given wavelength is close to constant. For all these reasons we will describe hereafter the measured signal (and S/N) as the value obtained at the center of the order and for a given extracted pixel (resolution bin), under the motivated assumption that this value varies only by a small amount along the echelle order.

By design, the 4-UT observing configuration is intended for spectroscopy of very faint targets. The lower resolution is a price to pay for the usage of an equivalent wider (square) fiber in the spectrograph that can carry the light of the four telescopes. Collecting the light from the four telescopes reduces the contribution of detector readout and (partially) dark noise when adequate slit configurations and binning levels are adopted. The two instrument modes MR42 (4$\times$2 binning) and MR84 (8$\times$4 binning) enable therefore very-low S/N spectroscopy and the study of fast transient events.

\begin{table*}
\caption{ESPRESSO's instrument modes and characteristics}             
\label{tab:modes}      
\centering          
\begin{tabular}{l c c c c c c}    

\hline\hline
Observing configuration & \multicolumn{4}{c}{1-UT} & \multicolumn{2}{c}{4-UT}\\
Instrument mode & UHR11 & HR11 & HR21 & HR42 & MR42 &  MR84 \\
\hline                    
Simultaneous reference & \multicolumn{6}{c}{Available (no simultaneous sky background on fiber B)}\\
Sky subtraction & \multicolumn{6}{c}{Available (no simultaneous reference on fiber B)}\\
Wavelength range  [nm]& \multicolumn{6}{c}{378.2--788.7}\\
Median resolving power $R=\lambda/\Delta\lambda$ & $>$190\,000\tablefootmark{a} & 138\,000 & 138\,000 & 130\,000 &  72\,500 & 70\,000\\
Aperture on sky [\arcsec] & 0.5 dia. & 1.0 dia. & 1.0 dia.  & 1.0 dia. &  4$\times$1.0 dia. & 4$\times$1.0 dia.\\
Spectral sampling [physical pixels] & 2.5 & 4.5 & 4.5 & 4.5 & 10 & 10 \\
Spatial sampling [slices$\times$physical pixels] & 2$\times$5  & 2$\times$9 & 2$\times$9 & 2$\times$9 & 2$\times$20 & 2$\times$20\\
Detector read-out speed [kpix\,s$^{-1}$]& 300 & 300  & 100 & 100 & 100 & 100 \\
Detector binning, spatial$\times$spectral [pixels]& 1$\times$1  & 1$\times$1  & 2$\times$1 & 4$\times$2  & 4$\times$2 & 8$\times$4\\
Effective spectral sampling [extracted pixels] & 2.5 & 4.5 & 4.5 & 2.25 & 5 & 2.5 \\
Detector read-out noise (RON) red/blue\tablefootmark{b} [e$^-$]& 5/8  & 5/8  & 2/3  & 2/4  & 2/4 & 2/5\\
Detector conversion factor red/blue [e$^-$/ADU] & 1.15/1.1  & 1.15/1.1  & 1.13/1.09 & 1.13/1.09  & 1.13/1.09 & 1.13/1.09\\
Total peak efficiency & 5\% & 10\% & 10\% & 10\% & 10\% & 10\% \\
\hline\hline
\multicolumn{7}{l}{$^{(a)}$To be considered as a lower limit, since many thorium lines used for the analysis are partially resolved at this resolution.}\\
\multicolumn{7}{l}{$^{(b)}$The 'FAST' read-out modes (HR11 and UHR11) of the blue detector are still undergoing optimisation.}
\end{tabular}
\end{table*}

The observations are prepared using the usual {\tt P2} tool\footnote{\url{www.eso.org/sci/observing/phase2/p2intro.html}}. On top of allowing the user to select from all the possible instrument configurations and defining observational constraints, it provides advanced functionalities such as target data retrieval from SIMBAD database, definition of ``blind offset'' from nearby stars, and monitoring of the field scanned by the reference fiber when in simultaneous-sky mode. For the computation of the exposure time and the expected S/N and RV precision, the user is referred to the ESPRESSO Exposure Time Calculator (ETC)\footnote{\url{www.eso.org/observing/etc/bin/gen/form?INS.NAME=ESPRESSO+INS.MODE=spectro}}. We shall nevertheless provide a benchmark: In a 1\,Min. integration in HR11 on a K2 star with $V=8$\,mag observed at an airmass of 1 and under median seeing conditions of 0.8\arcsec, an $S/N\sim115$ at blaze and per extracted pixel (or resolution bin) of 0.486\kms~is reached. This corresponds to a gain of about two magnitudes with respect to HARPS at the 3.6-m telescope. By using the relation between S/N and RV precision derived for non-rotating G and K-dwarfs, one estimates that this S/N delivers a photon-noise limited RV precision of about 50\cms. We refer to Section\,\ref{sec:snr} for further details. \\

An option available to the observer is the use fiber\,B. ESPRESSO provides two mutually exclusive possibilities: either to illuminate it with a simultaneous spectral reference to measure the instruments internal drift, or to use it to measure independently the sky background. In the first option, the secondary fiber is illuminated by the Fabry-P\'erot light. The many spectral lines are used to measure and correct for the minute-scale motions of the spectra on the CCDs induced by the short-term variations of pressure and temperature inside the spectrograph and the detector. This option becomes important when one wants to achieve an RV precision similar or better than the intrinsic stability of the instrument over the timescale of a night of about 1\ms \textit{rms}. If one aims at on-target RV photon-noise precision better than this value, then it is recommended to use the simultaneous reference. Conversely, if this precision is not required or one wants to perform any kind of measurement for which sub-\ms precision is not needed (e.g. transmission spectroscopy), then the use of the simultaneous reference is not necessary. Actually, in this latter case, and when high spectral fidelity observations are aimed, one needs rather to consider the contribution of the sky background. During solar activity minima and in a region of the sky devoid of starlight or moonlight, the V-band magnitude of sky spectra as seen through a 1\arcsec fiber is still of $V=20-21$\,mag. In presence of full Moon and thin clouds, the sky brightness can increase by many magnitudes. If we consider this, then it becomes clear that moon-light contamination may become an issue for precise RV measurements even at magnitudes as low as $V\le 11$\,mag (see e.g. \citet{cunha:2013} or \citet{arpita:2020}). It is therefore recommended to use fiber B to measure the sky background for `faint' targets. The numbers and magnitudes provided here are only for general guideline. A more detailed optimisation should be made for each program as a function of the specific science case.

\subsection{Data Flow}
\label{sec:dfs}
\subsubsection{Concept and elements}
For instruments installed at the VLT, the concept of Data Flow System (DFS) denotes a closed-loop software system that incorporates various software packages able to track the flow of scientific data with the highest possible level of quality. Its logical path follows the typical life-cycle of an astronomical observation: from the proposal preparation and handling, through observation preparation, instrument operation and control, to science data archive, reduction and quality control. Compared with other spectrographs hosted by ESO, ESPRESSO's DFS has some peculiarities.

\begin{figure}[h]
   \centering
   \includegraphics[width=\hsize, trim= 150 80 150 80, clip]{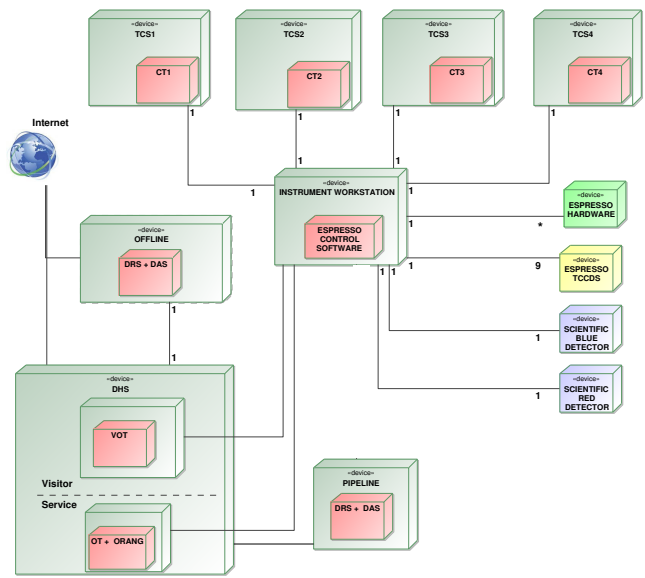}
   \caption{ESPRESSO final DFS deployment at Paranal. Figure from \citet{pepe:2014} and ESPRESSO-Project documentation.}%
   \label{fig:dfs}
    \end{figure}

The ESPRESSO DFS concept \citep{DiMarcantonio18} has been conceived since its preliminary design phases with the goal of maximizing operational efficiency, flexibility, and scientific output, while complying with the standard Paranal Observatory operational scheme. The main challenge derives from the requirement to operate ESPRESSO in a seamless way with any of the UT's or with all four UT's simultaneously. This must be possible not only with a predetermined schedule, but also ``on-the-fly''. The flexibility in ESPRESSO's operations has been tackled by adopting a new DFS deployment plan described in \citet{DiMarcantonio18} that is exceptional under various aspects because it has to cope with various telescope and instrument configurations while remaining operationally simple. Figure\,\ref{fig:dfs} shows  the  main  ESPRESSO  DFS elements and their  final  deployment. Besides the software packages already described, part of the software for the control of the CT devices has been incorporated in the VLT Telescope Control software to allow CT operations even if ESPRESSO is offline (thus avoiding conflicts, e.g., with VLTI operations).In addition to the standard DFS software packages, ESPRESSO is the first instrument to provide also a data analysis package able to extract relevant astronomical observables from the reduced data. The following DFS subsystems are specific to ESPRESSO:
\begin{itemize}
    \item the ETC hosted on ESO's web page\footnote{\url{www.eso.org/observing/etc/bin/gen/form?INS.NAME=ESPRESSO+INS.MODE=spectro}},
    \item the Control Software (CS) with the full suite of acquisition, observation and calibration templates able to control all vital parts of the instrument and the Coud\'e Train \citep{Calderone18},
    \item the Data Reduction Software (DRS) package (or ``pipeline'') capable of providing `science-ready', reduced data only minutes after the end of the individual observation,
    \item the Data Analysis Software (DAS) package that produces higher-level astronomical observables with no or limited supervision.
    \item DRS and DAS are distributed to the community.\footnote{\url{www.eso.org/sci/software/pipelines/index.html}}
\end{itemize}

DRS and DAS packages are delivered in an integrated way even though both can be employed independently. In standard operation, the DRS will be automatically triggered once raw FITS files are produced by the ESPRESSO Control Software (CS). In visitor mode, both the DRS and DAS can be used offline and stand-alone.

\subsubsection{Data Reduction Software}
\label{sec:drs}
The ESPRESSO Data Reduction Software (DRS), referred to as "pipeline", is a scientific software package whose purpose is to process the raw FITS frames generated by the instrument into reduced data products of various kinds, including science-grade high-resolution spectra of any astrophysical sources observed by ESPRESSO. An in-depth description of the ESPRESSO DRS is beyond the scope of this paper and will be given elsewhere (Lovis et al. in prep.). We limit ourselves here to a general overview.

The ESPRESSO DRS builds on a vast heritage of data reduction techniques and software. On the scientific side, it represents the next generation of a series of pipelines that started with the ELODIE spectrograph in the early 1990s \citep{baranne:1996}, then followed by CORALIE \citep{queloz:2000,pepe:2002}, HARPS \citep{mayor:2003}, and finally HARPS-N \citep{cosentino:2012}. Many of the data reduction steps and algorithms implemented in the ESPRESSO pipeline find their origins in the HARPS pipeline. However, several steps were significantly re-designed and improved with respect to HARPS to meet the more challenging requirements of ESPRESSO.

On the implementation side, the ESPRESSO pipeline is based on a software framework imposed by ESO to all VLT instrument pipelines, namely the Common Pipeline Library (CPL), a C-based toolkit offering basic functionalities for astronomical data manipulation and reduction\footnote{\url{www.eso.org/sci/software/cpl/index.html}}. The pipeline is structured as a set of top-level recipes, written in C, which handle the various raw data types produced by the instrument (all types of calibration and science frames). DRS recipes can be called either through the EsoRex command-line interface or through the EsoReflex GUI workflow environment, both developed and maintained by ESO. The ESPRESSO pipeline, including its source code and the DRS manual, is publicly available from the ESO pipeline repository\footnote{\url{www.eso.org/sci/software/pipelines/espresso/espresso-pipe-recipes.html}}.

Data processing is organized as a sequential reduction cascade in which each raw data type is associated to a dedicated DRS recipe. The pipeline generates intermediate data products at each stage, many of which are then used as inputs to the subsequent steps of the cascade. The final task is the reduction of science spectra. The main steps of the science processing are:

\begin{itemize}

\item Bias and dark subtraction
\item Subtraction of inter-order background
\item Optimal extraction of spectral orders using master flat-field as order profile, with masking of dead, hot and saturated pixels, and rejection of cosmic rays
\item Creation of extracted spectra (S2D) in (order, pixel) space with associated error and quality maps
\item Flat-fielding and de-blazing of S2D spectra
\item Wavelength calibration of S2D spectra using either a combination of ThAr and Fabry-Perot sources or the Laser Frequency Comb 
\item Computation of barycentric correction and shift of wavelength solutions to the Solar System barycenter
\item Instrumental drift computation and correction using fiber B (simultaneous reference mode)
\item Extraction of sky spectrum and creation of sky-subtracted S2D spectra using fiber B (simultaneous sky mode)
\item Creation of rebinned and merged 1D spectra (S1D) with associated error and quality maps
\item Creation of flux-calibrated S1D spectra using estimated absolute efficiency and extinction curves
\item Computation of the cross-correlation function (CCF) of the S2D spectrum with a stellar binary mask
\item Fit of the CCF with a Gaussian model to derive radial velocity and associated CCF parameters (FWHM, contrast, bisector span, asymmetry)
\item Computation of sky-subtracted CCFs with associated radial velocity and CCF parameters (simultaneous sky mode)

\end{itemize}

The ESPRESSO DRS has a modular design and can be adapted to the different modes of ESPRESSO, and to other echelle spectrographs like HARPS, through a set of configuration files. The DRS recipes themselves are the same for different modes and instruments.

\subsubsection{Data Analysis Software}
The ESPRESSO Data Analysis Software (DAS) is designed to perform scientific measurements on reduced data and to provide the users with high-level science-grade observables and products like lists of lines, continuum models and best-fit values of physical parameters. Like the DRS, it is structured as a set of recipes that can be launched individually (using the ESO's command-line interface EsoRex) or collectively as a work-flow (through the ESO Reflex environment). Each recipe performs a specific operation and may take as input both both DRS products and static calibration (list of reference lines, etc.).\\

\begin{figure*}[h]
   \centering
   \includegraphics[width=\hsize, trim= 70 40 50 40, clip]{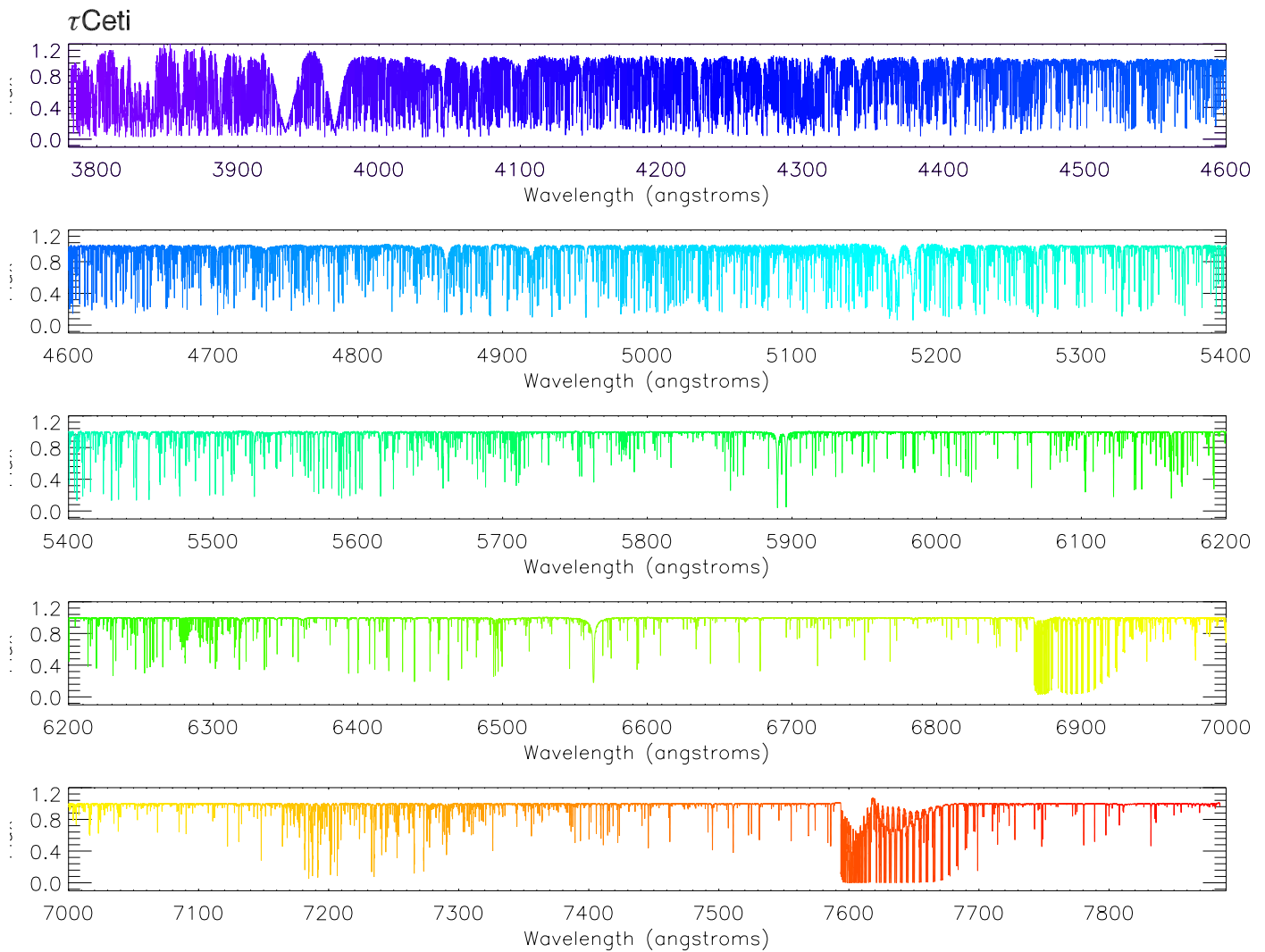}
   \caption{Output of the {\tt star\,II} workflow after normalizing and co-adding 40 HR11 ESPRESSO spectra of \object{$\tau$\,Cet}.}
   \label{fig:taucet-spectrum}
\end{figure*}

To cope with the non-linear nature of the analysis procedure, six different Reflex work-flows are included in the DAS, each with its own specific cascade of recipes. Three work-flows are devoted to the analysis of stellar spectra:
\begin{itemize}
    \item {\tt star\_I} combines multiple exposures of the same stellar spectrum  correcting for radial velocity (recipe {\tt espda\_coadd\_spec}), determines the equivalent width of selected absorption lines (recipe {\tt espda\_compu\_eqwidth}), and compares the equivalent widths with static calibrations to determine the effective temperature and the metallicity of the star (recipe {\tt espda\_compu\_starpar});
    \item {\tt star\_II} combines multiple exposures of the same stellar spectrum \textit{without} correcting for radial velocity (recipe {\tt espda\_coadd\_spec}), estimates the continuum level order by order (recipe {\tt espda\_fit\_starcont}), generates a synthetic spectrum corresponding to a given set of stellar and broadening parameters (recipe {\tt espda\_synth\_spec}), and computes and fits the cross-correlation function between the synthetic spectrum and the observed one to estimate the radial velocity of the star (recipe {\tt espda\_rv\_synth}) to finally compare the observed spectrum with the synthetic spectrum. As an example of the {\tt star\_II} workflow, we display in Figure\,\ref{fig:taucet-spectrum} a normalized spectrum of \object{$\tau$\,Cet}, computed by co-adding a sub-sample of 40 ESPRESSO HR11 2D-spectra.
    \item {\tt star\_III} estimates and corrects for the radial velocity by cross-correlating the spectrum in the order-wavelength space with a predefined mask (recipe {\tt espda\_compu\_radvel}) and measures the integrated fluxes in given pass bands to estimate the stellar activity indexes (recipe {\tt espda\_compu\_rhk}). 
\end{itemize}

Two workflows are devoted to the analysis of QSO spectra:
\begin{itemize}
    \item {\tt qso\_I} combines multiple exposures of the same QSO spectrum (recipe {\tt espda\_coadd\_spec}), lets the user mask the spectral regions which should not be considered in the analysis (recipe {\tt espda\_mask\_spec}), detects the absorption lines (recipe {\tt espda\_create\_linelist}), fits the continuum level by provisionally fitting and removing the absorption lines (recipe {\tt espda\_fit\_qsocont}), and finally identifies the detected lines and combines them into absorption systems, i.e.~different ionic transitions at the same red shift (recipe {\tt espda\_iden\_syst}); 
    \item {\tt qso\_II} fits the identified lines with Voigt profiles (recipe {\tt espda\_fit\_line}) and revises the continuum level taking into account the more precise line fitting (recipe {\tt espda\_fit\_qsocont}). 
\end{itemize}

Additionally, a {\tt coadd} workflow is provided, including only recipes {\tt espda\_coadd\_spec} and {\tt espda\_mask\_spec}. This is meant as a fast tool to visualize the co-added spectrum and check the signal-to-noise ratio.

The combination of several work-flows allows for a flexible analysis, while maintaining a great degree of automation in the procedure. Most recipes are equipped with interactive Python scripts (automatically called by Reflex) to plot and interact with the data, changing recipe parameters and prompting the re-execution of the recipe.  All products are provided in FITS format and can be easily handed to other analysis tools.

Algorithmically, the DAS includes both long-tested solutions like the line-finding technique implemented in ARES \citep{sousa07} and brand-new methods, like the drizzling-like approach to coaddition and the simultaneous continuum-line fitting in QSO spectra \citep{cupani16,cupani:2019}. The strict requirements imposed by the ESPRESSO science cases pushed for novel solutions that benefit to the entire community and are growing beyond the scope of the instrument (see e.g.~the Astrocook data analysis package \citep{cupani18} and the ARESv2 package \citep{sousa:2015}). 


\section{On-sky Performance}
\label{sec:perf}

\subsection{Spectroscopic performance}
\subsubsection{Spectral range}
The wavelength range of ESPRESSO was originally specified by ESO to cover at least the full visible range from 380 to 680\,nm, similarly to the HARPS spectrograph, with an optional extension to 350\,nm on the blue side and to 760\,nm on the red side. While the extension to the red side could easily be implemented, the blue extension would have produced considerable technical challenges. The implemented spectral range covers continuously the domain from 378.2\,nm to 788.7\,nm. This coverage is provided for all instrument modes (HR, UHR and MR), albeit at different resolving power.

\begin{figure}[h]
   \centering
   \includegraphics[width=\hsize]{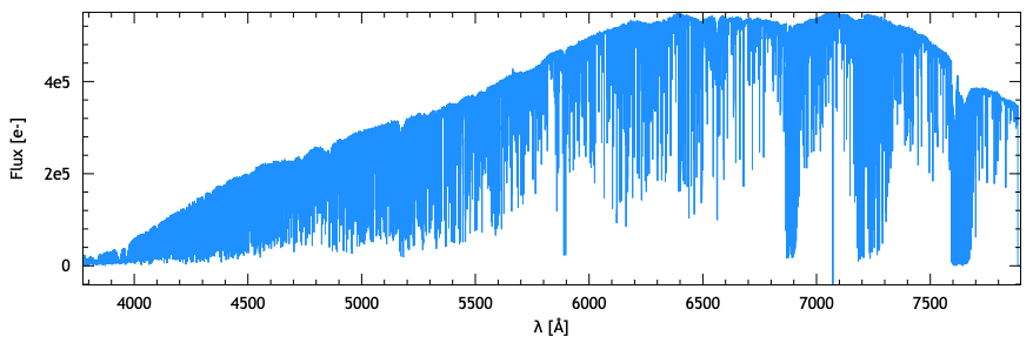}
   \includegraphics[width=\hsize]{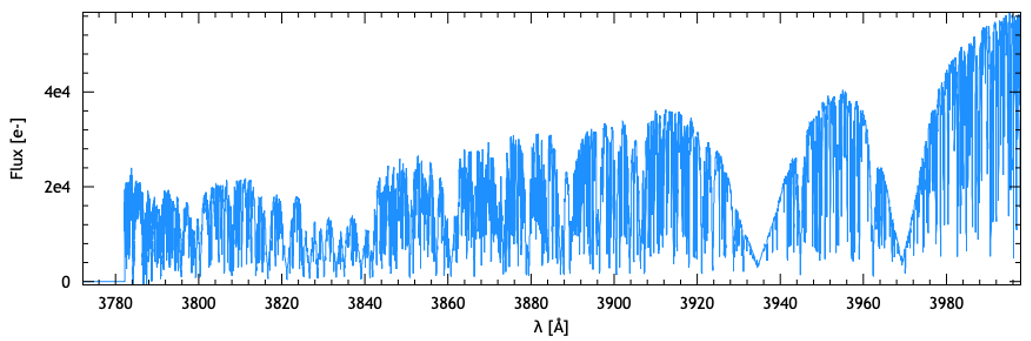}
   \includegraphics[width=\hsize]{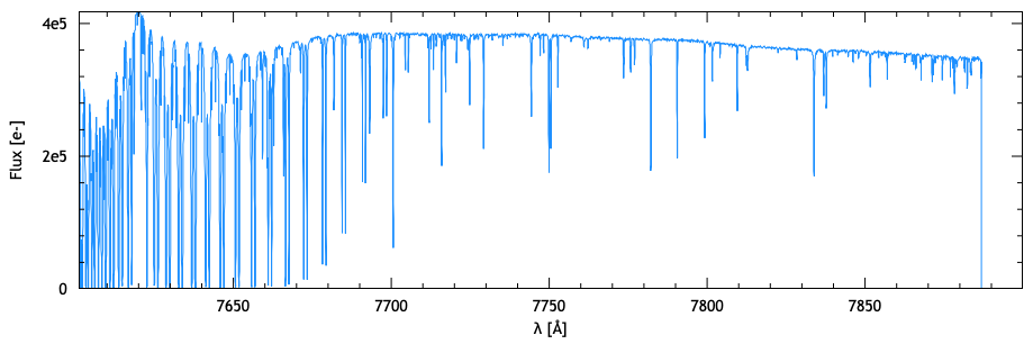}
   \caption{Top panel: Full extracted and wavelength calibrated spectrum of $\tau$-Cet observed with ESPRESSO in UHR-mode. Middle panel: Zoom on the blue end of the spectrum emphasizing the calcium doublet on the right side. Bottom panel: Zoom on the red end of the spectrum showing the deep and narrow sequence of telluric absorption lines on the left side.}
   \label{fig:spectrum}
\end{figure}

In order to illustrate the spectral range, richness and quality, we show in Figure\,\ref{fig:spectrum}, top panel, the extracted and wavelength-calibrated 1D-spectrum of $\tau$-Cet (HD\,10700) as observed in the UHR mode, while the lower two panels provide a zoom on the very blue and the very red end of the spectrum, respectively.

\subsubsection{Resolving power}
The resolving power $R=\lambda/\Delta\lambda$ of a spectrograph is defined as the ratio between the wavelength $\lambda$ and the full-width at half maximum $\Delta\lambda$  of a \textit{non-resolved} spectral line. In astronomical spectrographs, the slit width or fiber diameter is the result of the trade-off between resolving power and slit-efficiency. In order to contain slit losses, the on-sky projected slit width/fiber size is usually chosen to be \textit{comparable to} or larger than the median seeing of the observing site. In opposition, the spectrograph's optical design (size, dispersion) is adapted to produce the desired resolving power. As a consequence, the resolving power is not determined by the diffraction limit of the optics, but simply by the slit (fiber) width. Transferred to ESPRESSO, this implies that the spectral resolution $\Delta\lambda$ covered by the slit/fiber projected on the detector is directly proportional to the fiber width and the resolving power inversely proportional. The different modes of ESPRESSO produce therefore a different resolving power, as mentioned in Table\,\ref{tab:modes}.

However, the geometrical image of the optical fiber is convolved with the 'local' Instrumental Profile ($IP$) of the spectrograph that represents the impact of optical aberrations and diffraction on the image of a perfectly monochromatic source feeding the spectrograph. As a consequence, the fiber image on the detector will be slightly blurred and the resolving power reduced. Of course, the effect is larger, in relative terms, when the $IP$ is comparable in width to the fiber diameter. In other words, the narrower the fiber - and thus the higher the resolving power - the larger will be the reduction in resolving power compared to its theoretical value computed from geometrical considerations alone.

\begin{figure}[h]
   \centering
   \includegraphics[width=\hsize]{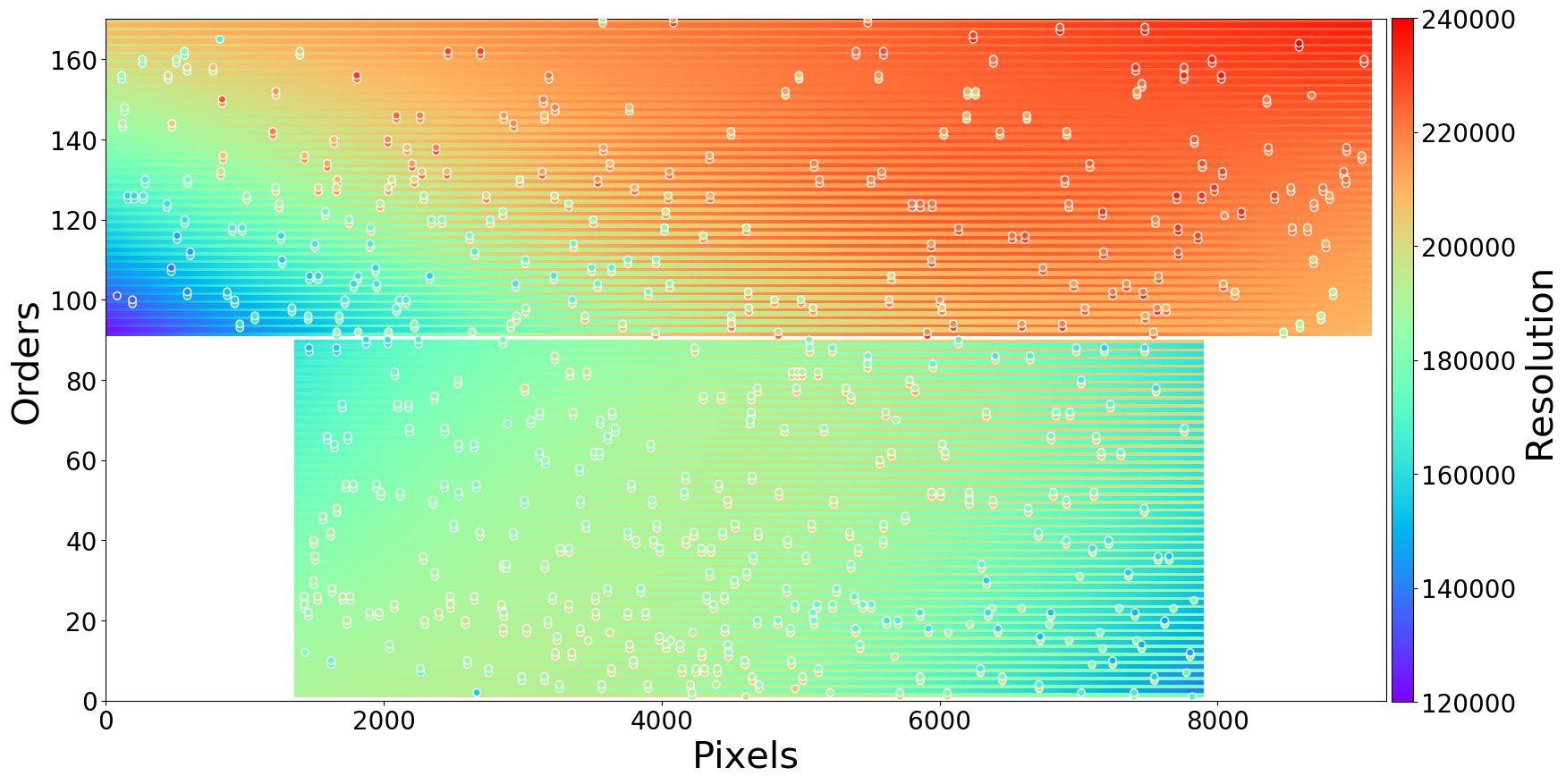}
   \includegraphics[width=\hsize]{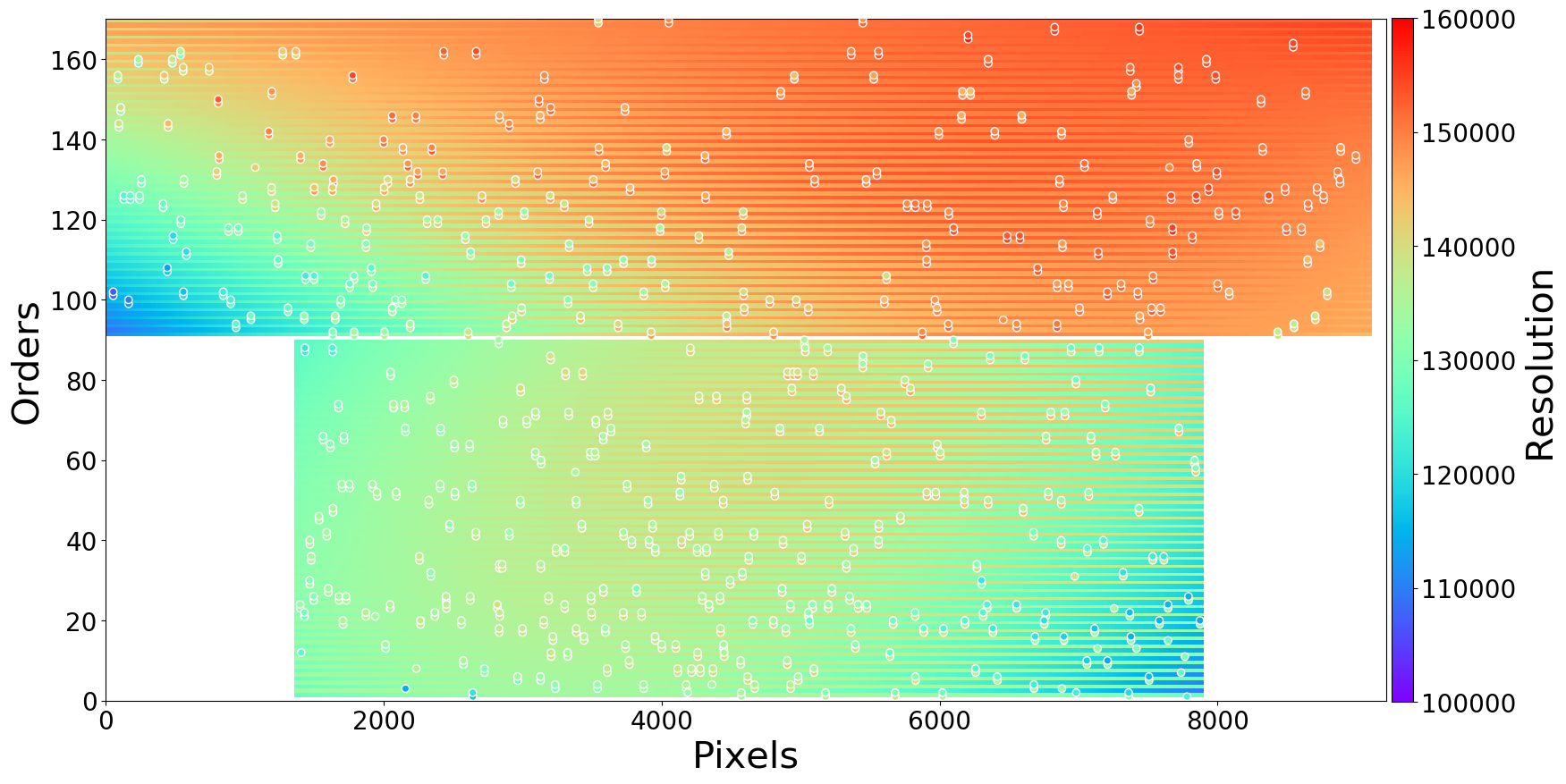}
   \includegraphics[width=\hsize]{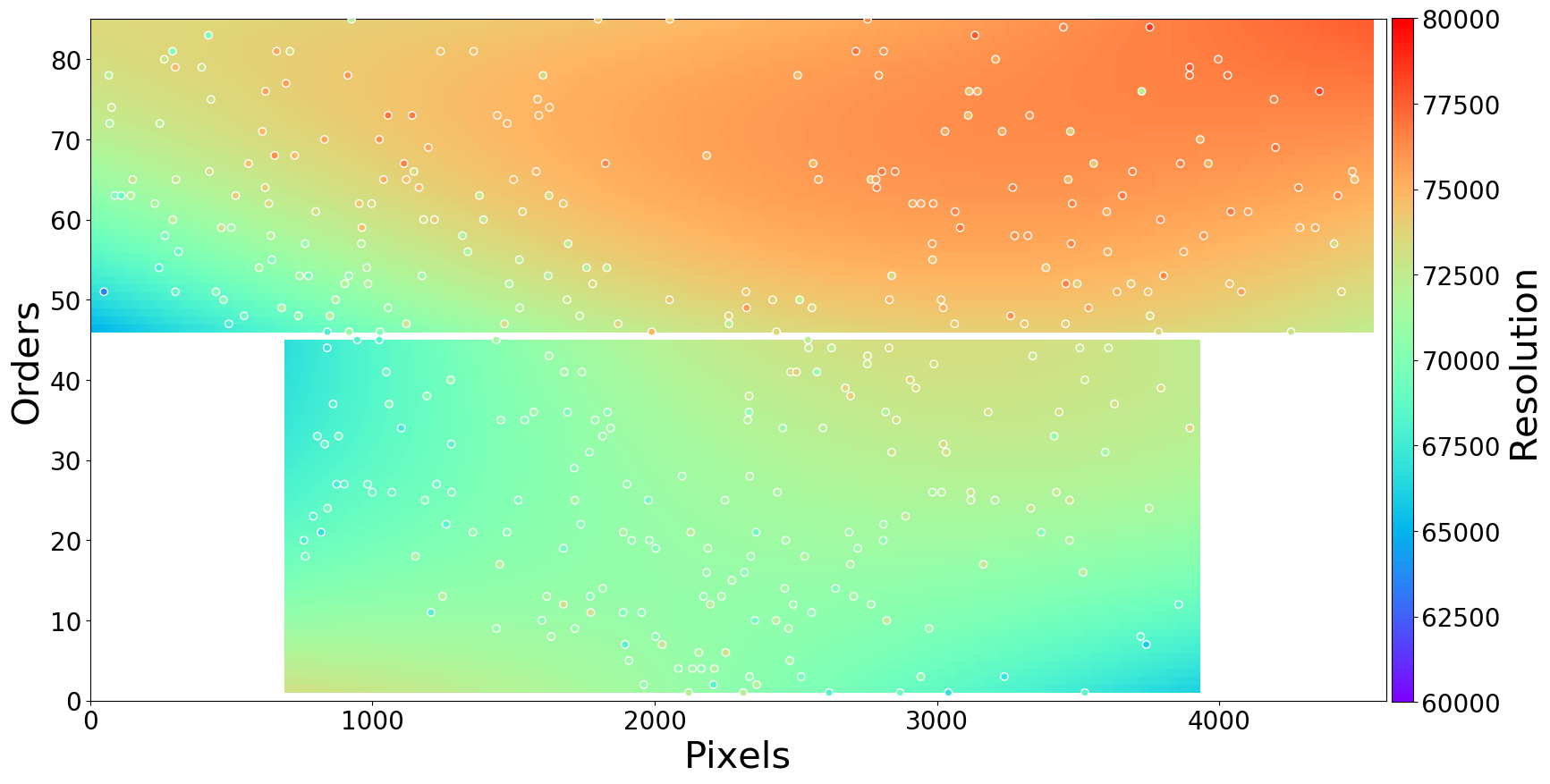}
   \caption{From top to bottom, 2D-maps of ESPRESSO's resolving power for its three instrument modes UHR, HR and MR. For each mode the two CCDs are represented (top = red chip, bottom = blue chip). The y-axis does not represent the physical diffraction order number but denotes a continuous (DRS-internal) numbering of every individually extracted slice of each order sorted by increasing wavelength (1\,=\,bluest wavelengths). The white dots show the location of thorium lines used for the computation of the resolving power. Since in the MR mode the two slices of a given order are merged together by the extraction algorithm, only half the number of orders are present. Their lines have graphically been increased in width to represent the same detector area as the two other modes.}
   \label{fig:resolution}
\end{figure}

In order to study the effectively achieved resolving power in the various modes and across the full spectrum, we have used Thorium-Argon lamp calibration frames, assuming that most of the Thorium emission lines are not or are only marginally resolved\footnote{From measurements using the (unresolved) LFC lines we know actually that most of the thorium lines are partially resolved. Since we cannot use the LFC over the full wavelength range, we prefer to provide to the reader (safe) lower limits for the resolving power.} by the spectrograph. For all non-saturated Thorium lines with sufficient signal we measured the line width and converted the FWHM into resolving power. By 2-D interpolation, we produce a 2-D map of the resolving power as a function of 'extracted pixel' (spectral direction, i.e. main dispersion) and echelle order (spatial direction, i.e. cross-dispersion). The three images in Figure\,\ref{fig:resolution} represent, from top to bottom, the UHR, the HR, and the MR mode. Each image is composed of one larger frame (top) showing the red chip and a smaller frame (bottom) showing the blue chip. The wavelength increases from bottom to top (cross-dispersion direction) and from left to right (main-dispersion direction along the order). For each order of the HR and UHR modes both slices are shown side-by-side, and, as it can be remarked, the resolving power is slightly different. In the MR mode the two (adjacent) slices must be extracted simultaneously by the DRS, resulting in a resolution map of one line per order. The map show also the position of the thorium line used for determining the resolving power and interpolate the maps.

In HR mode, the standard mode, a median resolving power of $R = $\,138\,000 is measured (requirements: $R = $\,120\,000). In the MR mode (4-UT configuration) the resolving power is reduced by about a factor of 2, i.e. $R = $\,72\,500, due to the larger fiber diameter (requirement: $R = $\,30\,000). The UHR mode achieves a median resolving power of the order of $R > $\,190\,000 in the blue arm, while exceeding $R = $\,220\,000, and thus being up to specification, in the red arm.

At first glance, a similar global behaviour is observed in all modes. The blue spectrum (lower chip) is slightly less well resolved and a trend to higher resolving power is observed along every order from left (blue end) to right (red end). Slight variations of resolving power are indeed expected from optical-design considerations and are known to depend on manufacturing parameters of each component and effective alignment of the spectrograph. 

The number of free parameters and manufacturing unknowns is so large that a detailed understanding of the observed map is close to impossible. For this reason, the spectrograph was aligned in such a way as to achieve a resolution map as homogeneous as possible while trying to maximize median resolving power. The following facts attest to the success of this strategy. First, the maps look very similar in all modes; this means that the impact of image quality of the spectrograph is not the dominant factor in the measured resolving power. Second, the resolving power in MR and HR mode are both compliant with the original requirements. On the other hand, the resolving power in UHR mode is not compliant with the original requirement, at least on the blue chip. It must be recalled, however, that the resolving power has been determined using thorium lines that are partially resolved in this mode. Finally, we note that ESPRESSO is today the only spectrograph in the southern hemisphere on an 8-m class telescope delivering in a single exposure the entire visible spectrum at a resolving power of $R>$190\,000.

\subsubsection{Detection efficiency}
We define the (total) photon-detection efficiency as the ratio between the number of photo-electrons per wavelength bin at $\lambda = 550$\,nm measured on the detector at the blaze of the echelle order, and the number of photons received from a point-like source \textit{before} entering the atmosphere assuming photometric conditions and observation airmass $AM=1$. Since in a spectrograph, the throughput strongly depends on the seeing, ESPRESSO's requirements were specified such to reach a peak efficiency of 10\% at the center of the echelle order for a seeing of 0.65\arcsec as measured in the focal plane of the telescope.

\begin{figure}[h]
   \centering
   \includegraphics[width=\hsize, trim= 60 60 60 60, clip]{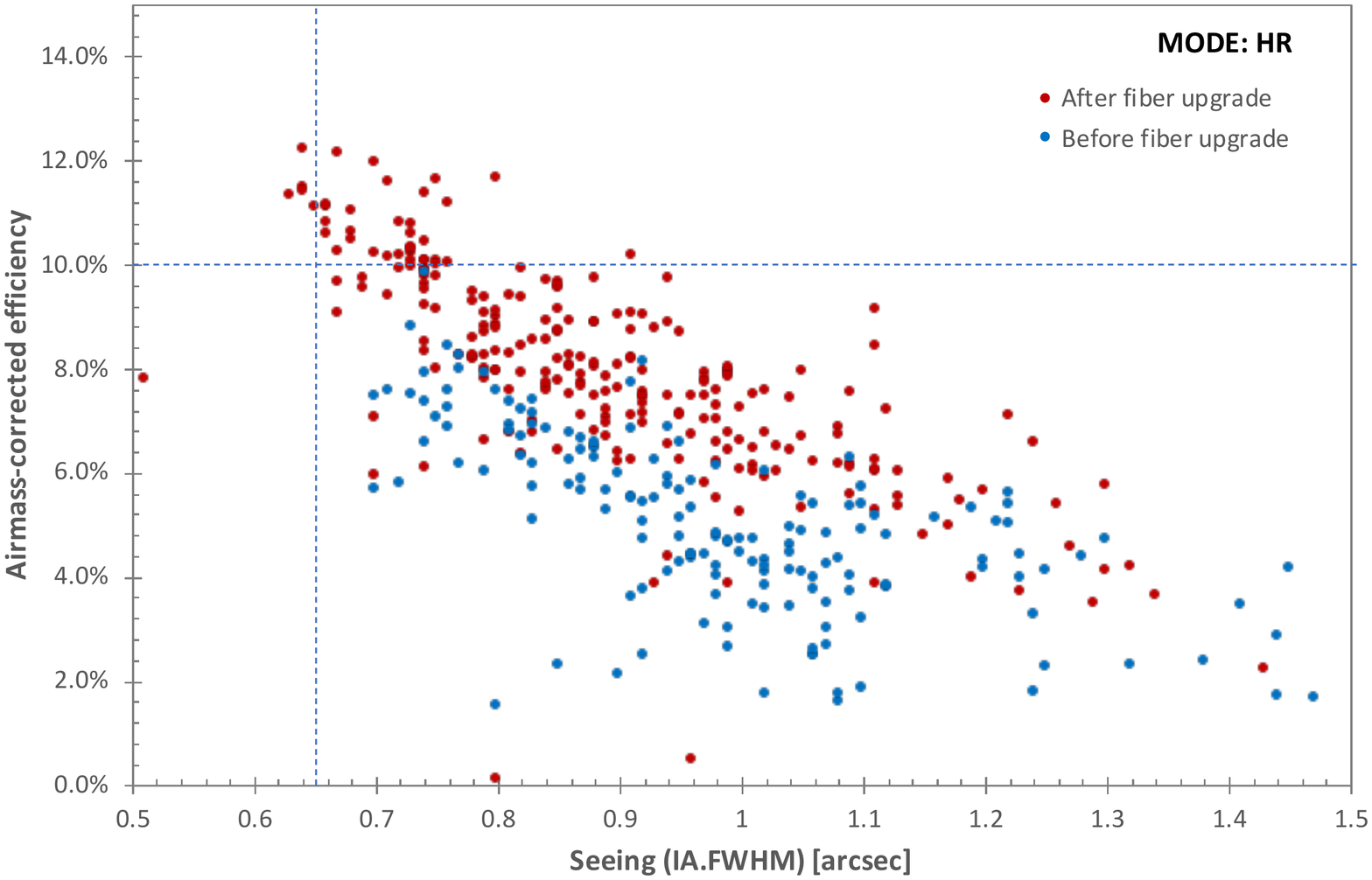}
   \caption{Photon-detection efficiency at 550 nm and airmass 0 as a function of telescope seeing (image quality $IA.FWHM$) before (blue dots) and after (red dots) the fiber-upgrade mission.}
   \label{fig:efficiency_vs_seeing}
\end{figure}

Figure\,\ref{fig:efficiency_vs_seeing} shows the measured detection efficiency at 550\,nm for airmass $AM=0$ (no absorption) as a function of measured image quality $IA.FWHM$\footnote{FITS-header keyword available in the raw images. $IA.FWHM$ gives the FWHM measured on the individual sub-apertures of the Shack-Hartmann scaled to zenith and at 550 nm. DIMM seeing is at zenith at 500nm, instead.}. The $IA.FWHM$ represents the ``seeing'' value at the Nasmyth focus of the telescope as computed by ESO's standard 'Image Analysis' tool. Blue data points show the measured efficiency of ESPRESSO shortly after first commissioning, while the red data points were obtained in the months following the fiber-upgrade mission of June 2019. The plot demonstrates that after the fiber upgrade the photon-detection efficiency at airmass 0 and for a seeing of 0.65\arcsec~ is close to 12\%, about 40\% higher than prior to the fiber exchange. Introducing an airmass of 1 the values drop by 10-20\% depending on the wavelength, such that the peak photon-detection efficiency still marginally reaches the specified 10\% level.

\begin{figure}[h]
   \centering
   \includegraphics[width=\hsize, trim= 60 60 60 60, clip]{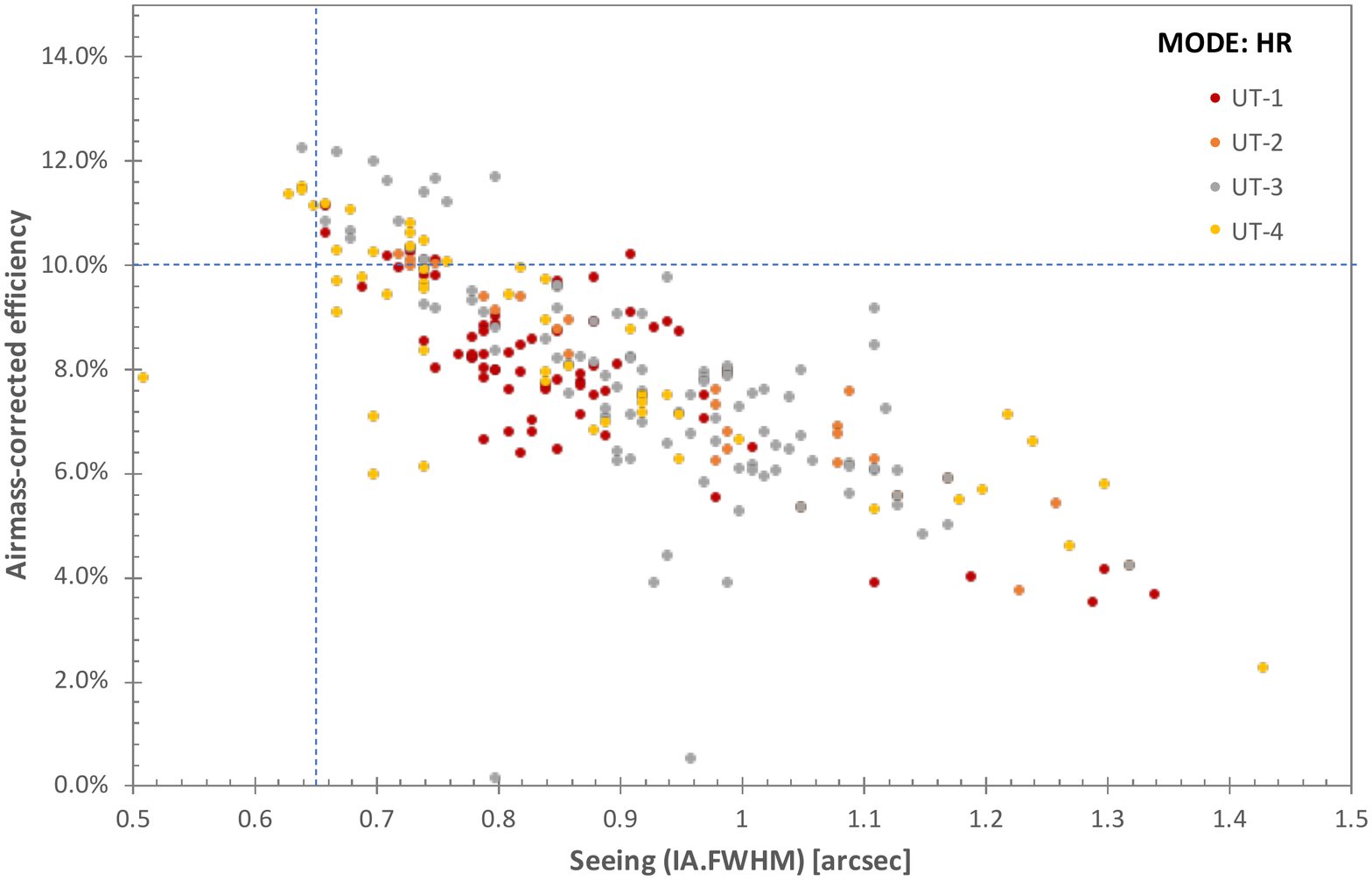}
   \includegraphics[width=\hsize, trim= 60 60 60 60, clip]{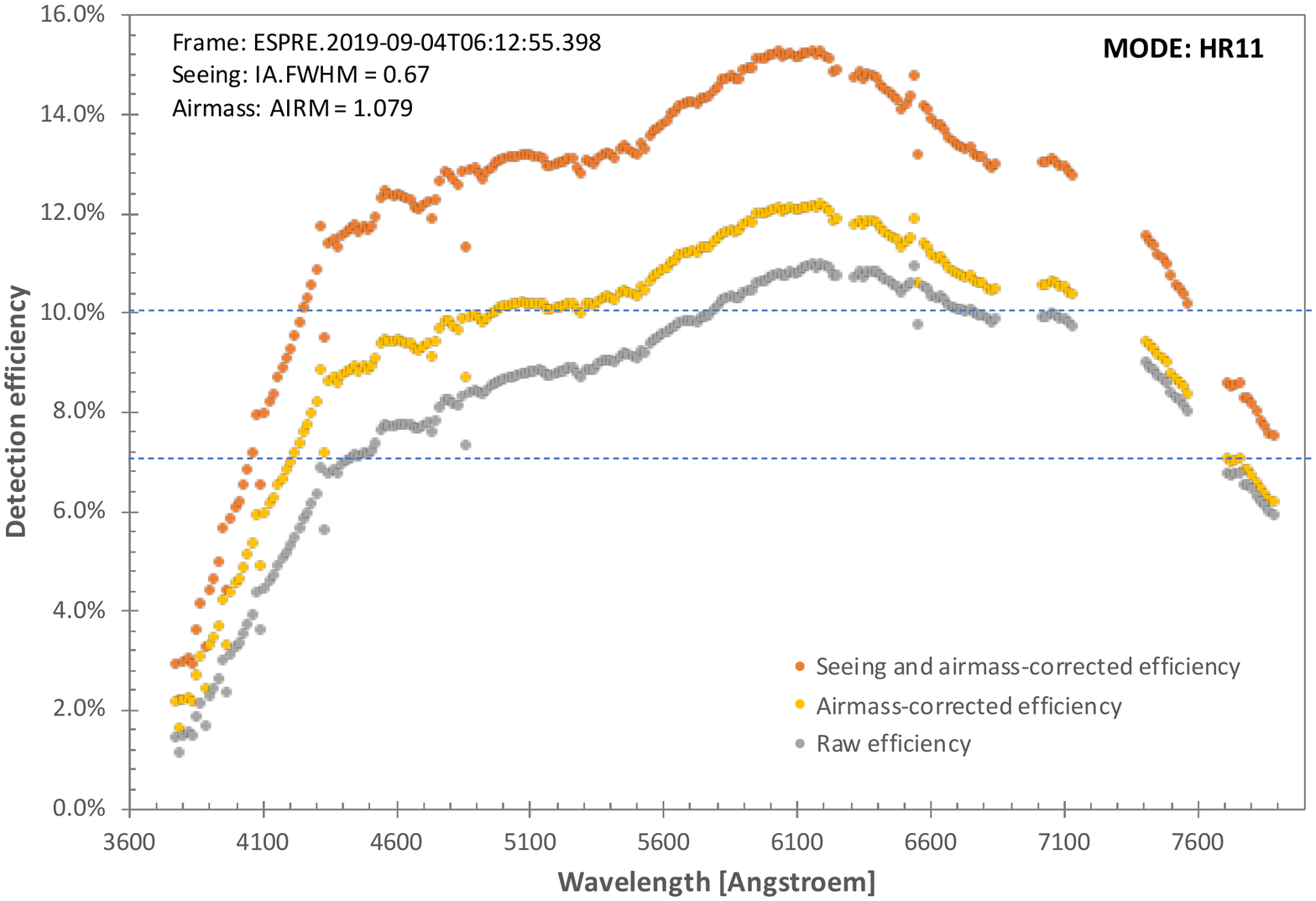}
   \caption{Bottom: Airmass-corrected efficiency (no atmpspheric absorption) for the HR-mode as a function of seeing  (image quality $IA.FWHM$). The plot combines the measurements from all telescopes. Top: Total photon-detection efficiency for the HR-mode in binning 1x1 as a function of wavelength. An example for an image quality of $IA.FWHM$ of 0.67\arcsec and airmass of 1.079 is shown. }
   \label{fig:efficiency_HR}
\end{figure}

In the following we present the photon-detection efficiency for all modes and as a function of wavelength, as obtained after the fiber upgrade. Figure\,\ref{fig:efficiency_HR} shows the detection efficiency for the HR mode as a function of wavelength as measured by the pipeline (raw), corrected for atmospheric absorption (airmass-corrected) and corrected for slit losses (seeing and airmass-corrected). The raw efficiency reaches 10\% at 550\,nm and exceeds it in the range between 550 and 680\,nm. The same plot shows however that the requirement that the photon-detection efficiency must not drop below 7\% inside the specified minimum wavelength range (380 to 680\,nm) of ESPRESSO is not fully reached. Indeed, for wavelengths bluewards of 430 nm the photon-detection efficiency goes below 7\% and reaches at the very blue end of the spectrum a value of about 2\%.

\begin{figure}[h]
   \centering
   \includegraphics[width=\hsize, trim= 60 60 60 60, clip]{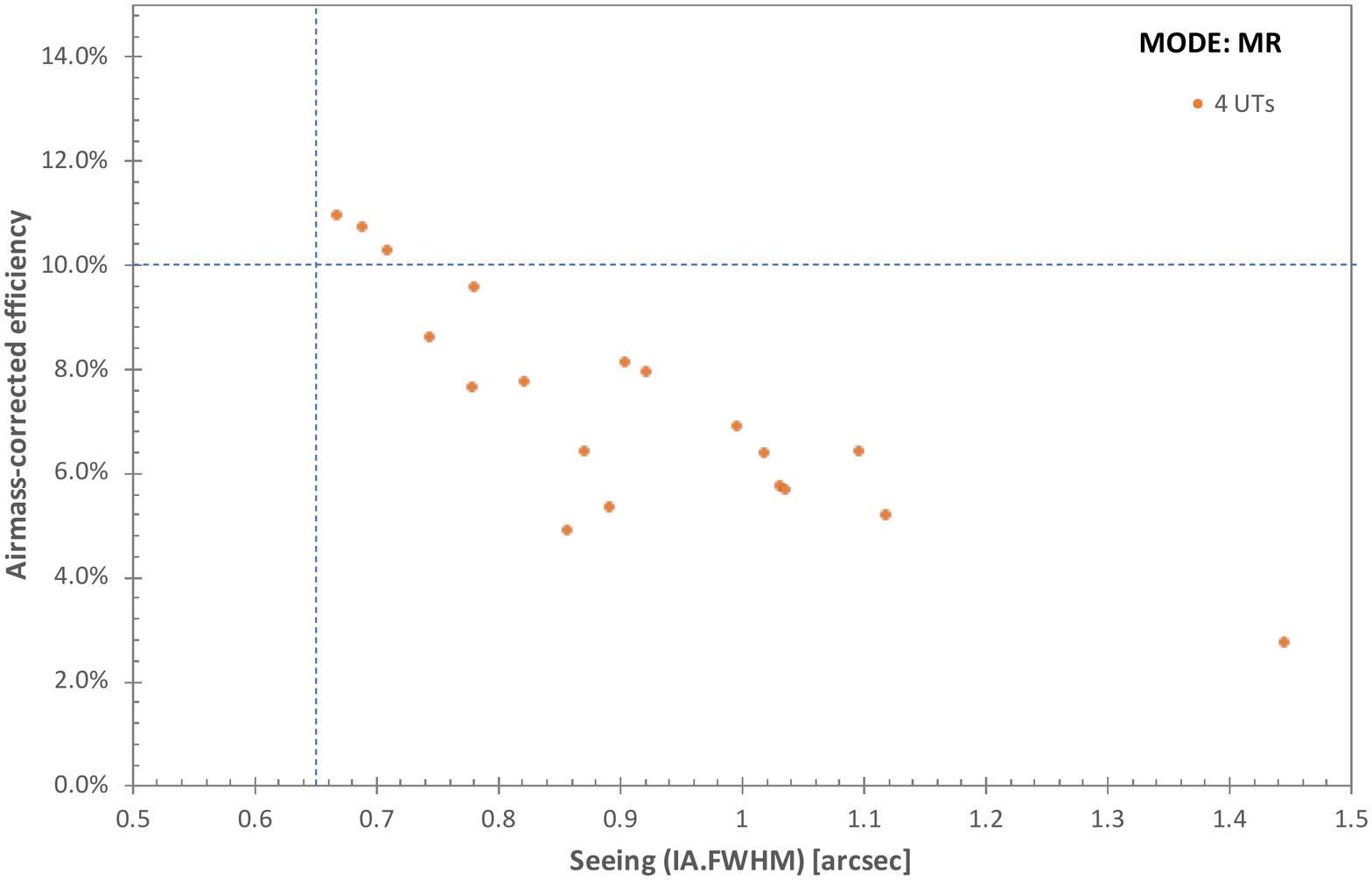}
   \includegraphics[width=\hsize, trim= 60 60 60 60, clip]{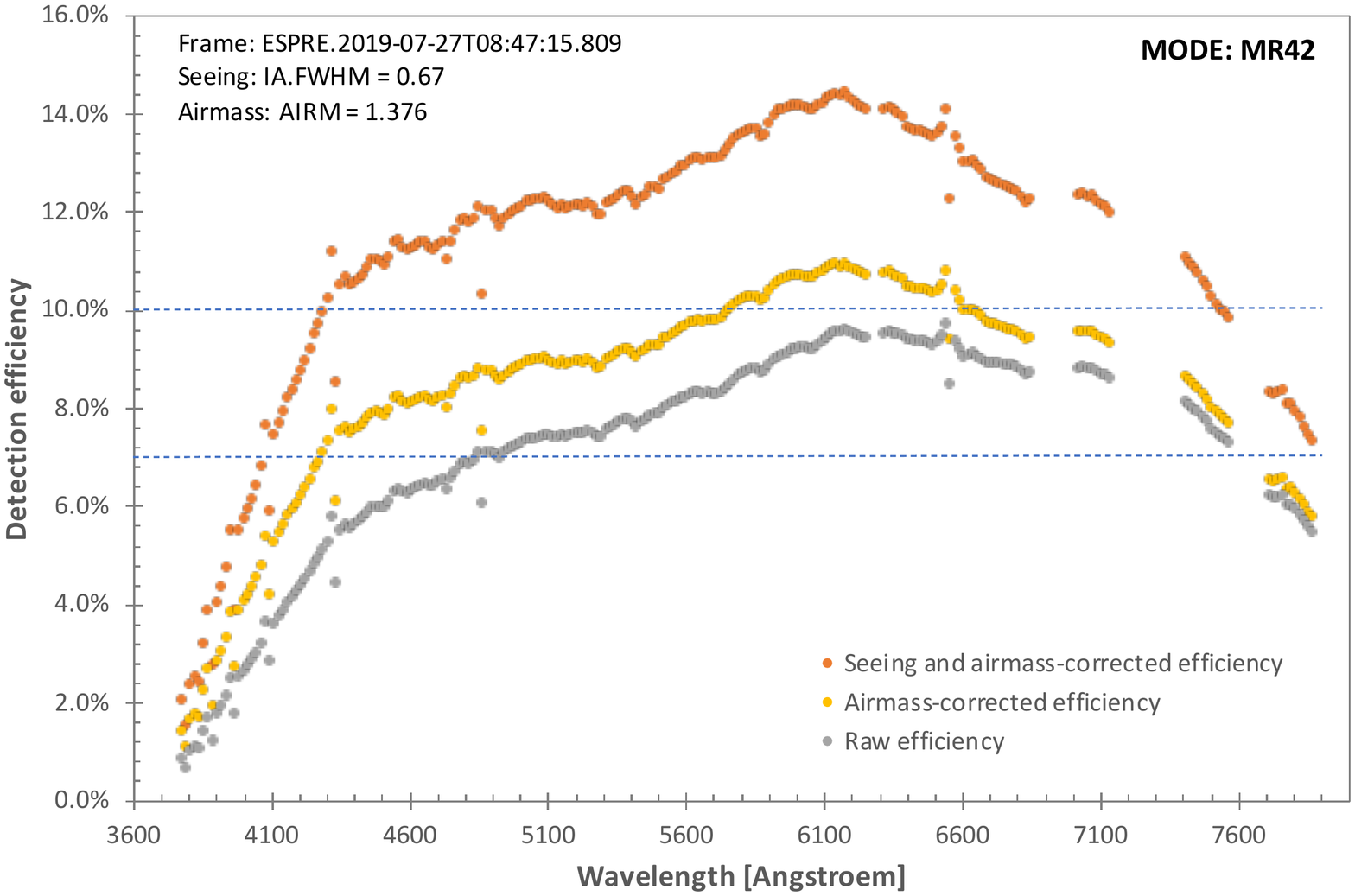}
   \caption{Bottom: Airmass-corrected efficiency (no atmospheric absorption) for the MR-mode as a function of seeing (image quality $IA.FWHM$). Top: Total photon-detection efficiency for the MR-mode in binning 4x2 as a function of wavelength. An example for an image quality of $IA.FWHM$ of 0.67\arcsec and airmass of 1.376 is shown. }
   \label{fig:efficiency_MR}
\end{figure}

The MR mode efficiency is specified to be identical to that of the HR mode, and indeed the two modes provide compatible results within measurement uncertainty. Figure\,\ref{fig:efficiency_MR} shows the efficiency as a function of wavelength and 'as-delivered' seeing, respectively. The same considerations as for the HR mode apply, with the only difference that much fewer observations are available at the time of writing and none could be obtained for nominal (photometric) conditions.

\begin{figure}[h]
   \centering
   \includegraphics[width=\hsize, trim= 60 60 60 60, clip]{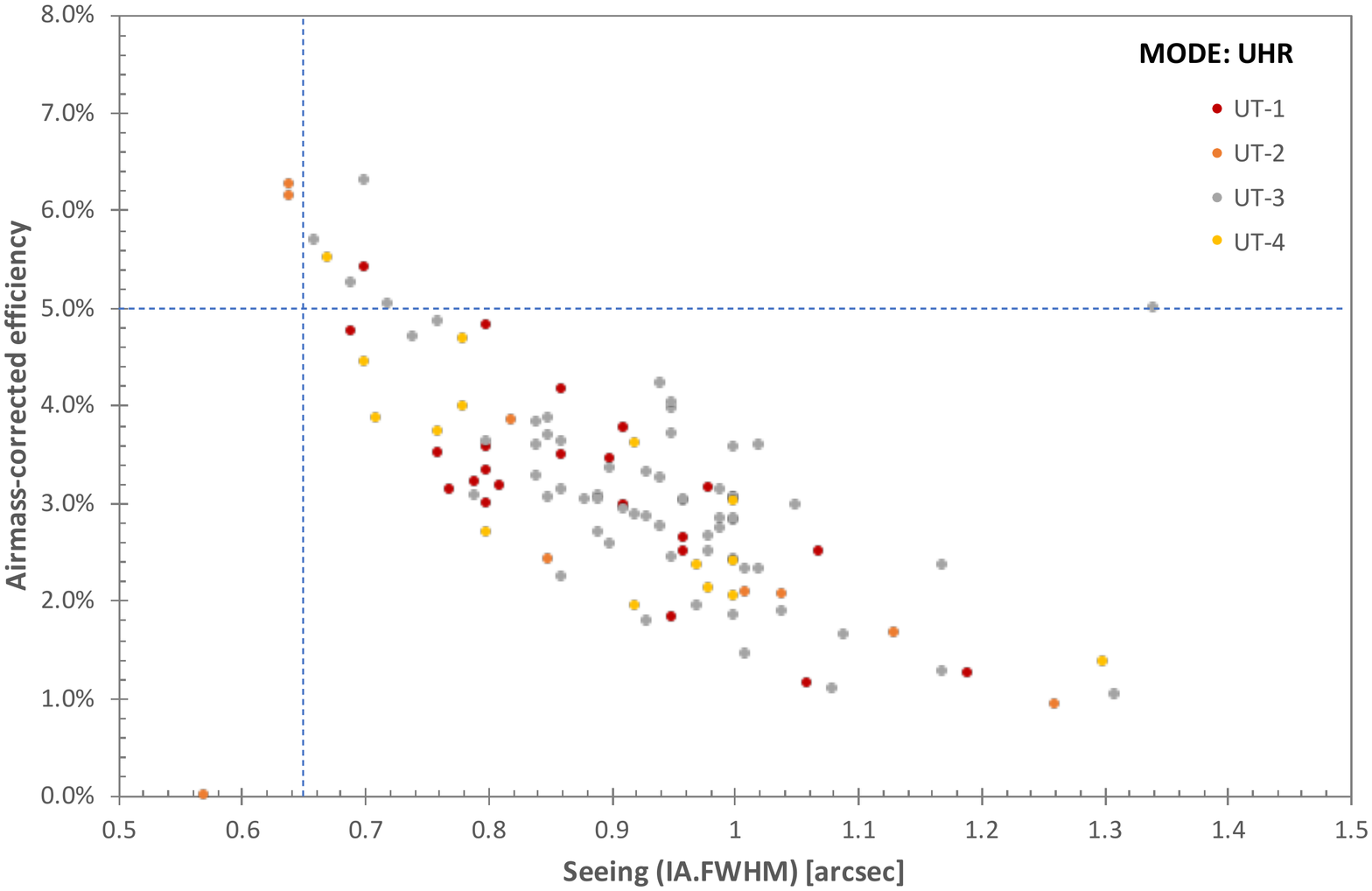}
   \includegraphics[width=\hsize, trim= 60 60 60 60, clip]{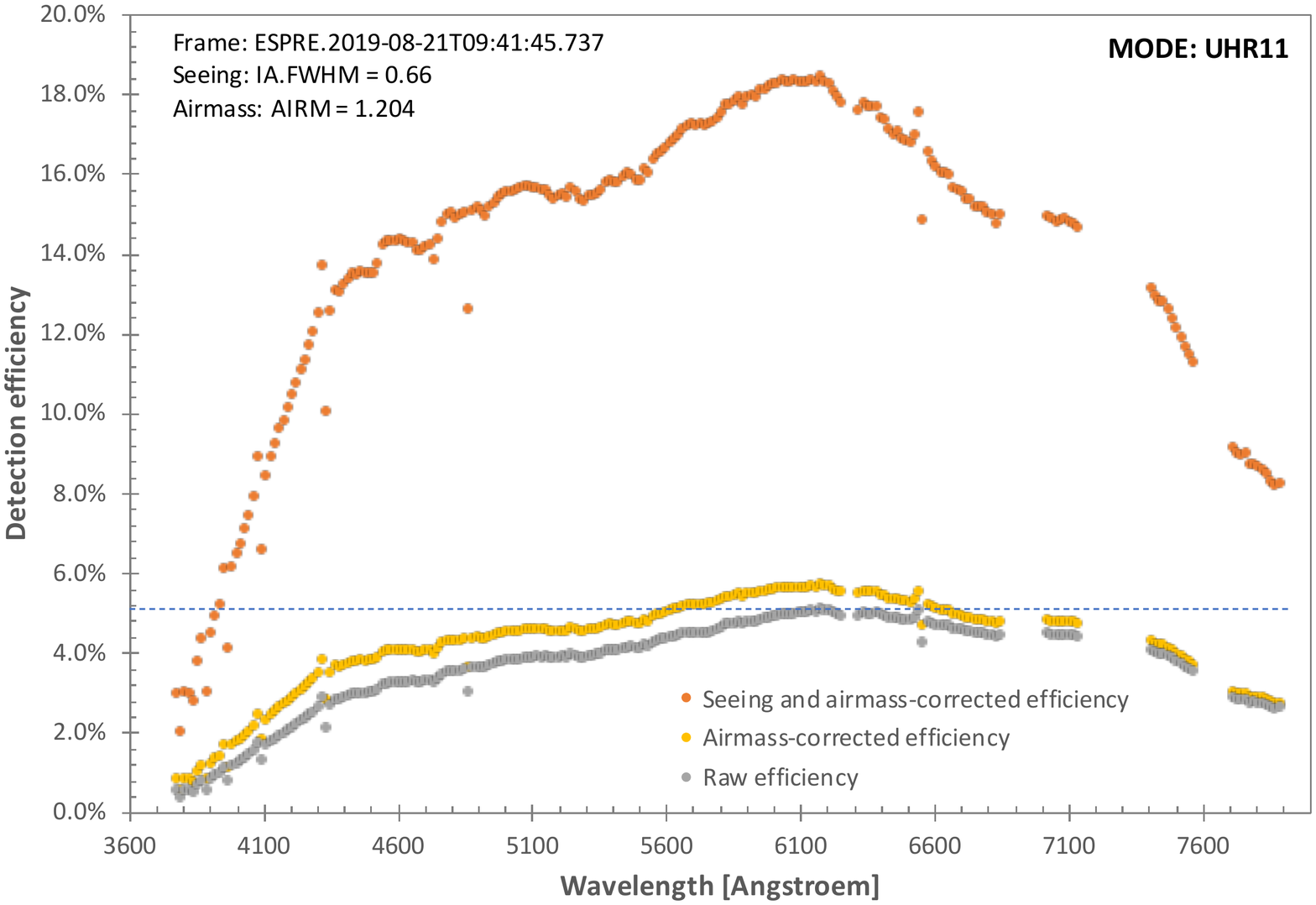}
   \caption{Bottom: Airmass-corrected efficiency (no atmpspheric absorption) for the UHR-mode as a function of seeing (image quality $IA.FWHM$). The plot combines the measurements on all telescopes. Top: Total photon-detection efficiency for the UHR-mode in binning 1x1 as a function of wavelength. An example for an image quality of $IA.FWHM$ of 0.66\arcsec and airmass of 1.204 is shown. }
   \label{fig:efficiency_UHR}
\end{figure}

Figure\,\ref{fig:efficiency_UHR} shows eventually the efficiency of the UHR mode as a function of wavelength and seeing, respectively. Considering the fiber diameter of 0.5\arcsec, the obtained photon-detection efficiency exceeds the expectations.\\

The plots of the efficiency versus the seeing for the HR and the UHR modes show measurements obtained with all the UTs of the VLT. The fact that no systematic efficiency difference is observed in between the telescopes demonstrates that the requirement of identical detection efficiency for each telescope is met. Furthermore, the fact that the MR-mode efficiency is very close to the HR-mode efficiency demonstrates that all four 'channels' of the MR mode (4-UT configuration) have similar -- and thus necessarily nominal -- efficiency. Moreover, we observed with one UT using the MR fiber link and cycling though all the four UTs individually, obtaining consistent figures for the throughput.

\subsubsection{Signal-to-noise ratio and precision}
\label{sec:snr}
Internal RV precision, i.e. the fundamental precision that can be obtained if only photon noise limits the measurement, is estimated using the algorithms described in \citet{bouchy:2001}. It shall be recalled that the fundamental RV precision not only depends on the number of photons and the signal-to-noise ratio of the measurement, but also on the number and effective width of the spectral lines that are used to measure the RV. For this reason, the RV precision depends, on the one hand, on the resolving power of the spectrograph, and, on the other hand, on the spectral type and the rotational velocity of the star (among other broadening), if a stellar spectrum is observed. By consequence, the different observing configurations and instrument modes of ESPRESSO will not deliver identical fundamental precision under identical observational conditions. In the following, we will only discuss the HR mode in more detail, since it is the standard mode designed for precise radial-velocity measurements.

\begin{figure}[h]
   \centering
   \includegraphics[width=\hsize, trim= 60 60 60 60, clip]{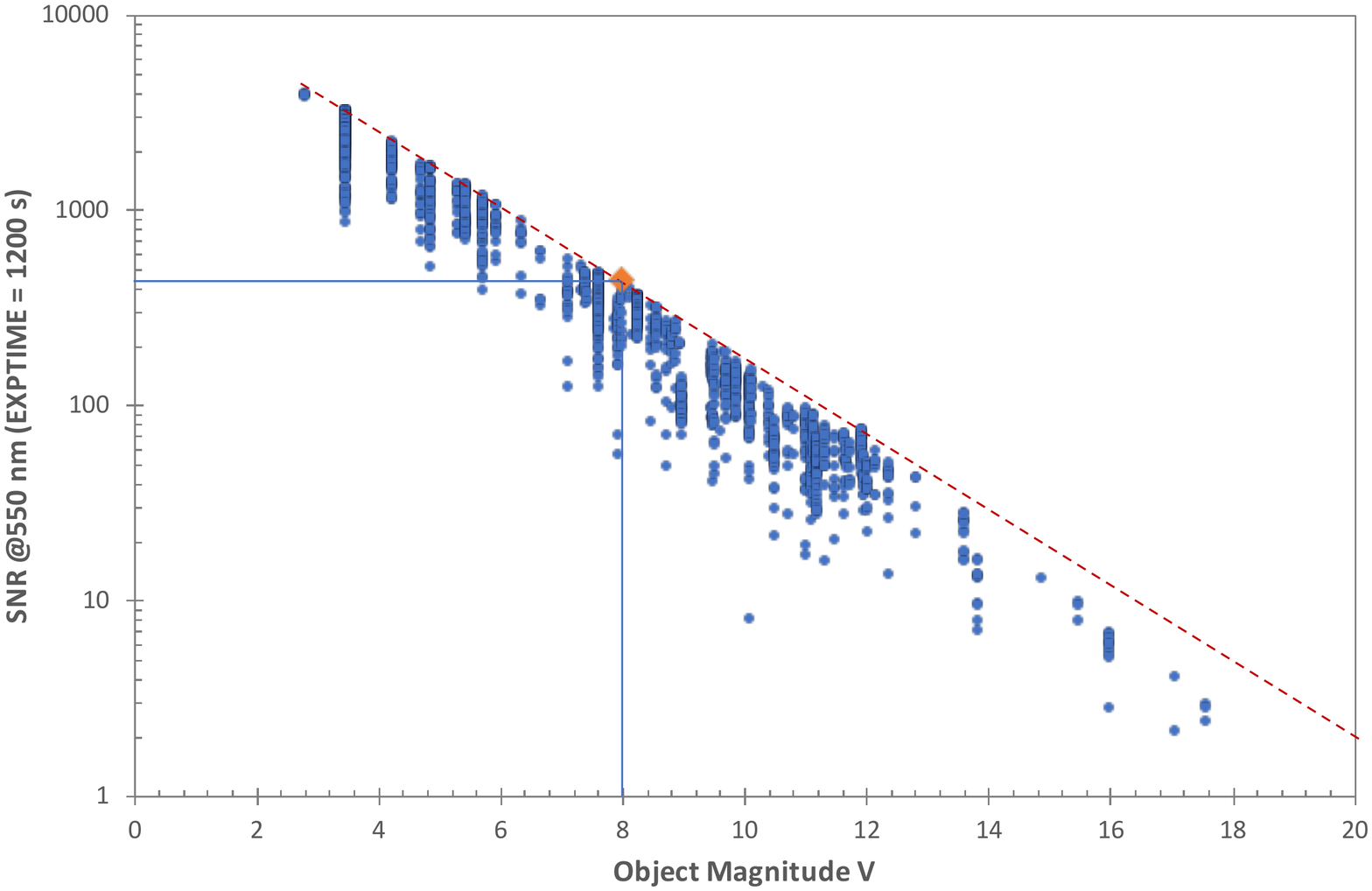}
   \caption{Measured S/N as a function of target magnitude for the HR11 mode (exposure time normalized to 1\,200\,s). The upper envelope given by the dashed line gives an indication of the S/N reached under optimum astroclimatic conditions.}
   \label{fig:snr_vs_mag}
\end{figure}

In order to estimate the S/N and the RV precision, and provide benchmarks for the standard HR11 mode, we analysed GTO RV data from July to December 2019 (after the fiber upgrade). Figure\,\ref{fig:snr_vs_mag} shows the S/N obtained at 550\,nm wavelength in an extracted-pixel bin of 9\,m\AA~(0.49\kms, no binning in spectral direction) as a function of target magnitude after normalizing them to the exposure time of 20\,Min. Taking the upper envelope described by the dashed line we measure a signal-to-noise ratio of the order of $S/N=430$ for $V = 8$\,mag and $S/N=70$ for $V = 12$\,mag. Starting at $V = 14$\,mag a significant departure from the photon-noise limited behaviour is observed.

\begin{figure}[h]
   \centering
   \includegraphics[width=\hsize, trim= 40 0 50 30, clip]{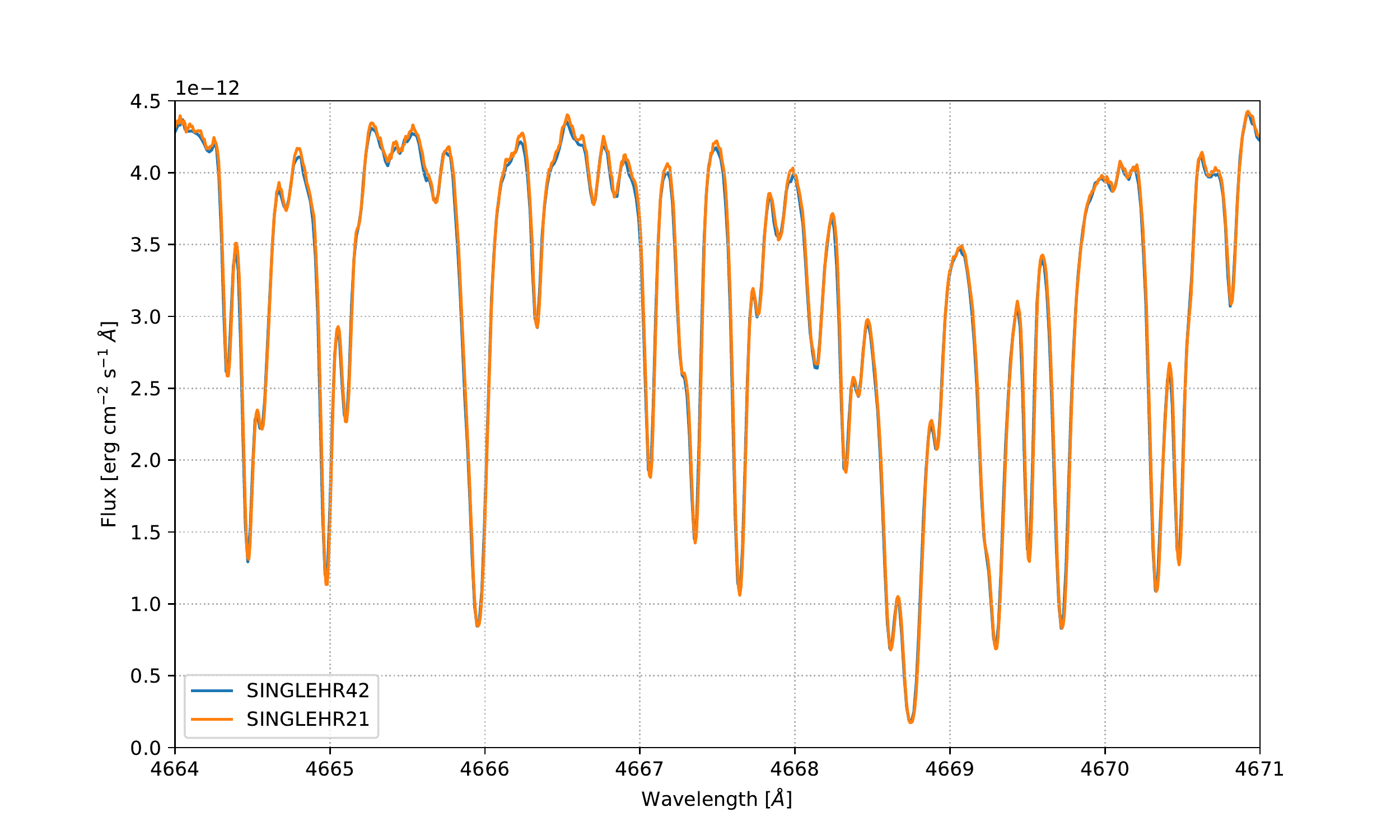}
   \caption{Comparison of the spectra of \object{HD\,85512} obtained in the HR42 and HR21 mode, respectively}
   \label{fig:HR42_HR21_comparison}
\end{figure}

\begin{figure}[h]
   \centering
   \includegraphics[width=\hsize, trim= 40 0 50 30, clip]{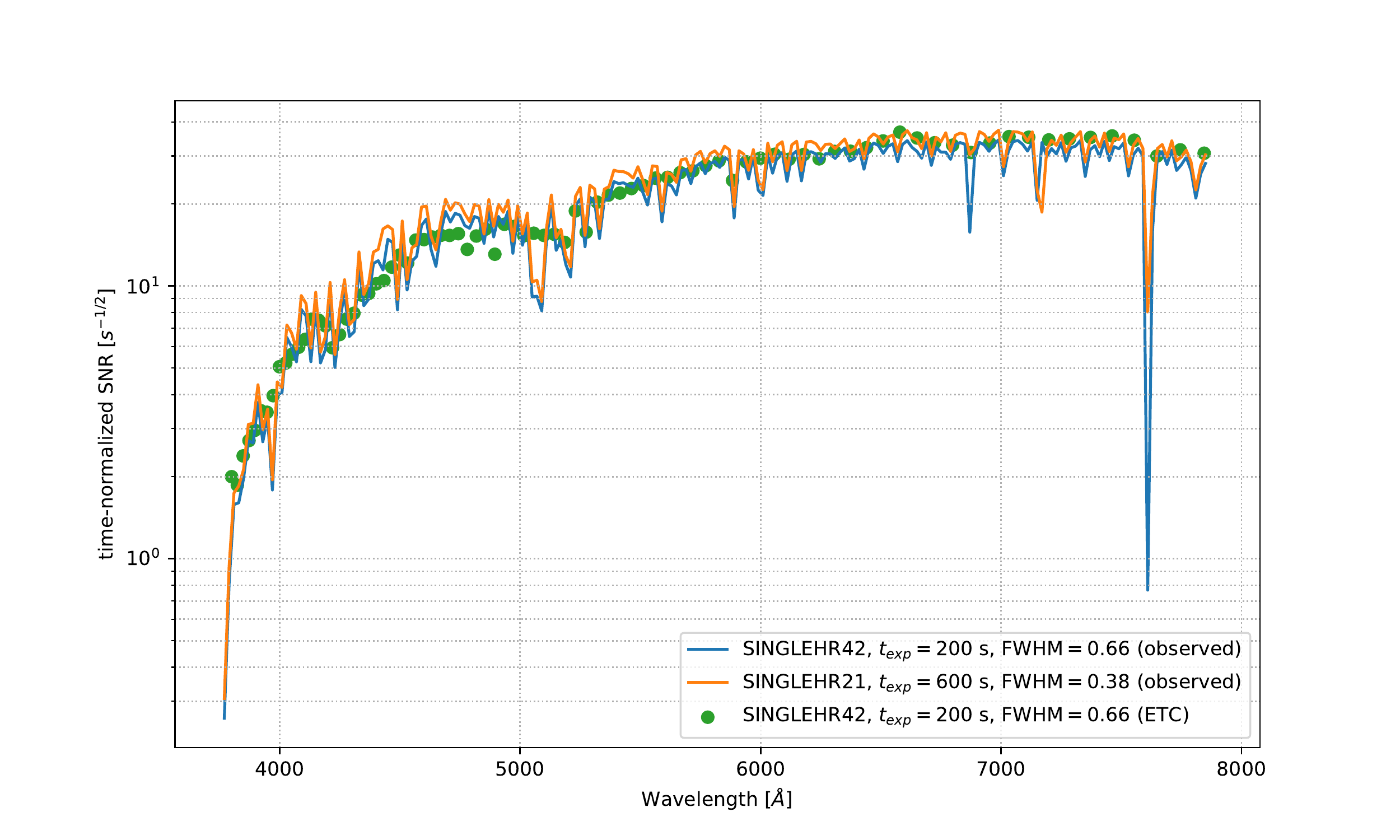}
   \caption{Comparison of the S/N for \object{HD\,85512} as a function of wavelength as computed by the ETC with the values measured in the HR42 and the HR21 modes, respectively.}
   \label{fig:etc_comparison}
\end{figure}

We consider thus $V = 14$\,mag to be the 'limiting magnitude' (transition between photon-noise and detector read-out noise limit) in the HR11 mode. While the S/N obtained in the HR21 and HR42 binning modes is identical up to this limiting magnitude, the limiting magnitude itself is shifted to about $V \sim 16$\,mag and $V \sim 18$\,mag, respectively. While the resolving power in the HR42 mode is reduced by about 5\% compared to the HR21 or HR11 modes, this difference is barely noted, since stellar absorption lines are resolved to some extent. Figure\,\ref{fig:HR42_HR21_comparison} shows a comparison of a small portion of the spectrum of \object{HD\,85512} observed with ESPRESSO in the HR21 and HR42 modes, respectively. The two spectra match almost perfectly after re-scaling, demonstrating the quality of the spectral extraction in the new HR42 mode and the negligible difference in spectral performance of the two modes. Furthermore, the spectral-energy distribution (SED), or the S/N measured as a function of wavelength, perfectly matches with the one predicted by the official ETC (Figure\,\ref{fig:etc_comparison}).

\subsection{S/N performance at faint magnitudes}
On the night of July 7th, 2019, i.e. after the fiber upgrade, we  observed the QSO \object{J16585649-5240283} of $V = 17.07$\,mag in HR21 mode with an integration time of 1'800\,s. The image quality measured at the Nasmyth platform of the telescopes was of $IA.FWHM = 1.1$\arcsec, while the DIMM seeing monitor indicated a seeing of the order of 0.9\arcsec~ during the exposure. The airmass was $AM=1.15$ on average. The average S/N measured at 550\,nm yields an $S/N=6.6$ in a spectral bin (extracted pixel) of 9\,m\AA~ (0.49\,\kms), and thus perfectly in line with the extrapolated envelope shown in Figure\,\ref{fig:snr_vs_mag}. In order to compare it with the technical requirements for the 1-UT configuration, this value has to be converted to an exposure of $texp=$\,3\,600\,s and a spectral bin of 24\,m\AA, where it yields an $S/N=15$, well above the specified value of $S/N=10$.\\

\begin{figure}[h]
   \centering
   \includegraphics[width=\hsize]{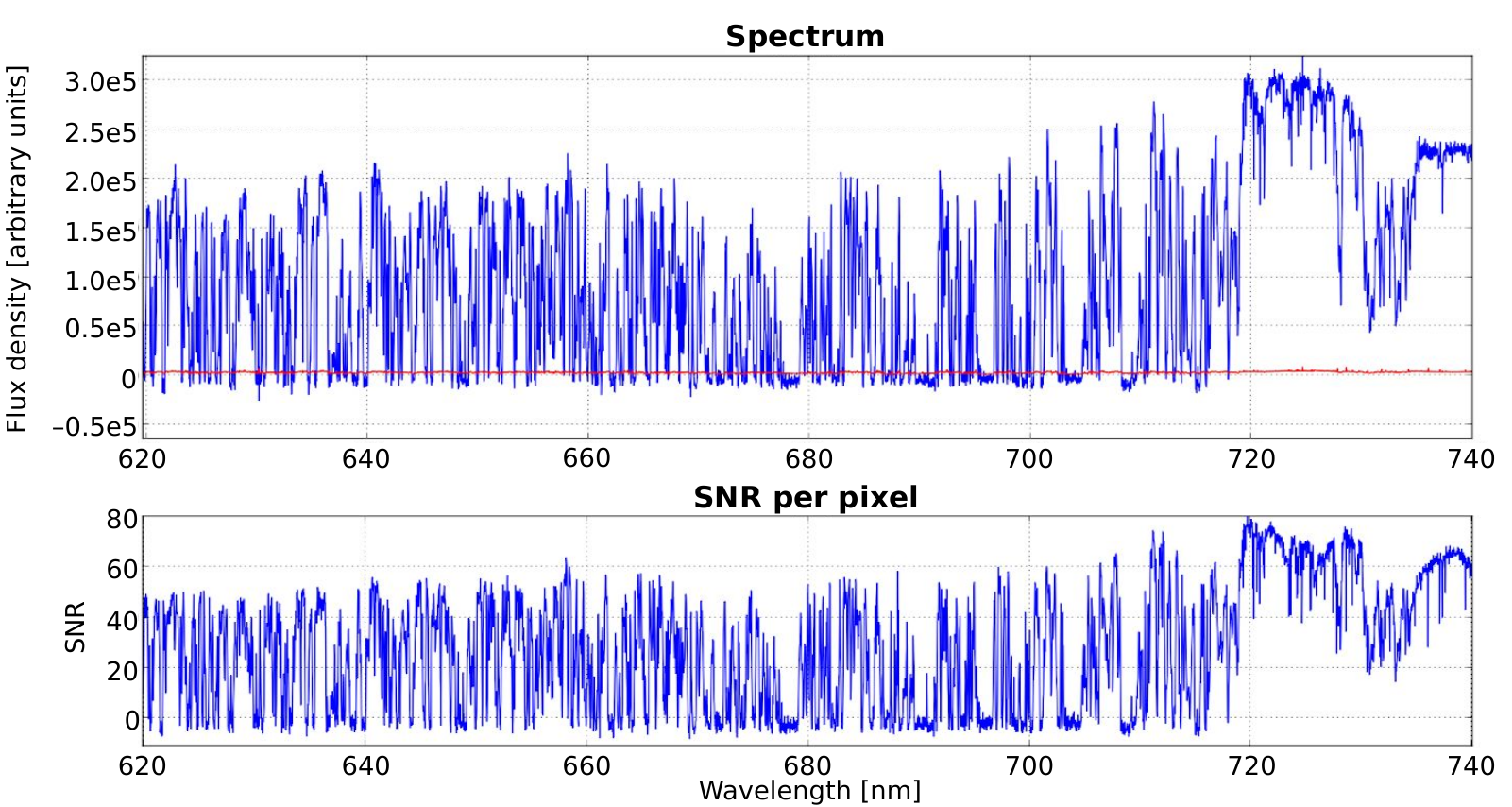}
   \caption{Reduced spectrum of the QSO J013539.27-212628.4 ($z=4.94$, $R=19.55$\,mag, \citet{li:2018}) showing the Lyman-alpha emission and a portion of the Lyman forest. Reduction was carried out at the telescope with the ESPRESSO DRS (see Section\,\ref{sec:drs}) and the spectra were resampled by the ESPRESSO DAS (see Section\,\ref{sec:drs}, \citet{cupani16} and \citet{cupani:2019}) using a fixed log-wavelength grid with a spectral bin of 2\,\kms. }
   \label{fig:qso_spectrum}
\end{figure}

On the night of December 1st, 2018, we observed the QSO \object{J013539.27-212628.4} of $V \sim 20$\,mag with an integration time of 1 hour in the MR mode with four UTs (Figure\,\ref{fig:qso_spectrum}). The as-delivered image quality at the Nasmyth platform of the telescopes was $IA.FWHM = 0.63$\arcsec, while the seeing monitor indicated a remarkable DIMM seeing of the order of 0.35\arcsec! The airmass was of $AM = 1.13$ on average. Data reduction was carried out at the telescope with the ESPRESSO DRS (see Section\,\ref{sec:drs}) and resampled by the ESPRESSO DAS (see Section\,\ref{sec:drs}, \citet{cupani16} and \citet{cupani:2019}). The average S/N measured at 550\,nm in HR84 mode yields, after subtraction of the background, $S/N = 8.5$ within a spectral bin specified for the 4-UT configuration of 48\,m\AA~ (2.63\,\kms); the extracted spectral bin (4 physical pixels) is about 35.5 mA~ (1.94\,\kms) and contains a measured $S/N$ of about 7.3. The S/N is computed in each spectral bin as the ratio between the collected flux and its formal error, propagated through reduction and resampling.

The reported measurement was obtained before the fiber-link intervention, thanks to which a gain of up to 40\% in throughput was achieved for the MR mode. We extrapolate from this result that after the fiber upgrade an $S/N = 9.7$ is achieved, very close to the specified value of 10 (spectral bin of 48\,m\AA, seeing = 0.8\arcsec). Finally, we also would like to point towards successful science-verification (SV) observations performed in August 2019 on a very faint lensed galaxy \citep{vanzella:2020} with $V \sim 21.5$\,mag and $texp = $\,3\,600\,s yielding to the significant detection of relevant emission lines despite the faint magnitude of the source and the high spectral resolution of the measurement.

\subsection{Radial-velocity performance}
\begin{figure*}[h]
   \centering
   \includegraphics[width=\hsize, trim= 75 60 95 60, clip]{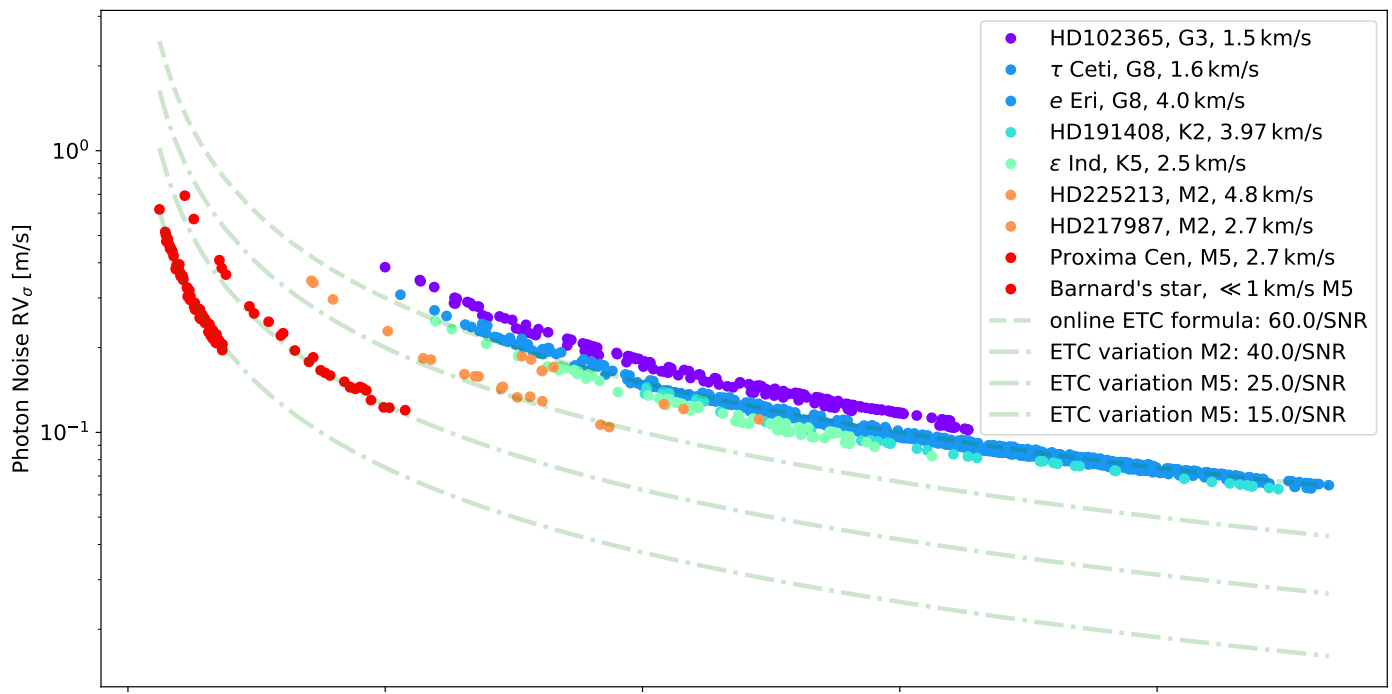}
   \caption{Internal (photon-noise limited) RV precision as a function of S/N. The dots show the 'on-spectra' determined RV error values for various standard RV targets for which the internal figure capture provides the spectral type and the $v \sin i$ value. The measured internal RV errors very nicely follow the scaled $1/(S/N)$ relation computed with the ETC and represented by the dash-dotted curves.}
   \label{fig:rv_vs_snr}
\end{figure*}

 In Figure\,\ref{fig:rv_vs_snr} we plot the \textit{measured} on-sky internal (photon-noise limited) precision in the standard HR mode as a function of measured $S/N$ for several RV-standards. At $S/N = 430$, 10\cms precision is reached for M stars, while for K and G dwarfs the RV precision lies between 10 and 20\cms, depending on the rotational level of the star, thus slightly worse than the specified value.\\

\begin{figure}[h]
   \centering
   \includegraphics[width=\hsize, trim= 60 60 60 60, clip]{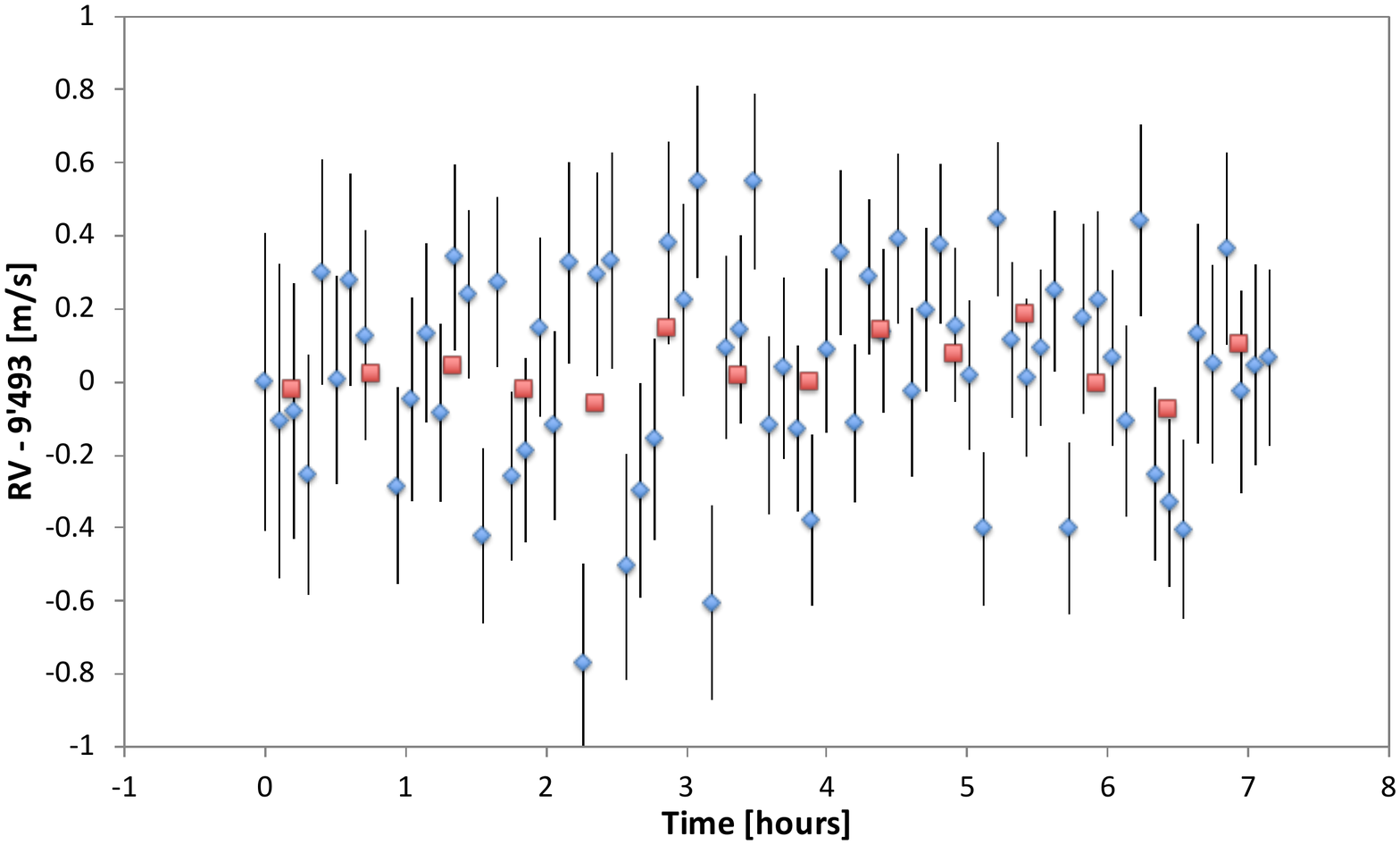}
   \caption{RV series (blue diamonds) of 7 hours on HD 85512. The red squares represent the binned data (5 data points each).}
   \label{fig:hd85512_night}
\end{figure}

In order to determine short- and medium-term RV repeatability, we observed, during the commissioning and in the frame of our GTO program, RV-standard stars taken from the HARPS high-precision program \citep{pepe:2011}. In Figure\,\ref{fig:hd85512_night} we show a time series recorded during commissioning 2A (March 1$^{st}$, 2018) on the $V = 7.65$\,mag K5-dwarf \object{HD\,85512}, a non-rotating, very quiet star. The sequence covers about 7 consecutive hours of observations in HR mode, binning 2x1, during which an average photon-noise error of 25\cms was obtained on each single exposure of 300\,s. The measured raw dispersion on the data is of 28\,\cms \textit{rms}, very close to the photon-noise uncertainty. When the exposures are binned in groups of 5 data points, the dispersion is reduced to 8\cms, again similar to the estimated average photon noise of 11\cms.

The dependence of the measurement of the RV from the telescope was tested using observations on the same star. Although this star hosts a planet ($K\sim0.8$\ms, $P\sim58$\,Days), it is one of the most quiet stars observable with ESPRESSO. Its magnitude is such that we can achieve a photon noise smaller than 20\cms in 15\,Min. of exposure. During a night with stable seeing and clear sky the star was observed in rapid sequence with all the four UTs. The exposure time of 900\,s allowed us to average the stellar acoustic pulsation (p-modes). As the data points in Figure\,\ref{fig:rv_UT} show, the RV variation between telescopes is within the 18\cms photon noise in this case. The RV variation observed when repeating the observation with UT1 and UT3 is attributed to stellar effects \citep{dumusque:2012}.
 
In order to test the long-term ``precision'' (measurement repeatability) we collected measurement series of RV-standard stars observed during the first year of operations. We note that the series include the moment of the fiber-upgrade in June 2019 and are likely to be affected by an instrumental RV offset. We advice ESPRESSO users to distinguish ESPRESSO data taken before and after June 2019 by introducing a free offset parameter, since it cannot be guaranteed that no instrumental RV-offset was introduced by this intervention on the spectrograph.

\begin{figure}[h]
   \centering
   \includegraphics[width=\hsize, trim= 45 20 80 30, clip]{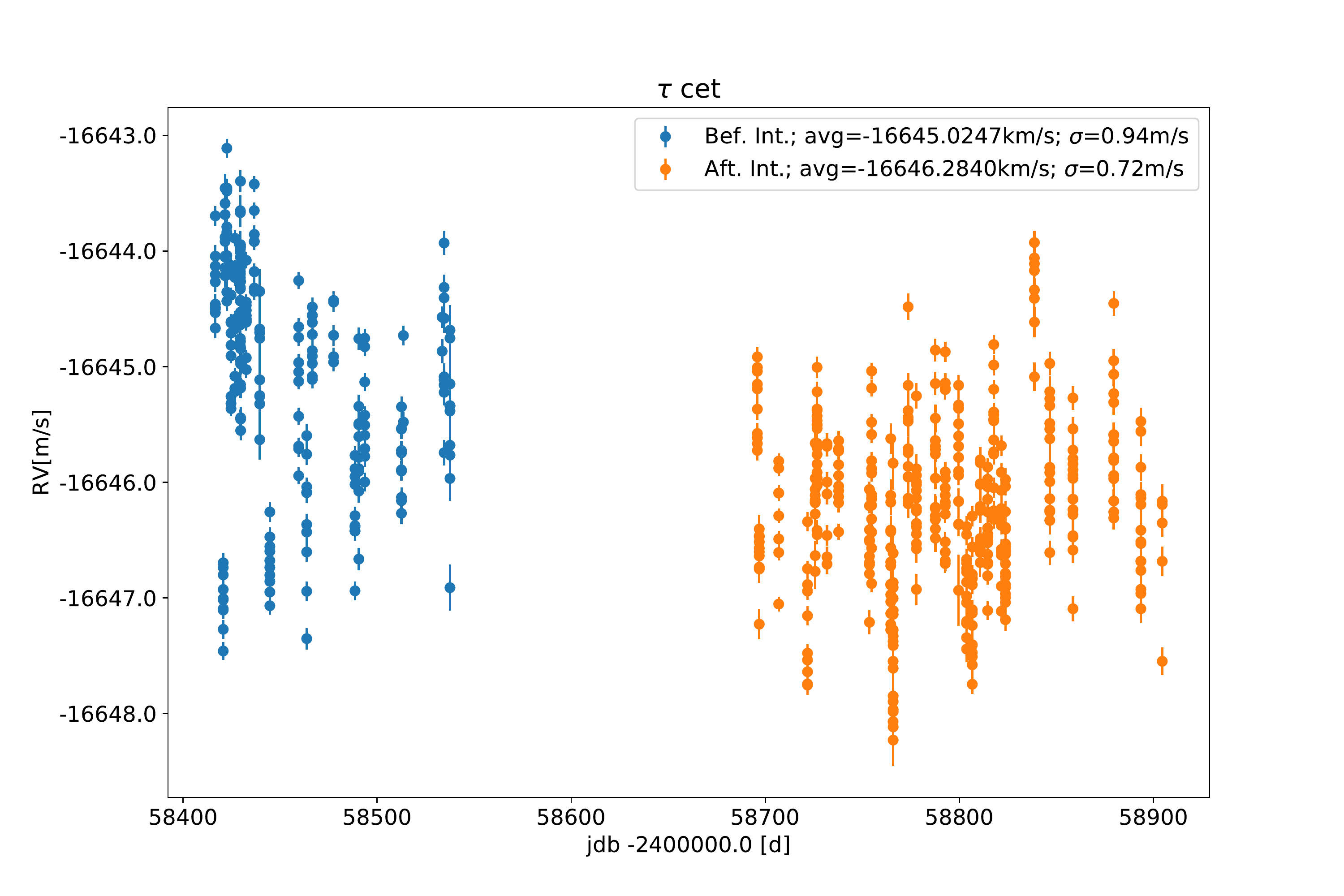}
   \caption{Radial velocities obtained with ESPRESSO on \object{$\tau$\,Cet} covering approximately one and half year of observations. Blue data points have been taken before, orange data points after the fiber upgrade of June 2019.}
   \label{fig:RV_repeatability_all}
\end{figure}

\begin{figure}[h]
   \centering
   \includegraphics[width=\hsize]{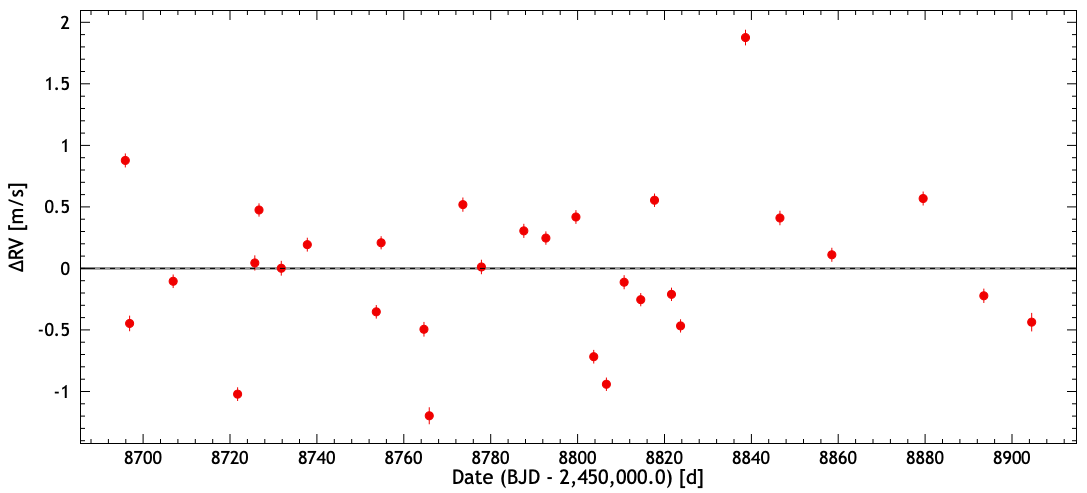}
   \caption{Radial velocities obtained with ESPRESSO on \object{$\tau$\,Cet}. Only data obtained after the fiber upgrade are shown. Each data point corresponds to a nightly bin. In this case the dispersion is of 0.58\ms}
   \label{fig:RV_repeatability_after}
\end{figure}

Figure\,\ref{fig:RV_repeatability_all} shows the radial velocities of \object{HD\,10700}, one of the two most stable stars of the GTO Working-Group 1 program (blind search for rocky planets). Data points taken before the fiber upgrade are shown in blue, those after the intervention in orange. The raw scatter of the un-binned data obtained before the fiber upgrade is 0.94\ms and after the fiber upgrade of 0.72\ms. We suspect that the higher dispersion in the early lifetime of the instrument is due to some 'instabilities' introduced by operational issues (air-condition failures, thermal instabilities of the detector cryostat, etc.). The scatter of the radial velocities of \object{$\tau$\,Cet} (\object{HD\,10700}) decreases to 0.5\ms when nightly bins are considered (Figure\,\ref{fig:RV_repeatability_after}). In summary, a long-term ($\sim1$\,year) RV precision of \textit{better} than 0.5\ms is demonstrated, despite the intrinsic measurement limitations set by the astrophysical targets themselves. ESPRESSO delivers (non-photon-noise limited) RVs better than HARPS when compared over a similar time span and the so-far tested RV performance is compatible with the 10\cms-precision requirement.

The RV repeatability in the MR mode has been analyzed on a one-hour time scale. It has been found to be better than 50\cms, consistent with the photon noise, and significantly better than the RV-precision requirement for this mode of 5\ms. The test was repeated after the fiber-link exchange but with only three UTs (UT1 was out of operations), obtaining a short-term repeatability of better than 40\cms, consistent with photon noise.

The RV repeatability of the UHR mode was tested to better than 50\cms (13\cms photon noise) on the short term (night). The measurement is dominated by the 5 minutes acoustic oscillations of the observed star (\object{HD\,10700}, exposure time of 1\,Min.). There was no requirement specified for the RV-precision in the UHR mode, but we expect it to be comparable to the HR mode.

\subsection{Equivalence of Unit Telescopes}
\begin{figure}[h]
   \centering
   \includegraphics[width=\hsize, trim= 10 0 30 40, clip]{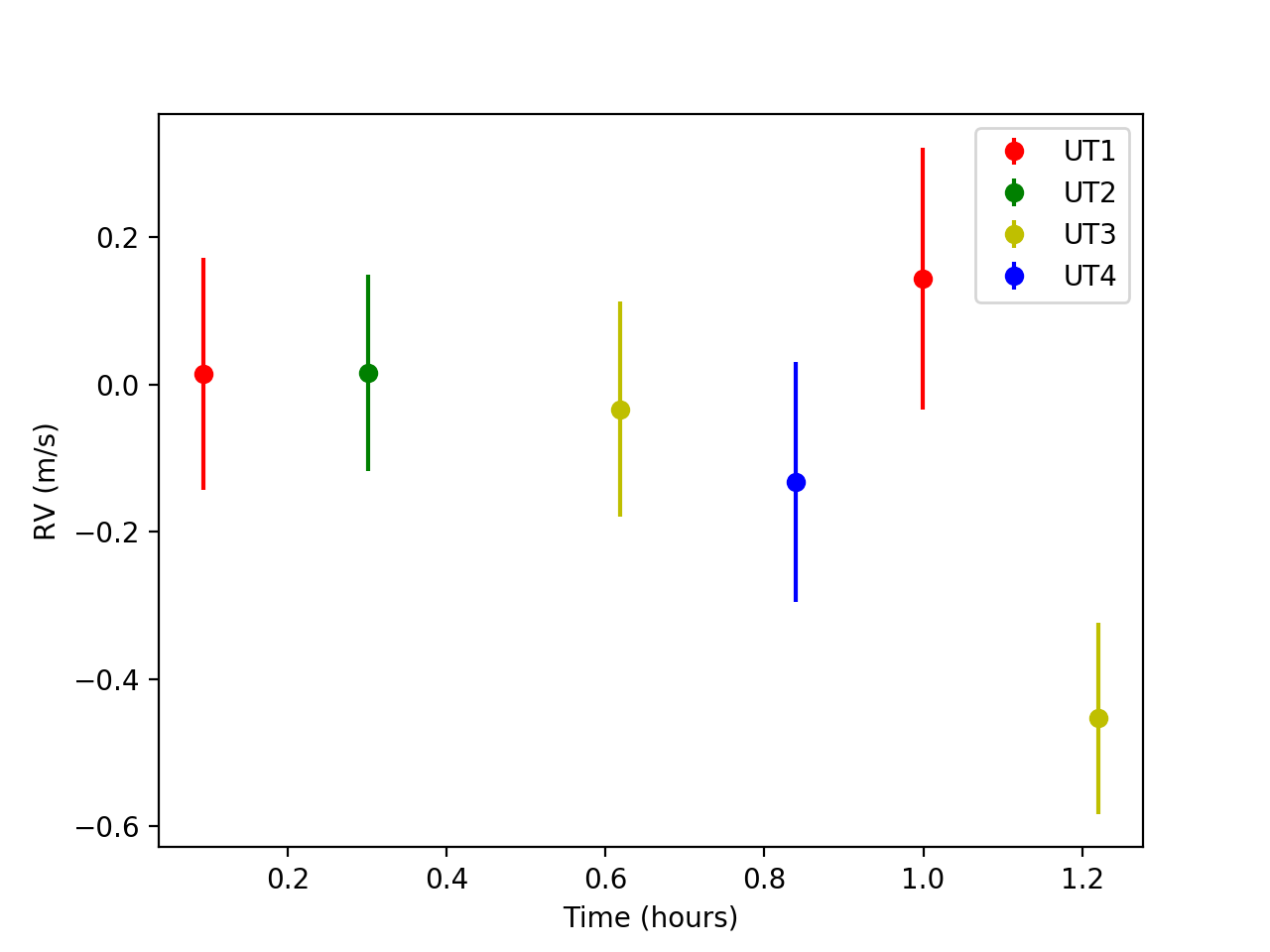}
   \caption{Radial velocitiy series on the quiet star \object{HD\,85512} (K6V, $V=7.65$\,mag) obtained with ESPRESSO while cycling through the various UTs. Individual exposures of 600\,s are shown. Error bars are essentially constant, demonstrating that the UTs have similar throughputs. The RV dispersion is compatible with the photon-noise induced RV error of about 18\cms \textit{rms} and with possible residual stellar p-modes that can be present even on late-type stars at this level of precision.}
   \label{fig:rv_UT}
\end{figure}

The fact that ESPRESSO can operate with any of the four UTs calls for a very particular requirement, namely that any performance figure of ESPRESSO is independent from the telescope used. While transmission may vary between one telescope and another, due to e.g., the different ages of the mirrors coatings (typically re-coated every two years), these differences are comparable to the absolute and chromatic variability introduced by the varying astro-climatic conditions (seeing, transparency and airmass, clouds). Despite these possible variations, we have conducted on-sky tests to demonstrate that both transmittance and RVs measured with ESPRESSO do not dependent on the UT on which the observations is carried out. Figure\,\ref{fig:rv_UT} shows RV measurements of the same star when switching from one telescope to the other. Error bars show that the photonic RV-uncertainty ($\sim 18$\cms), and thus the telescope transmittance, remain essentially constant. Since RV precision depends linearly on signal-to-noise ratio (S/N), this test demonstrates also that the throughput of ESPRESSO varies from one UT to the other at a level below the effect induced by the variability of observing conditions. A similar statement holds for the RV measurement: While cycling through the telescopes (first four measurements) no significant RV offset is observed. Returning to UT-1 (fifth measurement), the same RV is obtained within the error bar as for the very first measurement. The second measurement on UT-2 (sixth measurement) shows a 2-sigma deviation from the previous measurements. The RV dispersion of the series remains nevertheless compatible with the photonic error of ($\sim 18$\cms). The test was repeated on other stars and under different conditions, and similar results were obtained. They demonstrate that, if RV-offsets are induced by changing the UT, they must be smaller than 20\cms \textit{rms}.

Transmission and RV stability measurements are part of ESPRESSO's calibration plan and are monitored routinely. As for any of the ESO instruments, ``health'' and performance of ESPRESSO are constantly recorded and published on dedicated web pages.\footnote {\url{www.eso.org/observing/dfo/quality/ESPRESSO/common/score\_overview.html}}.


\section{Guaranteed-time observations with ESPRESSO}
\label{sec:obs}
In exchange for funding and building the ESPRESSO instrument, the ESPRESSO Consortium was awarded 273\,nights of observations with the VLT (known as Guaranteed-Time Observation, GTO), following a well-established and successful scheme of ESO for funding second-generation instruments for the La Silla and Paranal Observatory. Although formally granted to the Principal Investigator representing the Consortium, the latter has to present observing proposals to ESO prior to start of operations. The GTO proposals have to be approved by ESO's Observing Program Committee (OPC) before execution. All of the GTO proposals must follow the science program proposed by the Consortium to ESO at the time of the signature of the construction agreement. In the case of ESPRESSO, the proposed program foresaw to dedicate 80\% of the granted GTO to the search and characterisation of exoplanets and 10\% to the study of the (possible) variability of fundamental physical constants. The remaining 10\% of the GTO had been reserved for "other science cases" of interest for the Consortium that would become topical at the time of the start of operations. In the following, we briefly describe the objectives, the contents and the organisation of the ESPRESSO Consortium's GTO program. 

\subsection{Exoplanets}

The search for extrasolar planets is one of the fastest-expanding subjects in modern astrophysics. It was the detection of \object{51\,Peg\,b}, the first extrasolar planet around a solar-type star, by \citet{mayor:1995}, that engaged a whole new community in the hunt for exoplanets. \object{51\,Peg\,b} was detected through the RV variation its orbit induced in the parent star, that became one of the most efficient techniques for the detection and characterisation of exoplanets. After providing a significant wealth of information on hot-Jupiter characteristics, the RV technique has progressively revealed the existence of planets with lower mass down to the Earth- and super-Earth mass regime (including around our neighboring M-dwarf \object{Proxima\,Cen} by \cite{anglada:2016}. Over the past two decades, we have learned that at least 50\% of the stars in the solar neighborhood host a planet, and that most of these can be found in multi-planetary systems (e.g. \citet{lovis:2011}). Much of what is known about the exoplanet population comes from the detailed analysis of RV surveys (e.g. \citet{udry:2019}). The efficiency of the RV technique is only challenged by that of space missions like $Kepler$ (e.g., \citet{borucki:2011} and \citet{batalha:2013}), which delivered the largest number of transit candidates for the count of more than 4\,000\,planets known today. 

The combination of transits and spectroscopy played a fundamental role in the detailed characterisation of individual planets. Together, i.e. by the technique called "transit spectroscopy", they provide information on the parent star, precise measurement of planetary masses, radii and thus planetary densities, and allow the investigation of the planetary atmospheres. High-fidelity spectroscopy and RVs are essential tools to achieve these goals. When officially proposed to ESO (before the first results provided by the $Kepler$ mission and well before the discovery of Earth-mass planets around stars in the solar neighborhood), ESPRESSO had already been conceived to pursue the following scientific objectives in the exoplanetary field: the discovery and characterisation of Earth-mass planets (possibly) within the habitable zones of their parent stars of spectral types GKM, the characterisation of the atmospheres of individual exoplanets, and the precise mass determination of low-mass transiting planets announced by the $Kepler$ and $TESS$ missions. In what follows, we describe these three sub-programs. 

\subsubsection{Blind search for rocky planets in the habitable zone}
One of the main science drivers for ESPRESSO is the detection of Earth-mass planets in the habitable zone (HZ) around GKM stars. It was this driver that set the most stringent constraints on spectral resolution, thermo-mechanical and illumination stability, ultimately aiming at 10\cms of RV precision. This enables the search for Earth-mass planets inside solar-type stars' HZ, provided that these stars are 'quiet' and sufficiently bright, thus pointing towards late-G, K and M-dwarfs of the solar neighborhood. Their proximity make them particularly suitable for follow-up studies, such as for direct detection and characterization of their atmospheres. Starting in 2010, we performed several preparatory studies to define our target sample. We investigated the possible contamination by faint, spatially unresolved companions \citep{cunha:2013} or the impact of micro-tellurics on precise RVs \citep{cunha:2014}, among others. To avoid being biased by previous RV surveys, we built our WG-1 catalog from scratch, performing a spectroscopic characterization for the stars that had not yet been observed at high spectral resolution. We obtained observing time on HARPS, HARPS-N, and UVES to complement HARPS observations retrieved from the ESO archive. For all candidates we performed a detailed spectroscopic characterization, calculated the rotational velocity, and chromospheric activity indexes, among several other studies. The final target list contained 77 close-by, quiet, non-rotating stars covering a wide range of right ascension values and being thus observable the entire year. For these stars the photon noise attained in a 15-Min. integration time (or 30\,Min. for the faintest targets) allows us to detect the RV signal of a one Earth-mass planet orbiting inside the HZ. The details on the target selection, spectral characterization and further studies, along with reduced data access, are presented in \citet{hojjatpanah:2019}.

During the first year and a half of observations, we narrowed down our sample to $\sim45$ promising targets, of which 25 will be followed intensively and on which we expect to gather an average of 60-70 visits until the end of the GTO. The remaining 30 targets were discarded due to the presence of stellar companions, or simply to avoid competition with other programs already following them. 

It is worth mentioning that for several stars of this sub-program the RV scatter is inferior to 1\ms; we stress that this value is computed without any data post-processing and with the RVs provided by the publicly available pipeline. For \object{$\tau$\,Cet}, the RV scatter of the nightly-binned RVs is of 50\cms, as reported in a previous section. In Suarez-Mascareno et al. (2020, A\&A, in press) the first paper of this sub-program, 63 ESPRESSO RV-points are used to independently confirm the planet \object{Proxima Cen\,b}, characterize the stellar activity signal on the star and precisely determine the planetary. Once the planetary and stellar activity models are removed, the residuals scatter is at the level set by photon-noise of $\sigma_{RV}=30$\cms, demonstrating unprecedented precision.

\subsubsection{Characterization of planetary atmospheres}
The study of planetary atmospheres from the ground (see e.g. \cite{snellen:2008,redfield:2008,charbonneau:2002}) represents a breakthrough in our understanding of what exoplanets are made of (chemical composition), which is essential to determine planetary evolution and may be key to disentangle planet formation processes, for example, whether a planet was born in the orbit it is observed in or whether it has migrated over a large distance. This is the case of warm and hot exoplanets with short orbital periods that lie very close to their parent stars. These exoplanets are a natural laboratory for chemistry and quite likely, formation studies. Their high atmospheric temperatures prevent condensation and sinking of refractory species, thus making the atmospheric composition more representative of the whole chemical content. Before ESPRESSO, various atomic and molecular species have been detected at optical wavelengths in Jupiter-mass planets with expected equilibrium temperatures typically above $\sim$1\,000\,K (sodium, potassium, calcium, magnesium, neutral and ionised iron and titanium, titanium oxide, and hydrogen, see for example \citet{hoeijmakers:2019}, \citet{casasayasbarris:2019}, and references therein). Water vapour, helium and possibly methane and carbon-bearing species are also seen in hot massive exoplanets in the near-infrared (for example, see \citet{nortmann:2018}, \citet{sotzen:2020}, and references therein). Some warm Neptunes (e.g., \object{GJ\,436b}, \object{GJ\,3470b}) and the super-Earth \object{GJ\,1214b}, on the contrary, show cloudy atmospheres, which do not allow the determination of their composition. The least dense planets at small orbital distances from their host stars present atmospheres undergoing evaporation processes. These pioneering results have been possible through transit spectroscopy with the HST, Spitzer and ground-based facilities, including a pathfinder ESO program using HARPS and HARPS-N \citep{wyttenbach:2017}. Complementing these studies, high-resolution spectroscopy with HARPS has also shown to be a promising method to detect exoplanet atmospheres in reflection \citep{martins:2015}. It seems that there is a great variety of planetary atmospheres, and even though studies of individual outstanding exoplanets represent a highlight per s\'e, large number of observations are needed to address relevant questions on planet diversity, formation and evolution. The ESPRESSO GTO program will provide optical spectra of more than 30 exoplanetary atmospheres (from hot to warm planets) over the GTO duration and will contribute to the direct measurement of their atmospheric composition and structure.

One of the first ESPRESSO-GTO results is first the detection of an asymmetric absorption signature due to neutral iron during the primary transit of the ultra-hot giant planet WASP-76b \citep{ehrenreich:2020}. This feature is blue-shifted in the trailing limb (probably owing to the combination of planetary rotation and atmospheric wind), but no signature appears in the leading limb. We interpreted this as an evidence of the condensation of gaseous iron, which is injected from the hot (day) side atmosphere, in the cool (night) side of the planet. In coming papers, the ESPRESSO-GTO team will show that this scenario is likely to happen in more planets with similar temperatures as that of WASP-76b, and will report robust detections of atomic elements in the atmospheres of warm planets, like lithium, that were not solidly established before (Tabernero et al. in prep.; Borsa et al. in prep.).

\subsubsection{RV follow-up of \textit{K2} and \textit{TESS} transiting planets}

The \textit{Kepler} space mission has revolutionized our knowledge of exoplanets by discovering thousands of small transiting planets with radii below 4\,R$_\oplus$. It thereby confirmed that systems of short-period super-Earths and Neptunes are ubiquitous in our Galaxy; in fact, those appear to be a very frequent outcome of planet formation. However, the picture that emerged from the Kepler results has remained largely incomplete: the faintness of the host stars has been hampering a comprehensive characterisation of the Kepler systems through ground-based follow-up. In particular, host star fundamental parameters such as stellar radii are only poorly known in many cases, leading to inaccurate planetary radii \citep{fulton:2017,vaneylen:2018}. Also, planet masses could be reliably measured only in a limited number of systems having sufficiently bright host stars (e.g., \citet{buchhave:2016}). This situation improved with the advent of the K2 mission, which scanned the ecliptic plane in search for new transiting planets, orbiting typically brighter stars than the original Kepler discoveries. Since 2018, another space mission is further revolutionising the exoplanet field: the Transiting Exoplanet Survey Satellite (\textit{TESS}). \textit{TESS} is currently performing an all-sky transit search from space and has already discovered hundreds of new transiting planets orbiting brighter stars than \textit{K2}. Intense follow-up efforts using HARPS-like instruments have been ongoing for almost a decade to characterize super-Earths and Neptunes found by these missions, however still with a limited scope; only the more massive planets on shorter-period orbits are amenable to robust mass measurements.

ESPRESSO represents a breakthrough in this respect: Its 2-magnitude gain and improved RV precision with respect to HARPS-like instruments will for the first time enable a thorough exploration of the rocky planet population detected by \textit{K2} and \textit{TESS}. These rocky worlds have radii below $\sim$1.8\,R$_\oplus$ and their expected masses are typically below 6-8\,M$_\oplus$, although their mass distribution remain largely unconstrained as of today. ESPRESSO observations will be able to constrain the internal composition of these objects, i.e. their iron/rock/water mass fraction, under the reasonable assumption that they do not have any H/He envelope. Moreover, ESPRESSO will explore how the properties of this population vary with stellar irradiation, stellar mass, planetary architecture, and even stellar composition \citep{santos:2017}, shedding new light on the formation and evolution pathways of these systems. Ultimately, ESPRESSO will be able to characterize transiting rocky planets within the habitable zone of their host star with unprecedented details.

The ESPRESSO-GTO \textit{K2}/\textit{TESS} planet sample has been defined based on the following criteria: a) confirmed/validated planet candidate; b) planet radius $<2.0$\,R$_\oplus$; c) host star $V<14.5$\,mag (spectral-type dependent); and d) no precise mass measurement available. In addition, we intend to probe the rocky-to-gaseous transition at a given stellar irradiation level. This will allow us to constrain the evaporation processes that transform sub-Neptunes into naked rocky cores under the influence of XUV irradiation from the host star. According to the results of \citet{fulton:2017}, a particularly interesting irradiation range to carry out this experiment is $\sim50-200$\,$\times$ Earth's irradiation levels, where both gas-rich sub-Neptunes and gas-poor rocky planets appear to co-exist. Mass and density measurements from ESPRESSO will thus probe the mass and density thresholds leading to significant evaporation of planetary envelopes. The target sample for this experiment is the same as the one defined above, complemented in the following way: 1) adding the larger planets that are in the same systems as the rocky planets defined above, and 2) adding a few more sub-Neptunes and Neptunes ($<4$\,R$_\oplus$) in other systems, which are under irradiation levels of 50-200\,$\times$ Earth's level. The main objective of the program is to measure precise masses and bulk densities for all these objects. In total, we aim at characterising 50-100 small planets by the end of the GTO period.\\

To illustrate the potential of ESPRESSO and this sub-program we refer to the case of the planetary system \object{$\pi$\,Men}. \object{$\pi$\,Men} is a bright, ``naked-eye'' G0V-star ($V=5.6$\,mag) known to host a sub-stellar companion on a long-period and very eccentric orbit of $P_{\rm b}\sim$2\,000\,days. On September 6, 2018, The NASA Transiting Exoplanet Survey Satellite (\textit{TESS}; \citealt{ricker15}) announced the discovery of a second planet, the transiting super-Earth/sub-Neptune \object{$\pi$\,Men\,c} ($P_{\rm c}=$\,6.27 days; $R_{\rm c}=2$\rearth). Following the announcement, \cite{huang18} and \citet{gandolfi18} independently detected the spectroscopic orbit of \object{$\pi$\,Men\,c} by analyzing archival RVs of HARPS and UCLES, and confirmed its planetary nature.

The brightness of the star makes \object{$\pi$\,Men} a perfect target for testing the performance of ESPRESSO in measuring the mass and bulk density of low-mass planets. From September 5$^{th}$, 2018, to March 25$^{th}$, 2019, we recorded therefore a total of 275 spectra and covered a time span of 201 days. The spectra were acquired with a typical exposure time of 120\,s, providing median $S/N=243$ per extracted pixel at $\lambda=$\,500\,nm and a median RV (internal) precision of 25\cms. During each observing night we collected series of multiple spectra at a rate of 2 to 12 consecutive exposures, leading to a typical in-night RV precision of the binned data points of about 10\cms. Furthermore, \textit{TESS} re-observed the host star during cycle 1 (sectors 4,8,11-13) from October 2018 to July 2019, collecting 19 additional transits of planet\,c in short-cadence mode.

A detailed and complete analysis of the system architecture of \object{$\pi$\,Men} is presented in a parallel paper by Damasso et\,al. (this same edition of the journal). The authors combined the exquisite RVs of ESPRESSO with the \textit{TESS} photometry and with the \textit{Gaia} and \textit{HIPPARCOS} astrometry. Besides refining the mass of the transiting super-Earth and reducing the error bars on its orbital parameters, the authors also constrained the mutual orbital inclination of the two planets, the real mass of the long-period planet\,b and the system architecture in general.

\subsection{Varying fundamental constants}
The standard model of particle physics depends on many ($\sim27$) independent numerical parameters that determine, inter alia, the strengths of the different forces and the relative masses of all known fundamental particles.  There is no theoretical explanation for their actual value, but  nevertheless they determine the properties of atoms, cells, stars and the whole Universe. They are commonly referred to as the fundamental constants of nature, and our only definition for them is: any parameter that can not be calculated from first principles in a given theoretical framework, but must be determined by experiment. Nevertheless, most of the modern extensions of the standard model predict a variation of these constants at some level \citep{uzan:2011,bonifacio:2014,Martins:2017}.

For instance, in any theory involving more than four space-time dimensions, the constants we observe are merely four-dimensional shadows of the truly fundamental high dimensional constants. The four dimensional constants will then be seen to vary as the extra dimensions change slowly in size during their cosmological evolution. The most common example in modern cosmology are the quintessence models for the dark energy, in which the underlying dynamical scalar field is naturally expected to couple to the electromagnetic sector of the theory and lead to a variation of the fine-structure constant \citep{Carroll:1998}; thus tests of stability of the latter are a competitive cosmological probe \citep{CMartins:2015}. Earth-based laboratories can measure the local drift rate of these constants. For the fine structure constant $\alpha$, this has been constrained to a few parts per $10^{-17}$ yr${}^{-1}$ \citep{rosenband:2008}. This is more than six orders of magnitude slower that the typical cosmological drift rate (viz. $10^{-10}$ yr${}^{-1}$), showing that any such scalar field must be evolving very slowly.

Astronomical observations have a great potential for probing fundamental constants stability at very large distances and in the early Universe, and therefore identifying any slow but not-trivial evolution. In fact, the transition frequencies of the narrow metal absorption lines observed in the spectra of distant quasars are sensitive to $\alpha$, as first pointed out by \citet{bahcall:1967}, and those of the rare molecular hydrogen clouds are sensitive to $\mu$, the proton-to-electron mass ratio, as first pointed out by \citet{thompson:1975}.

With the advent of 10-m class telescopes, observations of spectral lines in distant QSOs gave the first hints that the value of the fine structure constant might change over time, being lower in the past by about 6 ppm \citep{webb:1999,murphy:2004}. The addition of other 143 VLT/UVES absorbers  have revealed a $4\sigma$  evidence for a dipole-like variation in $\alpha$ across the sky at the 10\,ppm level \citep{webb:2011,king:2012}. Several other constraints from higher-quality spectra of individual absorbers exist \citep{Molaro:2013,Evans:2014,Murphy:2016,Kotus:2016, Bainbridge:2017,Murphy:2017} but none of them  directly supports or strongly conflicts with the $\alpha$ dipole evidence, and a possible systematic producing opposite values in the two hemispheres is not easy to identify.

In order to probe $\mu$, the H$_2$ absorbers need to be at a redshift $z>2-2.5$ to place the Lyman and Werner H$_2$ transitions redwards of the atmospheric cut-off. Only five systems have been studied so far, with no current indication of variability at the level of ~10\,ppm \citep{rahmani:2013}. At lower redshifts precise constraints on $\mu$-variation are available from radio- and millimeter-wave spectra of cool clouds containing complex molecules like ammonia and methanol \citep{levshakov:2013}. Other techniques involving radio spectra typically constrain combinations of constants by comparing different types of transitions (e.g. electronic, hyperfine, rotational, etc.).

Extraordinary claims require extraordinary evidence, and a confirmation of variability with high statistical significance---in practice, at the level of 1\,ppm---is of crucial importance. Only a high resolution spectrograph that combines a large collecting area with extreme wavelength precision can provide definitive clarification. A relative variation in $\alpha$ or $\mu$ of 1\,ppm leads to velocity shifts of about 20\ms between typical combinations of transitions. ESPRESSO is therefore ideal for this task. About 10\% of the GTO time will be used to collect spectra of a sample of  carefully selected, relatively bright and well studied QSOs to significantly improve current measurements of $\alpha$  as well as $\mu$ with the aim to break the 1\,ppm precision.

The proposed observations, a first version of which is summarized in \citet{Leite:2016}, will test the universality of physical laws in an unexplored regime, which directly impacts fundamental physics and theoretical cosmology. While the impact of confirmed spacetime variations is self-evident, more stringent upper bounds on these variations are also important and will lead to improved constraints on the parameter space of various theoretical paradigms that predict their variability \citep{Martins:2017,Alves:2017,Martins:2019}.

\subsection{``Other science cases''}
The remainder 10\% of the GTO was not allocated to a specific science case but left available for ad-hoc proposals that would arise during the execution of the GTO program. Upcoming ideas are discussed and reviewed internally within the ESPRESSO-GTO Science Team before every call for proposal for the following observing period at ESO. Several programs have been or are being executed this way and have led to results, which will be partially presented in the next section. 

\subsubsection{Are very metal poor stars binaries?}
The most metal-poor stars in the Galaxy are  the most ancient fossil records of the chemical composition and  can provide clues on the pre-Galactic phases and on the nature of stars which synthesized the first metals. Masses and yields of Pop.\,III stars can be inferred from the observed elemental ratios in the most metal poor stars \citep{heger:2010}. One crucial question  is to explain the mere presence of very metal poor  low-mass stars. Indeed, for a long time the  first stars formed have been thought to be very massive but the recent discovery of several very metal-poor stars with [Fe/H]${\le -5.0}$  have shown an entirely new picture \citep{caffau:2011}.\\
There are  about   a dozen of  stars  known with [Fe/H]$< -4.5$  reaching down to [Fe/H]$\approx -7$, i.e. 10 million times below solar.  All but two show  high-C overabundance (CEMP stars). We are carrying on ESPRESSO observations  in order to probe their binary nature by means of accurate RV measurements. The observations will allow us to understand if  C  comes from a companion AGB or is related to a different kind of objects such as Weak SNe or Spinstars and providing unique information on the nature of the first stars in the Universe and on the first elemental enrichment.   

The vast majority of the known stars of ultra-low metallicity are known to be enhanced in carbon, and belong to the low-carbon band (A(C)$\le7.6$). It is generally, although not universally, accepted, that this peculiar chemical composition reflects the chemical composition of the gas cloud out of which these stars were formed. The first ultra metal-poor star discovered, HE\,0107--5240, is also enhanced in carbon and belongs to the {\it low-carbon band}. It has recently been claimed that its peculiar composition could be due to mass-transfer to a formerly AGB companion \citep{arentsen:2019}. ESPRESSO observations obtained in 2018 demonstrate unambiguously that the radial velocity of HE\,0107--5240 has increased from 2001 to 2018 and that the star is indeed a binary \citep{bonifacio:2020}. However, binarity does not necessarily imply mass exchange and the nature of the companion remains undetermined. The companion  could be either a white dwarf or an un-evolved star. The binarity of such an object holds important bearings on the formation of small mass from a cloud without elements able to cool the gas and favour the first condensations. High-resolution simulations of star formation from primordial gas have shown the importance of the fragmentation processes \citep{greif:2011}. This provides a way of producing low-mass stars even in the absence of efficient cooling mechanisms, favouring the formation of binary systems. 

\subsubsection{Asteroseismology of K-dwarf stars}
So far, no p-modes oscillations were detected on solar-type stars with spectral type later than K1V (e.g. \object{$\alpha$\,Cen\,B} \citep{carrier:2003,kjeldsen:2005}). An estimate of the oscillation-mode amplitude, frequency domain, frequency of the maximum amplitude $\nu_{max}$ and large splitting $\Delta\nu$ in a mid-K dwarf with an effective temperature $T_{\rm eff}$ lower than 5\,000\,K will allow to refine the scaling relation from the Sun \citep{kjeldsen:2011} at the cool edge of the main sequence. The frequency at which a solar-like oscillator presents its strongest amplitude $\nu_{\rm max}$ scales with the surface gravity $\log g$ and effective temperature $T_{\rm eff}$. The detection of p-modes oscillations in K dwarfs is still out of reach for photometric space missions and the level of radial-velocity precision required is out of reach for any spectrograph other than ESPRESSO.  

\object{HD\,40307} is one of the brightest mid-K dwarf and is well located for long time-series acquisition from the southern hemisphere. According to HARPS monitoring \citep{diaz:2016}, \object{HD\,40307} is a quiet star with log$R'_{hk}$ of -4.94, which presents a magnetic cycle with a period of about 10 years and hosts a four-planets system. According to its effective temperature (4\,977\,K), luminosity (0.228\,L$_{\odot}$) and mass (0.77\,M$_{\odot}$), \object{HD\,40307} is expected to have p-modes oscillations with an amplitude of 6-10 {\cms} and a typical period of 2.5-3\,Min., and a large separation of modes $\Delta\nu \sim 220-230$\,$\mu$Hz. 

\begin{figure}[h]
\begin{center}
  \includegraphics[width=\hsize]{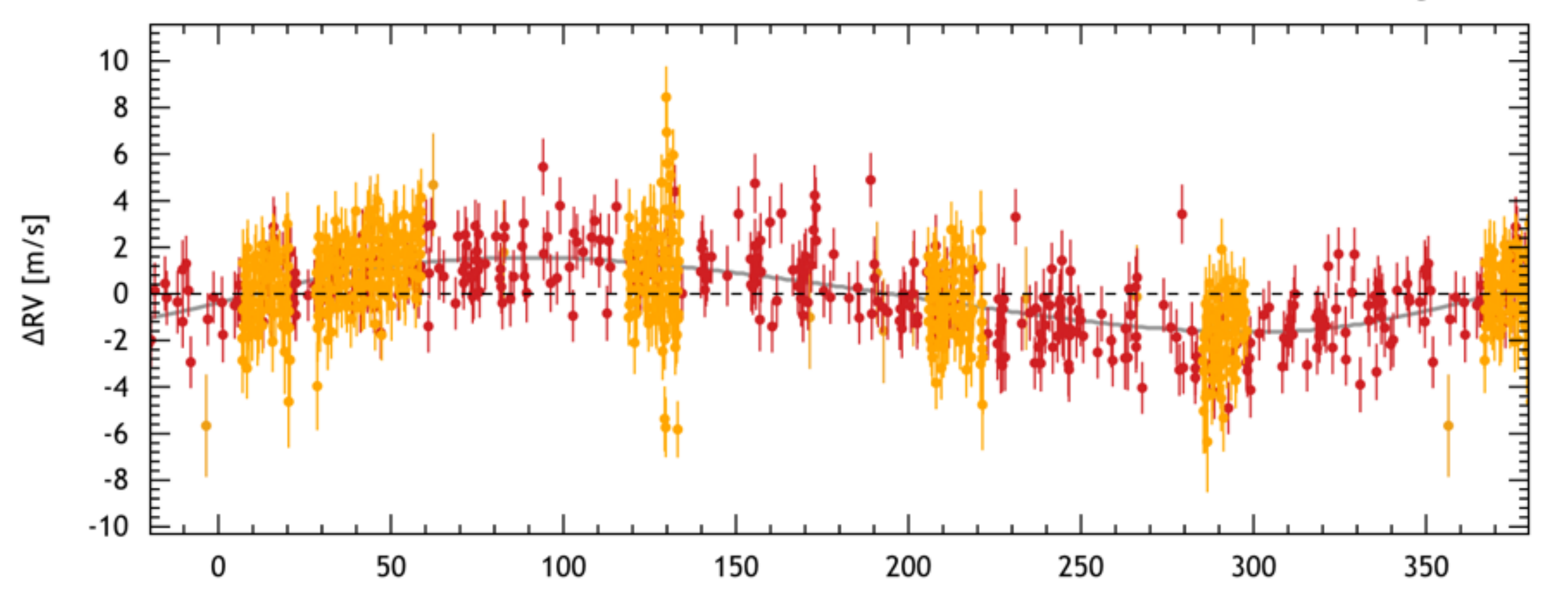}
\caption{ESPRESSO (yellow) and HARPS (red) radial velocities of \object{HD\,40307} phase-folded to 4.31-days period of the inner super-Earth hosted by the star.}
\label{hd40307_rv_phasefolded}
\end{center}
\end{figure}

\begin{figure}[h]
\begin{center}
\includegraphics[width=8cm, trim= 20 0 30 0, clip]{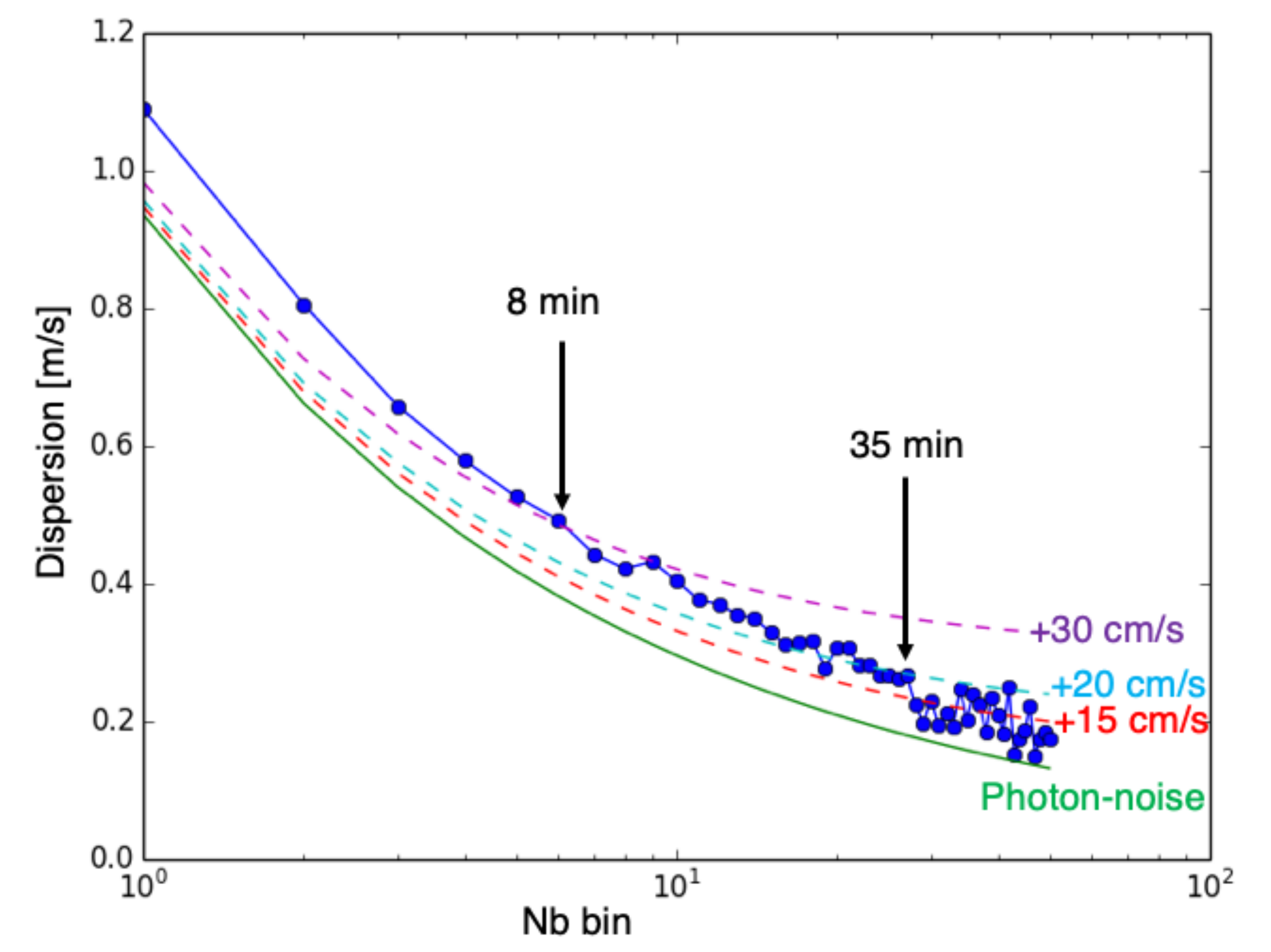}
\caption{Dispersion of the RV time series on \object{HD\,40307} as function of the number of binned data points.}
\label{hd40307_rms_bin}
\end{center}
\end{figure}

In December 2018, we conducted with ESPRESSO an asteroseismological campaign on \object{HD\,40307} over 5 nights. The detailed analysis of the RV time series and spectroscopic activity indicators (cross-correlation function FWHM, bisector span, CaII, H$\alpha$, etc.) will be presented in Bouchy et al. (in prep.). During this campaign, a total of 1\,150 measurements were obtained. The exposure time was set to 30\,seconds. The duty cycle was close to 78\,seconds due to increased overhead time, which was slightly reduced since. The average photon-noise uncertainty per exposure is 95\cms. Figure\,\ref{hd40307_rv_phasefolded} shows the ESPRESSO and HARPS radial velocities phase-folded to the 4.31-days planet of the \object{HD\,40307} system. ESPRESSO's exposure time was 30\,times shorter than the HARPS ones (900\,s). The typical dispersion of ESPRESSO RVs is 110\cms and not limited by the photon-noise. Figure\,\ref{hd40307_rms_bin} presents the RV dispersion as a function of the number of binned data point. One can see that, by binning 6 data points (8\,Min. bin), the dispersion is reduced to less than 50\cms and the additional noise (instrumental and stellar) is at the level of 30\cms. By averaging 27 data points (35\,Min. bin), the dispersion is further reduced to below 20\cms. These values suggest that the binning process progressively damps the signals having a stellar origin. Hence, Figure\,\ref{hd40307_rms_bin} leads us to the statement that the residual instrumental noise must be at the level of only 10\cms.  

\begin{figure}[h]
\begin{center}
\includegraphics[width=\hsize]{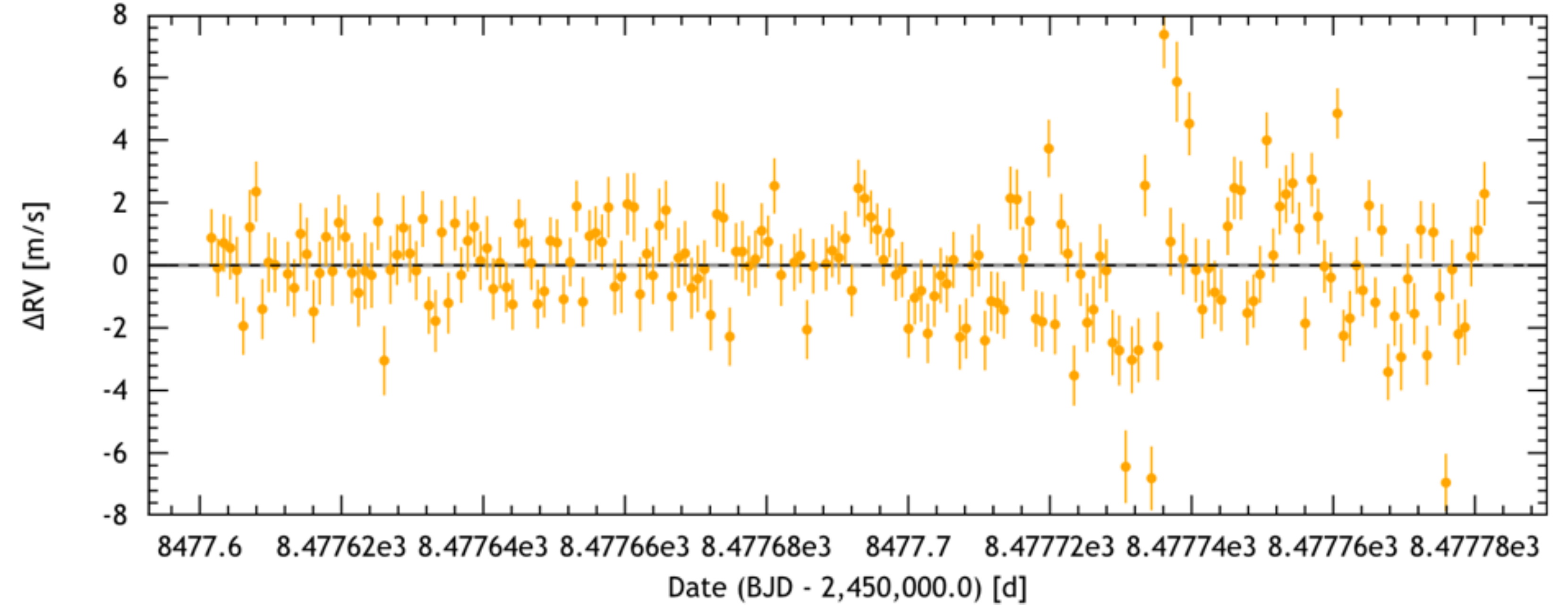}
\caption{ESPRESSO radial velocity sequence on \object{HD\,40307} of December 24, 2018, showing evidence of p-modes excitation and damping.}
\label{hd40307_zoom}
\end{center}
\end{figure}

We find weak evidences of p-modes detection at a level of only 3-4\cms, very close to our detection limit, slightly below the expected amplitude,  and with a large separation of modes close to 220\,$\mu$Hz. Furthermore, we suspect to see the excitation and damping of p-modes oscillations on the timescale of a few hours. Figure\,\ref{hd40307_zoom} shows the RV time series obtained on December 24, 2018. The RV jitter increases suddenly after BJD\,=\,8\,477.71 and with a modulation of 2.5\,Min. This pattern could not be linked to any (known) instrumental parameters and we think that it shows the onset of coherent pulsation modes in realtime, a snapshot of a ``stellar quake''.

\subsubsection{An ESPRESSO measurement of primordial Deuterium}
Observations of the abundances of the primordial elements produced in the Big Bang Nucleosynthesis (BBN) provide one of the earliest (first few minutes) probes of the cosmological scenario and any departure from the standard values and from homogeneity can be the smoking gun of new physics. Among the various elements and isotopes (He$^3$, He$^4$, Li) Deuterium is known to be a powerful diagnostic, in particular providing the best constraint on the baryon abundance, thanks to its sensitivity and monotonic dependence on the baryon-to-photon ratio, $\eta_{10}$.

Since almost twenty years, primordial Deuterium abundance has been measured with the ratio of D~I to H~I column densities in absorption systems observed in high-redshift quasar spectra. This is not an easy task: high-resolution data have been used to this end and not only a high signal-to-noise ratio (S/N) is required, but also a strict control on systematics. Systematics can be related to the process of measurement (continuum fitting, H~I Lyman-$\alpha$ line contamination, deblending of the velocity structure) but also to external possible correlations of the abundance with the column density/overdensity, metallicity and/or redshift of the absorber.

The relative large scatter observed in the D/H values throughout the era of accurate measurements has been typically ascribed to underestimates of the systematic errors. In a recent paper, \citet{cooke:2018} have selected only seven absorbers providing a significant measurement, which turns out to be compatible with a uniform primordial D/H~$ = (2.53\pm0.03) \times 10^{-5}$ and consistent within $2\sigma$ with the baryon density measurable from the Planck CMB observations.

ESPRESSO with its characteristics of stability and wavelength accuracy can provide measurements with much lower systematics. Furthermore, observations in 4UT configuration allows us to observe a bright quasar with a resolution $\sim 1.5-1.75$ times larger than those normally obtained with UVES@VLT or HIRES@Keck. A larger resolution is critical to improve the determination of the continuum level and the characterization of the velocity structure of the systems. 

In August 2018, we have observed the quasar PKS 1937-1009 for 5 hours with ESPRESSO in 4UT configuration. This is one of the brightest quasars ($V=16.95$\,mag) in the Southern sky with a relatively high emission redshift ($z_{\rm em}\simeq 3.787$), allowing us to observe its Lyman forest in the ESPRESSO wavelength range. In the spectrum of this object there are two Lyman Limit Systems ($\log(N_{\rm HI}) \sim 18$) for which a primordial Deuterium measurement was carried out at $z\simeq 3.256$ and at $z\simeq 3.572$ (see \cite{riemer:2015,riemer:2017} and references therein). The analysis of the ESPRESSO spectra (D'Odorico et al. in prep.) is improving our knowledge of the chemical abundances of the two systems, while the S/N in the Lyman forest will probably need to be increased to obtain a more accurate determination of the D/H abundances than previously reported.  

\subsubsection{Small-scale structure of the Intergalactic Medium}
A number of studies on the large-scale properties of the intergalactic medium have concentrated on the crucial problem of the physical scale of Lyman absorbers. Observations (e.g. \citet{dodorico:1998}, \citet{dodorico:2002}, \citet{rauch:2005}) have shown that their typical size is large and they must be part of the general large scale structure. Ionization arguments, analytical and Monte Carlo modelling of absorption in double lines of sight (\citet{findlay:2018} and references therein) and cosmological hydro-simulations all suggest that the absorbing structures are part of a filamentary cosmic web, expected to follow the Hubble  flow on large (Mpc) scales, i.e. at least on scales larger than the typical coherence length of these structures. On intermediate scales (of order 100\,kpc) the effects of gravitational collapse may become more pronounced, with the gas gaining kinetic energy in the galactic potential wells, while on the smallest (subkiloparsec) scales processes like stellar winds and supernova explosions should be the dominant sources of momentum.

\begin{figure}[h]
\begin{center}
\includegraphics[width=\hsize, trim= 10 0 30 0, clip]{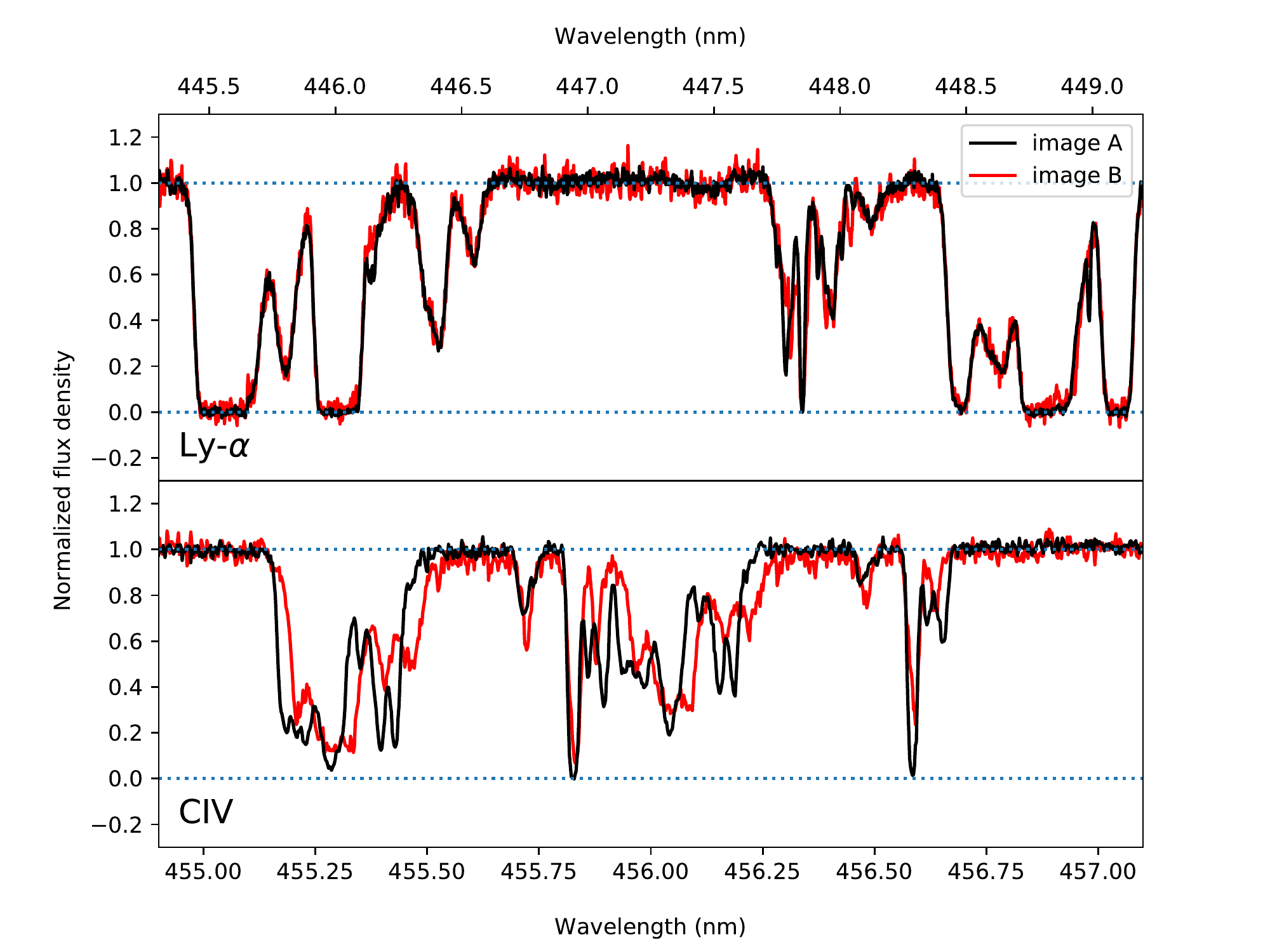}
\caption{Portion of spectrum of Q0142-100AB observed with ESPRESSO. Top: Region of Lyman-alpha forest (with a few metal interlopers), which shows a remarkable stability. Bottom: Complex CIV\,$\lambda\lambda$1548,1550 absorption system, which shows a different velocity structure along the two lines of sight (image A and B, black and red curves respectively).}
\label{fig:qso-lens}
\end{center}
\end{figure}

We have taken advantage of the unique capabilities of ESPRESSO to explore the kpc and sub-kpc scales, and gain insight into the feedback processes governing the formation of galaxies and also to verify whether systematic variations of the dynamical and ionization state in the Lyman forest might affect the Sandage Test \citep{loeb:1998,cristiani:2007,liske:2008} in future experiments. We observed with ESPRESSO in 4UT configuration a bright lensed QSO with emission redshift 2.72 and image split on the sky of 2,5\arcsec. The lines of sight of Q0142-100AB span in Ly-alpha redshift range from $z=2.7$ down to 2.2 and represent, due to the geometry of the lens, transverse scales from about 15 physical kpc down to the sub-kpc range. This is the brightest known lens in the Southern sky for which the Lyman forest is observable by ESPRESSO and the separation of at least two images of the lens is above 2\arcsec (in order to avoid fiber flux contamination). The Lyman forest on these scales shows a remarkable stability:  
pairs of absorptions observed in the two lines of sight display the same column density, Doppler parameter  and redshift, within the measurement error of the photon statistics (Figure\,\ref{fig:qso-lens}, top). In particular the velocity differences appear to be below the observational RMS limit of 1.1 km/s. On the contrary, metal lines, and the CIV species in particular (Figure\,\ref{fig:qso-lens}, bottom), show strong variations, indicating that they must have variations of structure on scales of the order of $10^2$\,pc (Cristiani et al. in prep.).

\section{Conclusions}
ESPRESSO@VLT offers to the astronomical community high-resolution, high-fidelity, full-range visible spectroscopy on a 10-m class telescope. The instrument's quite unique features open a new parameter space in ground-based observations that can address a large variety of scientific cases. In this paper we presented the ESPRESSO instrument and provided information on its characteristics and advice on how to carry out observations.

Using on-sky test carried out during the commissioning phase and the first year of operations, we demonstrated that all the main performance requirements have been met. First, high resolving power of $R\sim$140\,000 and $R>$190\,000 is provided in the 1-UT configuration over the full wavelength range from 378.2 to 788.7\,nm. In 4-UT configuration a resolving power of $R\sim$70\,000 is achieved, despite having to cope with the etendue of \textit{four} 8-m telescopes. Second, the overall throughput, from top of the atmosphere to the detector, peaks at 10\% in both the 1-UT and the 4-UT configurations. Third, short-term (night) radial-velocity measurement are limited by photon noise down to the 10\cms level. Over the long term (months), a repeatability of better than 50\cms \textit{rms} was demonstrated. Given the fact that photon noise and stellar jitter affect single RV-measurements, this value provides an upper limit for the instrumental precision. Last, we have demonstrated that throughput and RV measurement does not depend on the used telescope. This fact allows for a unique versatility of the VLT in its operations.

We presented our GTO program that is being carried out on ESPRESSO. Due to the time-series nature of the radial-velocity observations, and in particular the blind-search sub-program, we were not expecting to produce results shortly after the start of operations. Two other sub-programs aiming at RV follow-up of \textit{TESS} and \textit{K2}, and at observing planetary atmospheres through transit spectroscopy, however, immediately benefitted from the unique spectroscopic capabilities of ESPRESSO (e.g. \citet{ehrenreich:2020}). The simultaneous start of \textit{TESS} and ESPRESSO operations enabled furthermore exceptional synergies between spaced-based transit searches and ground-based radial-velocity observations. The results on $\pi$\,Men presented in Damasso et al. (this same edition of the journal) are the direct demonstration of the power of simultaneous high-precision photometric and spectroscopic observations.

\begin{acknowledgements}
ESPRESSO is a 25\,MEUR project that lasted about 10 years and involved dozens of collaborators. The large number of co-authors is a manifestation of this effort, but it is not able to represent adequately the passion of \textit{all} the people who made this project possible. It is for this simple reason that F.P would like to express his profound gratitude to  technical and administrative assistants, technicians, engineers, managers and science collaborators who helped navigating through this challenging adventure with flexibility, dedication and creativity. We cannot avoid problems in Life, but by working with Friends, all together, we can overcome them. 

F.P. and C.L. would like to acknowledge the Swiss National Science Foundation (SNSF) for supporting research with ESPRESSO through the SNSF grants nr. 140649, 152721, 166227 and 184618. The ESPRESSO Instrument Project was partially funded through SNSF's FLARE program for large infrastructures. This work has been carried out in part within the framework of the NCCR PlanetS supported by the Swiss National Science Foundation.

M.D. acknowledges financial support from Progetto Premiale 2015 FRONTIERA (OB.FU. 1.05.06.11) funding scheme of the Italian Ministry of Education, University, and Research. We acknowledge the computing centers of INAF - Osservatorio Astronomico di Trieste/Osservatorio Astrofisico di Catania, under the coordination of the CHIPP project, for the availability of computing resources and support.

This work was supported by FCT -- Funda\,c\~ao para a Ci\^encia e Tecnologia -- through national funds and by FEDER through COMPETE2020 - Programa Operacional Competitividade e Internacionaliza\,c\,~ao  -- by these grants: UID/FIS/04434/2019; UIDB/04434/2020; UIDP/04434/2020; PTDC/FIS-AST/32113/2017 \& POCI-01-0145-FEDER-032113; PTDC/FIS-AST/28953/2017 \& POCI-01-0145-FEDER-028953; PTDC/FIS-AST/28987/2017 \& POCI-01-0145-FEDER-028987.
V.A., S.C.C.B., O.D.S.D., J.P.F. and S.S. acknowledge support from FCT through work contracts nºs IF/00650/2015/CP1273, IF/01312/2014/CP1215/CT0004, DL 57/2016/CP1364/CT0004, DL 57/2016/CP1364/CT0005, IF/00028/2014/CP1215/CT0002.

J.L.-B. has been funded by the Spanish State Research Agency (AEI) Projects No.ESP2017-87676-C5-1-R and No. MDM-2017-0737 Unidad de Excelencia "Mar\'ia de Maeztu"- Centro de Astrobiolog\'ia (INTA-CSIC).

V.B. acknowledges support by the Swiss National Science Foundation (SNSF) in the frame of the National Centre for Competence in Research ``PlanetS''.

This project has received funding from the European Research Council (ERC) under the European Union's Horizon 2020 research and innovation program (project Four Aces, grant agreement No 724427).
R.R., C.A.P., J.I.G.H. and A.S.M. acknowledge financial support from the Spanish Ministry of Science and Innovation (MICINN) project AYA2017-86389-P.

J.I.G.H. acknowledges financial support from the Spanish MICINN under the 2013 Ram\'on y Cajal program RYC-2013-14875.

M.T.M. thanks the Australian Research Council for \textsl{Future Fellowship} grant FT180100194 which supported this work.

This publication makes use of the Data \& Analysis Center for Exoplanets (DACE), which is a facility based at the University of Geneva (CH) dedicated to extrasolar planets data visualisation, exchange and analysis. DACE is a platform of the Swiss National Centre of Competence in Research (NCCR) PlanetS, federating the Swiss expertise in Exoplanet research. The DACE platform is available at https://dace.unige.ch.

\end{acknowledgements}

\bibliographystyle{aa} 
\bibliography{references.bib} 

\begin{thebibliography}{125}
\expandafter\ifx\csname natexlab\endcsname\relax\def\natexlab#1{#1}\fi

\bibitem[{{Aliverti} {et~al.}(2016){Aliverti}, {Pariani}, {Moschetti}, \&
  {Riva}}]{aliverti:2016}
{Aliverti}, M., {Pariani}, G., {Moschetti}, M., \& {Riva}, M. 2016, Society of
  Photo-Optical Instrumentation Engineers (SPIE) Conference Series, Vol. 9908,
  {Integration and alignment through mechanical measurements: the example of
  the ESPRESSO front-end units}, 99087C

\bibitem[{{{\'A}lvarez} {et~al.}(2018){{\'A}lvarez}, {Lizon}, {Hughes}, {Pepe},
  {Riva}, {Megevand}, {Lothar}, {Iwert}, \& {Lo Curto}}]{alvarez:2018}
{{\'A}lvarez}, D., {Lizon}, J.-L., {Hughes}, I., {et~al.} 2018, in Society of
  Photo-Optical Instrumentation Engineers (SPIE) Conference Series, Vol. 10702,
  \procspie, 107026W

\bibitem[{Alves {et~al.}(2017)Alves, Silva, Martins, \& Leite}]{Alves:2017}
Alves, C., Silva, T., Martins, C., \& Leite, A. 2017, Phys.\ Lett.\ B, 770, 93

\bibitem[{{Anglada-Escud{\'e}} {et~al.}(2016){Anglada-Escud{\'e}}, {Amado},
  {Barnes}, {Berdi{\~n}as}, {Butler}, {Coleman}, {de La Cueva}, {Dreizler},
  {Endl}, {Giesers}, {Jeffers}, {Jenkins}, {Jones}, {Kiraga}, {K{\"u}rster},
  {L{\'o}pez-Gonz{\'a}lez}, {Marvin}, {Morales}, {Morin}, {Nelson}, {Ortiz},
  {Ofir}, {Paardekooper}, {Reiners}, {Rodr{\'\i}guez},
  {Rodr{\'\i}guez-L{\'o}pez}, {Sarmiento}, {Strachan}, {Tsapras}, {Tuomi}, \&
  {Zechmeister}}]{anglada:2016}
{Anglada-Escud{\'e}}, G., {Amado}, P.~J., {Barnes}, J., {et~al.} 2016, \nat,
  536, 437

\bibitem[{{Arentsen} {et~al.}(2019){Arentsen}, {Starkenburg}, {Shetrone},
  {Venn}, {Depagne}, \& {McConnachie}}]{arentsen:2019}
{Arentsen}, A., {Starkenburg}, E., {Shetrone}, M.~D., {et~al.} 2019, \aap, 621,
  A108

\bibitem[{{Arns}(2016)}]{arns:2016}
{Arns}, J.~A. 2016, Society of Photo-Optical Instrumentation Engineers (SPIE)
  Conference Series, Vol. 9908, {Performance characteristics of two volume
  phase holographic grisms produced for the ESPRESSO spectrograph}, 990861

\bibitem[{{Avila} {et~al.}(2016){Avila}, {Cabral}, {Coelho}, {Santos}, {Silva},
  {Delabre}, {Megevand }, {Abreu}, {Oliveira}, {Riva}, {Aliverti}, \&
  {Pariani}}]{avila:2016}
{Avila}, G., {Cabral}, A., {Coelho}, J.~P., {et~al.} 2016, Society of
  Photo-Optical Instrumentation Engineers (SPIE) Conference Series, Vol. 9912,
  {Alignment of the ESPRESSO Coud{\'e} train on the ESO VLT}, 99124L

\bibitem[{{Bahcall} \& {Schmidt}(1967)}]{bahcall:1967}
{Bahcall}, J.~N. \& {Schmidt}, M. 1967, \prl, 19, 1294

\bibitem[{{Bainbridge} \& {Webb}(2017)}]{Bainbridge:2017}
{Bainbridge}, M.~B. \& {Webb}, J.~K. 2017, \mnras, 468, 1639

\bibitem[{{Baldini} {et~al.}(2016){Baldini}, {Calderone}, {Cirami}, {Coretti},
  {Cristiani}, {Di Marcantonio}, {M{\'e}gevand}, {Riva}, \&
  {Santin}}]{baldini:2016}
{Baldini}, V., {Calderone}, G., {Cirami}, R., {et~al.} 2016, Society of
  Photo-Optical Instrumentation Engineers (SPIE) Conference Series, Vol. 9913,
  {Integration of the instrument control electronics for the ESPRESSO
  spectrograph at ESO-VLT}, 99132H

\bibitem[{{Baranne} {et~al.}(1996){Baranne}, {Queloz}, {Mayor}, {Adrianzyk},
  {Knispel}, {Kohler}, {Lacroix}, {Meunier}, {Rimbaud}, \&
  {Vin}}]{baranne:1996}
{Baranne}, A., {Queloz}, D., {Mayor}, M., {et~al.} 1996, \aaps, 119, 373

\bibitem[{{Batalha} {et~al.}(2013){Batalha}, {Rowe}, {Bryson}, {Barclay},
  {Burke}, {Caldwell}, {Christiansen}, {Mullally}, {Thompson}, {Brown},
  {Dupree}, {Fabrycky}, {Ford}, {Fortney}, {Gilliland}, {Isaacson}, {Latham},
  {Marcy}, {Quinn}, {Ragozzine}, {Shporer}, {Borucki}, {Ciardi}, {Gautier},
  {Haas}, {Jenkins}, {Koch}, {Lissauer}, {Rapin}, {Basri}, {Boss}, {Buchhave},
  {Carter}, {Charbonneau}, {Christensen-Dalsgaard}, {Clarke}, {Cochran},
  {Demory}, {Desert}, {Devore}, {Doyle}, {Esquerdo}, {Everett}, {Fressin},
  {Geary}, {Girouard}, {Gould}, {Hall}, {Holman}, {Howard}, {Howell},
  {Ibrahim}, {Kinemuchi}, {Kjeldsen}, {Klaus}, {Li}, {Lucas}, {Meibom},
  {Morris}, {Pr{\v{s}}a}, {Quintana}, {Sanderfer}, {Sasselov}, {Seader},
  {Smith}, {Steffen}, {Still}, {Stumpe}, {Tarter}, {Tenenbaum}, {Torres},
  {Twicken}, {Uddin}, {Van Cleve}, {Walkowicz}, \& {Welsh}}]{batalha:2013}
{Batalha}, N.~M., {Rowe}, J.~F., {Bryson}, S.~T., {et~al.} 2013, \apjs, 204, 24

\bibitem[{{Bonifacio} {et~al.}(2020){Bonifacio}, {Molaro}, {Adibekyan},
  {Aguado}, {Alibert}, {Allende Prieto}, {Caffau}, {Cristiani}, {Cupani}, {Di
  Marcantonio}, {D'Odorico}, {Ehrenreich}, {Figueira}, {Genova}, {Gonz{\'a}lez
  Hern{\'a}ndez}, {Lo Curto}, {Lovis}, {Martins}, {Mehner}, {Micela}, {Monaco},
  {Nunes}, {Pepe}, {Poretti}, {Rebolo}, {Santos}, {Saviane}, {Sousa},
  {Sozzetti}, {Suarez-Mascare{\~n}o}, {Udry}, \&
  {Zapatero-Osorio}}]{bonifacio:2020}
{Bonifacio}, P., {Molaro}, P., {Adibekyan}, V., {et~al.} 2020, \aap, 633, A129

\bibitem[{{Bonifacio} {et~al.}(2014){Bonifacio}, {Rahmani}, {Whitmore},
  {Wendt}, {Centurion}, {Molaro}, {Srianand}, {Murphy}, {Petitjean},
  {Agafonova}, {D'Odorico}, {Evans}, {Levshakov}, {Lopez}, {Martins},
  {Reimers}, \& {Vladilo}}]{bonifacio:2014}
{Bonifacio}, P., {Rahmani}, H., {Whitmore}, J.~B., {et~al.} 2014, Astronomische
  Nachrichten, 335, 83

\bibitem[{{Borucki} {et~al.}(2011){Borucki}, {Koch}, {Basri}, {Batalha},
  {Brown}, {Bryson}, {Caldwell}, {Christensen-Dalsgaard}, {Cochran}, {DeVore},
  {Dunham}, {Gautier}, {Geary}, {Gilliland}, {Gould}, {Howell}, {Jenkins},
  {Latham}, {Lissauer}, {Marcy}, {Rowe}, {Sasselov}, {Boss}, {Charbonneau},
  {Ciardi}, {Doyle}, {Dupree}, {Ford}, {Fortney}, {Holman}, {Seager},
  {Steffen}, {Tarter}, {Welsh}, {Allen}, {Buchhave}, {Christiansen}, {Clarke},
  {Das}, {D{\'e}sert}, {Endl}, {Fabrycky}, {Fressin}, {Haas}, {Horch},
  {Howard}, {Isaacson}, {Kjeldsen}, {Kolodziejczak}, {Kulesa}, {Li}, {Lucas},
  {Machalek}, {McCarthy}, {MacQueen}, {Meibom}, {Miquel}, {Prsa}, {Quinn},
  {Quintana}, {Ragozzine}, {Sherry}, {Shporer}, {Tenenbaum}, {Torres},
  {Twicken}, {Van Cleve}, {Walkowicz}, {Witteborn}, \& {Still}}]{borucki:2011}
{Borucki}, W.~J., {Koch}, D.~G., {Basri}, G., {et~al.} 2011, \apj, 736, 19

\bibitem[{{Bouchy} {et~al.}(2001){Bouchy}, {Pepe}, \& {Queloz}}]{bouchy:2001}
{Bouchy}, F., {Pepe}, F., \& {Queloz}, D. 2001, \aap, 374, 733

\bibitem[{{Buchhave} {et~al.}(2016){Buchhave}, {Dressing}, {Dumusque}, {Rice},
  {Vanderburg}, {Mortier}, {Lopez-Morales}, {Lopez}, {Lundkvist}, {Kjeldsen},
  {Affer}, {Bonomo}, {Charbonneau}, {Collier Cameron}, {Cosentino}, {Figueira},
  {Fiorenzano}, {Harutyunyan}, {Haywood}, {Johnson}, {Latham}, {Lovis},
  {Malavolta}, {Mayor}, {Micela}, {Molinari}, {Motalebi}, {Nascimbeni}, {Pepe},
  {Phillips}, {Piotto}, {Pollacco}, {Queloz}, {Sasselov}, {S{\'e}gransan},
  {Sozzetti}, {Udry}, \& {Watson}}]{buchhave:2016}
{Buchhave}, L.~A., {Dressing}, C.~D., {Dumusque}, X., {et~al.} 2016, \aj, 152,
  160

\bibitem[{{Cabral} {et~al.}(2019){Cabral}, {Abreu}, {Coelho}, {Avila}, {Riva},
  {Santos}, \& {Pepe}}]{cabral:2019}
{Cabral}, A., {Abreu}, M., {Coelho}, J., {et~al.} 2019, in IV International
  Conference on Applications of Optics and Photonics AOP

\bibitem[{{Cabral} {et~al.}(2014){Cabral}, {Abreu}, {Coelho}, {Gomes},
  {Monteiro}, {Oliveira}, {Santos}, {{\'A}vila}, {Delabre}, {Riva}, {Di
  Marcantonio}, {Coretti}, {Santos}, {Zerbi}, \& {M{\'e}gevand}}]{cabral:2014}
{Cabral}, A., {Abreu}, M., {Coelho}, J., {et~al.} 2014, Society of
  Photo-Optical Instrumentation Engineers (SPIE) Conference Series, Vol. 9147,
  {ESPRESSO Coud{\'e}-Train: complexities of a simultaneous optical feeding
  from the four VLT unit telescopes}, 91478Q

\bibitem[{{Cabral} {et~al.}(2013){Cabral}, {Coelho}, {Abreu}, {Monteiro},
  {Gomes}, {Santos}, {Oliveira}, {{\'A}vila}, {Delabre}, {M{\'e}gevand},
  {Zerbi}, {Di Marcantonio}, {Lovis}, \& {Santos}}]{cabral:2013}
{Cabral}, A., {Coelho}, J., {Abreu}, M., {et~al.} 2013, Society of
  Photo-Optical Instrumentation Engineers (SPIE) Conference Series, Vol. 8785,
  {Optical design of a Coud{\'e}-Train for a stable and efficient simultaneous
  feeding of the ESPRESSO spectrograph from the four VLT telescopes}, 87850L

\bibitem[{{Caffau} {et~al.}(2011){Caffau}, {Bonifacio}, {Fran{\c{c}}ois},
  {Sbordone}, {Monaco}, {Spite}, {Spite}, {Ludwig}, {Cayrel}, {Zaggia},
  {Hammer}, {Randich}, {Molaro}, \& {Hill}}]{caffau:2011}
{Caffau}, E., {Bonifacio}, P., {Fran{\c{c}}ois}, P., {et~al.} 2011, \nat, 477,
  67

\bibitem[{{Calderone} {et~al.}(2016){Calderone}, {Baldini}, {Cirami},
  {Coretti}, {Cristiani}, {Di Marcantonio}, {Landoni}, {M{\'e}gevand}, {Riva},
  \& {Santin}}]{calderone:2016}
{Calderone}, G., {Baldini}, V., {Cirami}, R., {et~al.} 2016, Society of
  Photo-Optical Instrumentation Engineers (SPIE) Conference Series, Vol. 9913,
  {The technical CCDs in ESPRESSO: usage, performances, and network
  requirements}, 99132K

\bibitem[{{Calderone} {et~al.}(2018{\natexlab{a}}){Calderone}, {Baldini},
  {Cirami}, {Coretti}, {Cristiani}, {Di Marcantonio}, \&
  {M{\'e}gevand}}]{calderone:2018}
{Calderone}, G., {Baldini}, V., {Cirami}, R., {et~al.} 2018{\natexlab{a}}, in
  Society of Photo-Optical Instrumentation Engineers (SPIE) Conference Series,
  Vol. 10707, \procspie, 107072G

\bibitem[{{Calderone} {et~al.}(2018{\natexlab{b}}){Calderone}, {Baldini},
  {Cirami}, {Coretti}, {Cristiani}, {Di Marcantonio}, \&
  {M{\'e}gevand}}]{Calderone18}
{Calderone}, G., {Baldini}, V., {Cirami}, R., {et~al.} 2018{\natexlab{b}}, in
  Society of Photo-Optical Instrumentation Engineers (SPIE) Conference Series,
  Vol. 10707, \procspie, 107072G

\bibitem[{{Carrier} \& {Bourban}(2003)}]{carrier:2003}
{Carrier}, F. \& {Bourban}, G. 2003, \aap, 406, L23

\bibitem[{Carroll(1998)}]{Carroll:1998}
Carroll, S.~M. 1998, Phys.\ Rev.\ Lett., 81, 3067

\bibitem[{{Casasayas-Barris} {et~al.}(2019){Casasayas-Barris}, {Pall{\'e}},
  {Yan}, {Chen}, {Kohl}, {Stangret}, {Parviainen}, {Helling}, {Watanabe},
  {Czesla}, {Fukui}, {Monta{\~n}{\'e}s-Rodr{\'\i}guez}, {Nagel}, {Narita},
  {Nortmann}, {Nowak}, {Schmitt}, \& {Zapatero Osorio}}]{casasayasbarris:2019}
{Casasayas-Barris}, N., {Pall{\'e}}, E., {Yan}, F., {et~al.} 2019, \aap, 628,
  A9

\bibitem[{{Cersullo} {et~al.}(2019){Cersullo}, {Coffinet}, {Chazelas}, {Lovis},
  \& {Pepe}}]{cersullo:2019}
{Cersullo}, F., {Coffinet}, A., {Chazelas}, B., {Lovis}, C., \& {Pepe}, F.
  2019, \aap, 624, A122

\bibitem[{{Charbonneau} {et~al.}(2002){Charbonneau}, {Brown}, {Noyes}, \&
  {Gilliland}}]{charbonneau:2002}
{Charbonneau}, D., {Brown}, T.~M., {Noyes}, R.~W., \& {Gilliland}, R.~L. 2002,
  \apj, 568, 377

\bibitem[{{Chazelas} {et~al.}(2012){Chazelas}, {Pepe}, \&
  {Wildi}}]{chazelas:2012}
{Chazelas}, B., {Pepe}, F., \& {Wildi}, F. 2012, Society of Photo-Optical
  Instrumentation Engineers (SPIE) Conference Series, Vol. 8450, {Optical
  fibers for precise radial velocities: an update}, 845013

\bibitem[{{Chazelas} {et~al.}(2010){Chazelas}, {Pepe}, {Wildi}, {Bouchy},
  {Perruchot}, \& {Avila}}]{chazelas:2010}
{Chazelas}, B., {Pepe}, F., {Wildi}, F., {et~al.} 2010, Society of
  Photo-Optical Instrumentation Engineers (SPIE) Conference Series, Vol. 7739,
  {New scramblers for precision radial velocity: square and octagonal fibers},
  773947

\bibitem[{{Cooke} {et~al.}(2018){Cooke}, {Pettini}, \& {Steidel}}]{cooke:2018}
{Cooke}, R.~J., {Pettini}, M., \& {Steidel}, C.~C. 2018, \apj, 855, 102

\bibitem[{{Cosentino} {et~al.}(2012){Cosentino}, {Lovis}, {Pepe}, {Collier
  Cameron}, {Latham}, {Molinari}, {Udry}, {Bezawada}, {Black}, {Born},
  {Buchschacher}, {Charbonneau}, {Figueira}, {Fleury}, {Galli}, {Gallie},
  {Gao}, {Ghedina}, {Gonzalez}, {Gonzalez}, {Guerra}, {Henry}, {Horne},
  {Hughes}, {Kelly}, {Lodi}, {Lunney}, {Maire}, {Mayor}, {Micela}, {Ordway},
  {Peacock}, {Phillips}, {Piotto}, {Pollacco}, {Queloz}, {Rice}, {Riverol},
  {Riverol}, {San Juan}, {Sasselov}, {Segransan}, {Sozzetti}, {Sosnowska},
  {Stobie}, {Szentgyorgyi}, {Vick}, \& {Weber}}]{cosentino:2012}
{Cosentino}, R., {Lovis}, C., {Pepe}, F., {et~al.} 2012, Society of
  Photo-Optical Instrumentation Engineers (SPIE) Conference Series, Vol. 8446,
  {Harps-N: the new planet hunter at TNG}, 84461V

\bibitem[{{Cristiani} {et~al.}(2007){Cristiani}, {Avila}, {Bonifacio},
  {Bouchy}, {Carswell}, {D'Odorico}, {D'Odorico}, {Delabre}, {Dekker},
  {Dessauges}, {Dimarcantonio}, {Garcia-Lopez}, {Grazian}, {Haehnelt},
  {Herreros}, {Israelian}, {Levshakov}, {Liske}, {Lovis}, {Manescau}, {Martin},
  {Mayor}, {Megevand}, {Molaro}, {Murphy}, {Pasquini}, {Pepe}, {Perez},
  {Queloz}, {Rebolo}, {Santin}, {Shaver}, {Span{\`o}}, {Udry}, {Vanzella},
  {Viel}, {Zapatero}, {Zerbi}, \& {Zucker}}]{cristiani:2007}
{Cristiani}, S., {Avila}, G., {Bonifacio}, P., {et~al.} 2007, Nuovo Cimento B
  Serie, 122, 1165

\bibitem[{{Cunha} {et~al.}(2013){Cunha}, {Figueira}, {Santos}, {Lovis}, \&
  {Bou{\'e}}}]{cunha:2013}
{Cunha}, D., {Figueira}, P., {Santos}, N.~C., {Lovis}, C., \& {Bou{\'e}}, G.
  2013, \aap, 550, A75

\bibitem[{{Cunha} {et~al.}(2014){Cunha}, {Santos}, {Figueira}, {Santerne},
  {Bertaux}, \& {Lovis}}]{cunha:2014}
{Cunha}, D., {Santos}, N.~C., {Figueira}, P., {et~al.} 2014, \aap, 568, A35

\bibitem[{{Cupani} {et~al.}(2018){Cupani}, {Calderone}, {Cristiani}, {Di
  Marcantonio}, {D'Odorico}, \& {Taffoni}}]{cupani18}
{Cupani}, G., {Calderone}, G., {Cristiani}, S., {et~al.} 2018, in Society of
  Photo-Optical Instrumentation Engineers (SPIE) Conference Series, Vol. 10707,
  \procspie, 1070723

\bibitem[{{Cupani} {et~al.}(2016){Cupani}, {D'Odorico}, {Cristiani},
  {Gonz{\'a}lez-Hern{\'a}ndez}, {Lovis}, {Sousa}, {Calderone}, {Cirami}, {Di
  Marcantonio}, \& {M{\'e}gevand}}]{cupani16}
{Cupani}, G., {D'Odorico}, V., {Cristiani}, S., {et~al.} 2016, in Society of
  Photo-Optical Instrumentation Engineers (SPIE) Conference Series, Vol. 9913,
  \procspie, 99131T

\bibitem[{{Cupani} {et~al.}(2019){Cupani}, {D'Odorico}, {Cristiani},
  {Gonz{\'a}lez Hern{\'a}ndez}, {Lovis}, {Sousa}, {Di Marcantonio}, \&
  {M{\'e}gevand }}]{cupani:2019}
{Cupani}, G., {D'Odorico}, V., {Cristiani}, S., {et~al.} 2019, in Astronomical
  Society of the Pacific Conference Series, Vol. 521, Astronomical Data
  Analysis Software and Systems XXVI, ed. M.~{Molinaro}, K.~{Shortridge}, \&
  F.~{Pasian}, 362

\bibitem[{{Di Marcantonio} {et~al.}(2018){Di Marcantonio}, {Sosnowska},
  {Cupani}, {D'Odorico}, {Lovis}, {Segovia}, {Sousa}, {Gonz{\'a}lez
  Hern{\'a}ndez}, {Calderone}, {Cirami}, {Modigliani}, {Lo Curto}, {Cristiani},
  {Molaro}, {Pepe}, \& {M{\'e}gevand}}]{DiMarcantonio18}
{Di Marcantonio}, P., {Sosnowska}, D., {Cupani}, G., {et~al.} 2018, in Society
  of Photo-Optical Instrumentation Engineers (SPIE) Conference Series, Vol.
  10704, \procspie, 107040F

\bibitem[{{D{\'\i}az} {et~al.}(2016){D{\'\i}az}, {S{\'e}gransan}, {Udry},
  {Lovis}, {Pepe}, {Dumusque}, {Marmier}, {Alonso}, {Benz}, {Bouchy},
  {Coffinet}, {Collier Cameron}, {Deleuil}, {Figueira}, {Gillon}, {Lo Curto},
  {Mayor}, {Mordasini}, {Motalebi}, {Moutou}, {Pollacco}, {Pompei}, {Queloz},
  {Santos}, \& {Wyttenbach}}]{diaz:2016}
{D{\'\i}az}, R.~F., {S{\'e}gransan}, D., {Udry}, S., {et~al.} 2016, \aap, 585,
  A134

\bibitem[{{D'Odorico} {et~al.}(1998){D'Odorico}, {Cristiani}, {D'Odorico},
  {Fontana}, {Giallongo}, \& {Shaver}}]{dodorico:1998}
{D'Odorico}, V., {Cristiani}, S., {D'Odorico}, S., {et~al.} 1998, \aap, 339,
  678

\bibitem[{{D'Odorico} {et~al.}(2002){D'Odorico}, {Petitjean}, \&
  {Cristiani}}]{dodorico:2002}
{D'Odorico}, V., {Petitjean}, P., \& {Cristiani}, S. 2002, \aap, 390, 13

\bibitem[{{Dumusque}(2012)}]{dumusque:2012}
{Dumusque}, X. 2012, PhD thesis, Observatory of Geneva

\bibitem[{{Ehrenreich} {et~al.}(2020){Ehrenreich}, {Lovis}, {Allart}, {Rosa
  Zapatero Osorio}, {Pepe}, {Cristiani}, {Rebolo}, {Santos}, {Borsa},
  {Demangeon}, {Dumusque}, {Gonz{\'a}lez Hern{\'a}ndez}, {Casasayas-Barris},
  {S{\'e}gransan}, {Sousa}, {Abreu}, {Adibekyan}, {Affolter}, {Allende Prieto},
  {Alibert}, {Aliverti}, {Alves}, {Amate}, {Avila}, {Baldini}, {Bandy}, {Benz},
  {Bianco}, {Bolmont}, {Bouchy}, {Bourrier}, {Broeg}, {Cabral}, {Calderone},
  {Pall{\'e}}, {Cegla}, {Cirami}, {Coelho}, {Conconi}, {Coretti}, {Cumani},
  {Cupani}, {Dekker}, {Delabre}, {Deiries}, {D'Odorico}, {Di Marcantonio},
  {Figueira}, {Fragoso}, {Genolet}, {Genoni}, {G{\'e}nova Santos}, {Hara},
  {Hughes}, {Iwert}, {Kerber}, {Knudstrup}, {Land oni}, {Lavie}, {Lizon},
  {Lendl}, {Lo Curto}, {Maire}, {Manescau}, {Martins}, {M{\'e}gevand},
  {Mehner}, {Micela}, {Modigliani}, {Molaro}, {Monteiro}, {Monteiro},
  {Moschetti}, {M{\"u}ller}, {Nunes}, {Oggioni}, {Oliveira}, {Pariani},
  {Pasquini}, {Poretti}, {Rasilla}, {Redaelli}, {Riva}, {Santana Tschudi},
  {Santin}, {Santos}, {Segovia Milla}, {Seidel}, {Sosnowska}, {Sozzetti},
  {Span{\`o}}, {Su{\'a}rez Mascare{\~n}o}, {Tabernero}, {Tenegi}, {Udry},
  {Zanutta}, \& {Zerbi}}]{ehrenreich:2020}
{Ehrenreich}, D., {Lovis}, C., {Allart}, R., {et~al.} 2020, Nature
  [\eprint[arXiv]{2003.05528}]

\bibitem[{Evans {et~al.}(2014)}]{Evans:2014}
Evans, T. {et~al.} 2014, Mon.\ Not.\ Roy.\ Astron.\ Soc., 445, 128

\bibitem[{{Findlay} {et~al.}(2018){Findlay}, {Prochaska}, {Hennawi},
  {Fumagalli}, {Myers}, {Bartle}, {Chehade}, {DiPompeo}, {Shanks}, {Lau}, \&
  {Rubin}}]{findlay:2018}
{Findlay}, J.~R., {Prochaska}, J.~X., {Hennawi}, J.~F., {et~al.} 2018, \apjs,
  236, 44

\bibitem[{{Frank} {et~al.}(2018){Frank}, {Kerber}, {Avila}, {Di Lieto}, {Lo
  Curto}, \& {Manescau}}]{frank:2018}
{Frank}, C., {Kerber}, F., {Avila}, G., {et~al.} 2018, in Society of
  Photo-Optical Instrumentation Engineers (SPIE) Conference Series, Vol. 10702,
  \procspie, 107026P

\bibitem[{{Fulton} {et~al.}(2017){Fulton}, {Petigura}, {Howard}, {Isaacson},
  {Marcy}, {Cargile}, {Hebb}, {Weiss}, {Johnson}, {Morton}, {Sinukoff},
  {Crossfield}, \& {Hirsch}}]{fulton:2017}
{Fulton}, B.~J., {Petigura}, E.~A., {Howard}, A.~W., {et~al.} 2017, \aj, 154,
  109

\bibitem[{{Gandolfi} {et~al.}(2018){Gandolfi}, {Barrag{\'a}n}, {Livingston},
  {Fridlund}, {Justesen}, {Redfield}, {Fossati}, {Mathur}, {Grziwa}, {Cabrera},
  {Garc{\'\i}a}, {Persson}, {Van Eylen}, {Hatzes}, {Hidalgo}, {Albrecht},
  {Bugnet}, {Cochran}, {Csizmadia}, {Deeg}, {Eigm{\"u}ller}, {Endl}, {Erikson},
  {Esposito}, {Guenther}, {Korth}, {Luque}, {Monta{\~n}es Rodr{\'\i}guez},
  {Nespral}, {Nowak}, {P{\"a}tzold}, \& {Prieto-Arranz}}]{gandolfi18}
{Gandolfi}, D., {Barrag{\'a}n}, O., {Livingston}, J.~H., {et~al.} 2018, \aap,
  619, L10

\bibitem[{{Genoni} {et~al.}(2016){Genoni}, {Riva}, {Pariani}, {Aliverti}, \&
  {Moschetti}}]{genoni:2016}
{Genoni}, M., {Riva}, M., {Pariani}, G., {Aliverti}, M., \& {Moschetti}, M.
  2016, Society of Photo-Optical Instrumentation Engineers (SPIE) Conference
  Series, Vol. 9908, {Spectrograph sensitivity analysis: an efficient tool for
  different design phases}, 99086Z

\bibitem[{{Gonz{\'a}lez Hern{\'a}ndez} {et~al.}(2018){Gonz{\'a}lez
  Hern{\'a}ndez}, {Pepe}, {Molaro}, \& {Santos}}]{gonzalez:2018}
{Gonz{\'a}lez Hern{\'a}ndez}, J.~I., {Pepe}, F., {Molaro}, P., \& {Santos},
  N.~C. 2018, {ESPRESSO on VLT: An Instrument for Exoplanet Research}, 157

\bibitem[{{Gracia Temich} {et~al.}(2018{\natexlab{a}}){Gracia Temich},
  {Rasilla}, \& {Salata}}]{gracia:2018-1}
{Gracia Temich}, F., {Rasilla}, J.~L., \& {Salata}, S. 2018{\natexlab{a}}, in
  Society of Photo-Optical Instrumentation Engineers (SPIE) Conference Series,
  Vol. 10702, \procspie, 1070273

\bibitem[{{Gracia Temich} {et~al.}(2018{\natexlab{b}}){Gracia Temich},
  {Rasilla}, {Salata}, {Pepe}, {Avila}, {M{\'e}gevand}, {Rebolo}, \&
  {Riva}}]{gracia:2018-2}
{Gracia Temich}, F., {Rasilla}, J.~L., {Salata}, S., {et~al.}
  2018{\natexlab{b}}, in Society of Photo-Optical Instrumentation Engineers
  (SPIE) Conference Series, Vol. 10706, \procspie, 1070628

\bibitem[{{Greif} {et~al.}(2011){Greif}, {Springel}, {White}, {Glover},
  {Clark}, {Smith}, {Klessen}, \& {Bromm}}]{greif:2011}
{Greif}, T.~H., {Springel}, V., {White}, S. D.~M., {et~al.} 2011, \apj, 737, 75

\bibitem[{{Heger} \& {Woosley}(2010)}]{heger:2010}
{Heger}, A. \& {Woosley}, S.~E. 2010, \apj, 724, 341

\bibitem[{{Hoeijmakers} {et~al.}(2019){Hoeijmakers}, {Ehrenreich}, {Kitzmann},
  {Allart}, {Grimm}, {Seidel}, {Wyttenbach}, {Pino}, {Nielsen}, {Fisher},
  {Rimmer}, {Bourrier}, {Cegla}, {Lavie}, {Lovis}, {Patzer}, {Stock}, {Pepe},
  \& {Heng}}]{hoeijmakers:2019}
{Hoeijmakers}, H.~J., {Ehrenreich}, D., {Kitzmann}, D., {et~al.} 2019, \aap,
  627, A165

\bibitem[{{Hojjatpanah} {et~al.}(2019){Hojjatpanah}, {Figueira}, {Santos},
  {Adibekyan}, {Sousa}, {Delgado-Mena}, {Alibert}, {Cristiani}, {Gonz{\'a}lez
  Hern{\'a}ndez}, {Lanza}, {Di Marcantonio}, {Martins}, {Micela}, {Molaro},
  {Neves}, {Oshagh}, {Pepe}, {Poretti}, {Rojas-Ayala}, {Rebolo}, {Su{\'a}rez
  Mascare{\~n}o}, \& {Zapatero Osorio}}]{hojjatpanah:2019}
{Hojjatpanah}, S., {Figueira}, P., {Santos}, N.~C., {et~al.} 2019, \aap, 629,
  A80

\bibitem[{{Huang} {et~al.}(2018){Huang}, {Burt}, {Vanderburg}, {G{\"u}nther},
  {Shporer}, {Dittmann}, {Winn}, {Wittenmyer}, {Sha}, {Kane}, {Ricker}, {Vand
  erspek}, {Latham}, {Seager}, {Jenkins}, {Caldwell}, {Collins}, {Guerrero},
  {Smith}, {Quinn}, {Udry}, {Pepe}, {Bouchy}, {S{\'e}gransan}, {Lovis},
  {Ehrenreich}, {Marmier}, {Mayor}, {Wohler}, {Haworth}, {Morgan}, {Fausnaugh},
  {Ciardi}, {Christiansen}, {Charbonneau}, {Dragomir}, {Deming}, {Glidden},
  {Levine}, {McCullough}, {Yu}, {Narita}, {Nguyen}, {Morton}, {Pepper},
  {P{\'a}l}, {Rodriguez}, {Stassun}, {Torres}, {Sozzetti}, {Doty},
  {Christensen-Dalsgaard}, {Laughlin}, {Clampin}, {Bean}, {Buchhave}, {Bakos},
  {Sato}, {Ida}, {Kaltenegger}, {Palle}, {Sasselov}, {Butler}, {Lissauer},
  {Ge}, \& {Rinehart}}]{huang18}
{Huang}, C.~X., {Burt}, J., {Vanderburg}, A., {et~al.} 2018, \apjl, 868, L39

\bibitem[{{Huke} {et~al.}(2018){Huke}, {Sch{\"a}fer}, {Reiners}, {Seemann},
  {Riva}, {Pepe}, {Chazelas}, {Mas{\l}owski}, {Kowzan}, {McCracken}, \&
  {Reid}}]{huke:2018}
{Huke}, P., {Sch{\"a}fer}, S., {Reiners}, A., {et~al.} 2018, in Society of
  Photo-Optical Instrumentation Engineers (SPIE) Conference Series, Vol. 10702,
  \procspie, 107029L

\bibitem[{{King} {et~al.}(2012){King}, {Webb}, {Murphy}, {Flambaum},
  {Carswell}, {Bainbridge}, {Wilczynska}, \& {Koch}}]{king:2012}
{King}, J.~A., {Webb}, J.~K., {Murphy}, M.~T., {et~al.} 2012, \mnras, 422, 3370

\bibitem[{{Kjeldsen} \& {Bedding}(2011)}]{kjeldsen:2011}
{Kjeldsen}, H. \& {Bedding}, T.~R. 2011, \aap, 529, L8

\bibitem[{{Kjeldsen} {et~al.}(2005){Kjeldsen}, {Bedding}, {Butler},
  {Christensen-Dalsgaard}, {Kiss}, {McCarthy}, {Marcy}, {Tinney}, \&
  {Wright}}]{kjeldsen:2005}
{Kjeldsen}, H., {Bedding}, T.~R., {Butler}, R.~P., {et~al.} 2005, \apj, 635,
  1281

\bibitem[{Kotus {et~al.}(2017)Kotus, Murphy, \& Carswell}]{Kotus:2016}
Kotus, S.~M., Murphy, M.~T., \& Carswell, R.~F. 2017, Mon.\ Not.\ Roy.\
  Astron.\ Soc., 464, 3679

\bibitem[{{Landoni} {et~al.}(2016){Landoni}, {Riva}, {Pepe}, {Aliverti},
  {Cabral}, {Calderone}, {Cirami}, {Cristiani}, {Di Marcantonio}, {Genoni},
  {M{\'e}gevand}, {Moschetti}, {Oggioni}, \& {Pariani}}]{landoni:2016}
{Landoni}, M., {Riva}, M., {Pepe}, F., {et~al.} 2016, Society of Photo-Optical
  Instrumentation Engineers (SPIE) Conference Series, Vol. 9913, {ESPRESSO
  front end guiding algorithms: from design phase to implementation and
  validation toward the commissioning}, 99133Q

\bibitem[{{Landoni} {et~al.}(2014){Landoni}, {Riva}, {Pepe}, {Conconi},
  {Zerbi}, {Cabral}, {Cristiani}, \& {Megevand}}]{landoni:2014}
{Landoni}, M., {Riva}, M., {Pepe}, F., {et~al.} 2014, Society of Photo-Optical
  Instrumentation Engineers (SPIE) Conference Series, Vol. 9147, {ESPRESSO
  front end exposure meter: a chromatic approach to radial velocity
  correction}, 91478K

\bibitem[{Leite {et~al.}(2016)Leite, Martins, Molaro, Corre, \&
  Cristiani}]{Leite:2016}
Leite, A., Martins, C., Molaro, P., Corre, D., \& Cristiani, S. 2016, Phys.\
  Rev.\ D, 94, 123512

\bibitem[{{Levshakov} {et~al.}(2013){Levshakov}, {Reimers}, {Henkel}, {Winkel},
  {Mignano}, {Centuri{\'o}n}, \& {Molaro}}]{levshakov:2013}
{Levshakov}, S.~A., {Reimers}, D., {Henkel}, C., {et~al.} 2013, \aap, 559, A91

\bibitem[{{Li} {et~al.}(2018){Li}, {Wolf}, {Bian}, {Onken}, {Schmidt},
  {Tisserand }, {Alonzi}, \& {Jeat Hon}}]{li:2018}
{Li}, Z., {Wolf}, C., {Bian}, F., {et~al.} 2018, arXiv e-prints,
  arXiv:1805.03429

\bibitem[{{Liske} {et~al.}(2008){Liske}, {Grazian}, {Vanzella}, {Dessauges},
  {Viel}, {Pasquini}, {Haehnelt}, {Cristiani}, {Pepe}, {Avila}, {Bonifacio},
  {Bouchy}, {Dekker}, {Delabre}, {D'Odorico}, {D'Odorico}, {Levshakov},
  {Lovis}, {Mayor}, {Molaro}, {Moscardini}, {Murphy}, {Queloz}, {Shaver},
  {Udry}, {Wiklind}, \& {Zucker}}]{liske:2008}
{Liske}, J., {Grazian}, A., {Vanzella}, E., {et~al.} 2008, \mnras, 386, 1192

\bibitem[{{Lizon} {et~al.}(2018){Lizon}, {Dekker}, {Manescau}, {Megevan},
  {Pepe}, \& {Riva}}]{lizon:2018}
{Lizon}, J.~L., {Dekker}, H., {Manescau}, A., {et~al.} 2018, in Society of
  Photo-Optical Instrumentation Engineers (SPIE) Conference Series, Vol. 10701,
  \procspie, 107012P

\bibitem[{{Lizon} {et~al.}(2016){Lizon}, {Iwert}, {Deiries}, {Decker},
  {Hinterschuster}, {Manescau}, {Megevand}, {Mueller}, {Pepe}, \&
  {Riva}}]{lizon:2016}
{Lizon}, J.~L., {Iwert}, O., {Deiries}, S., {et~al.} 2016, Society of
  Photo-Optical Instrumentation Engineers (SPIE) Conference Series, Vol. 9908,
  {An ultra-stable cryostat for the detectors of ESPRESSO}, 990866

\bibitem[{{Lo Curto} {et~al.}(2015){Lo Curto}, {Pepe}, {Avila}, {Boffin},
  {Bovay}, {Chazelas}, {Coffinet}, {Fleury}, {Hughes}, {Lovis}, {Maire},
  {Manescau}, {Pasquini}, {Rihs}, {Sinclaire}, \& {Udry}}]{locurto:2015}
{Lo Curto}, G., {Pepe}, F., {Avila}, G., {et~al.} 2015, The Messenger, 162, 9

\bibitem[{{Loeb}(1998)}]{loeb:1998}
{Loeb}, A. 1998, \apjl, 499, L111

\bibitem[{{Lovis} {et~al.}(2011){Lovis}, {S{\'e}gransan}, {Mayor}, {Udry},
  {Benz}, {Bertaux}, {Bouchy}, {Correia}, {Laskar}, {Lo Curto}, {Mordasini},
  {Pepe}, {Queloz}, \& {Santos}}]{lovis:2011}
{Lovis}, C., {S{\'e}gransan}, D., {Mayor}, M., {et~al.} 2011, \aap, 528, A112

\bibitem[{Martins(2017)}]{Martins:2017}
Martins, C. 2017 [\eprint[arXiv]{1709.02923}]

\bibitem[{Martins {et~al.}(2015)Martins, Pinho, Alves, Pino, Rocha, \& von
  Wietersheim}]{CMartins:2015}
Martins, C., Pinho, A., Alves, R., {et~al.} 2015, JCAP, 08, 047

\bibitem[{Martins \& Vacher(2019)}]{Martins:2019}
Martins, C. \& Vacher, L. 2019, Phys.\ Rev.\ D, 100, 123514

\bibitem[{{Martins} {et~al.}(2015){Martins}, {Santos}, {Figueira}, {Faria},
  {Montalto}, {Boisse}, {Ehrenreich}, {Lovis}, {Mayor}, {Melo}, {Pepe},
  {Sousa}, {Udry}, \& {Cunha}}]{martins:2015}
{Martins}, J.~H.~C., {Santos}, N.~C., {Figueira}, P., {et~al.} 2015, \aap, 576,
  A134

\bibitem[{{Mayor} {et~al.}(2003){Mayor}, {Pepe}, {Queloz}, {Bouchy},
  {Rupprecht}, {Lo Curto}, {Avila}, {Benz}, {Bertaux}, {Bonfils}, {Dall},
  {Dekker}, {Delabre}, {Eckert}, {Fleury}, {Gilliotte}, {Gojak}, {Guzman},
  {Kohler}, {Lizon}, {Longinotti}, {Lovis}, {Megevand}, {Pasquini}, {Reyes},
  {Sivan}, {Sosnowska}, {Soto}, {Udry}, {van Kesteren}, {Weber}, \&
  {Weilenmann}}]{mayor:2003}
{Mayor}, M., {Pepe}, F., {Queloz}, D., {et~al.} 2003, The Messenger, 114, 20

\bibitem[{{Mayor} \& {Queloz}(1995)}]{mayor:1995}
{Mayor}, M. \& {Queloz}, D. 1995, \nat, 378, 355

\bibitem[{{M{\'e}gevand} {et~al.}(2014){M{\'e}gevand}, {Zerbi}, {Di
  Marcantonio}, {Cabral}, {Riva}, {Abreu}, {Pepe}, {Cristiani}, {Rebolo Lopez},
  {Santos}, {Dekker}, {Aliverti}, {Allende}, {Amate}, {Avila}, {Baldini},
  {Bandy}, {Bristow}, {Broeg}, {Cirami}, {Coelho}, {Conconi}, {Coretti},
  {Cupani}, {D'Odorico}, {De Caprio}, {Delabre}, {Dorn}, {Figueira}, {Fragoso},
  {Galeotta}, {Genolet}, {Gomes}, {Gonz{\'a}lez Hern{\'a}ndez}, {Hughes},
  {Iwert}, {Kerber}, {Land oni}, {Lizon}, {Lovis}, {Maire}, {Mannetta},
  {Martins}, {Molaro}, {Monteiro}, {Moschetti}, {Oliveira}, {Zapatero Osorio},
  {Poretti}, {Rasilla}, {Santana Tschudi}, {Santos}, {Sosnowska}, {Sousa},
  {Tenegi}, {Toso}, {Vanzella}, \& {Viel}}]{megevand:2014}
{M{\'e}gevand}, D., {Zerbi}, F.~M., {Di Marcantonio}, P., {et~al.} 2014,
  Society of Photo-Optical Instrumentation Engineers (SPIE) Conference Series,
  Vol. 9147, {ESPRESSO: the radial velocity machine for the VLT}, 91471H

\bibitem[{{Molaro}(2009)}]{molaro2009}
{Molaro}, P. 2009, Astrophysics and Space Science Proceedings, 9, 389

\bibitem[{{Molaro} {et~al.}(2006){Molaro}, {Murphy}, \&
  {Levshakov}}]{molaro2006}
{Molaro}, P., {Murphy}, M.~T., \& {Levshakov}, S.~A. 2006, in IAU Symposium,
  Vol. 232, The Scientific Requirements for Extremely Large Telescopes, ed.
  P.~{Whitelock}, M.~{Dennefeld}, \& B.~{Leibundgut}, 198--203

\bibitem[{Molaro {et~al.}(2013)}]{Molaro:2013}
Molaro, P. {et~al.} 2013, Astron.\ Astrophys., 555, A68

\bibitem[{Murphy \& Cooksey(2017)}]{Murphy:2017}
Murphy, M.~T. \& Cooksey, K.~L. 2017, Mon.\ Not.\ Roy.\ Astron.\ Soc., 471,
  4930

\bibitem[{{Murphy} {et~al.}(2004){Murphy}, {Flambaum}, {Webb}, {Dzuba},
  {Prochaska}, \& {Wolfe}}]{murphy:2004}
{Murphy}, M.~T., {Flambaum}, V.~V., {Webb}, J.~K., {et~al.} 2004, {Constraining
  Variations in the Fine-Structure Constant, Quark Masses and the Strong
  Interaction}, ed. S.~G. {Karshenboim} \& E.~{Peik}, Vol. 648, 131--150

\bibitem[{Murphy {et~al.}(2016)Murphy, Malec, \& Prochaska}]{Murphy:2016}
Murphy, M.~T., Malec, A.~L., \& Prochaska, J.~X. 2016, Mon.\ Not.\ Roy.\
  Astron.\ Soc., 461, 2461

\bibitem[{{Nortmann} {et~al.}(2018){Nortmann}, {Pall{\'e}}, {Salz},
  {Sanz-Forcada}, {Nagel}, {Alonso-Floriano}, {Czesla}, {Yan}, {Chen},
  {Snellen}, {Zechmeister}, {Schmitt}, {L{\'o}pez-Puertas}, {Casasayas-Barris},
  {Bauer}, {Amado}, {Caballero}, {Dreizler}, {Henning}, {Lamp{\'o}n}, {Montes},
  {Molaverdikhani}, {Quirrenbach}, {Reiners}, {Ribas}, {S{\'a}nchez-L{\'o}pez},
  {Schneider}, \& {Zapatero Osorio}}]{nortmann:2018}
{Nortmann}, L., {Pall{\'e}}, E., {Salz}, M., {et~al.} 2018, Science, 362, 1388

\bibitem[{{Oggioni} {et~al.}(2016){Oggioni}, {Pariani}, {Moschetti}, {Riva},
  {Genoni}, {Aliverti}, \& {Landoni}}]{oggioni:2016}
{Oggioni}, L., {Pariani}, G., {Moschetti}, M., {et~al.} 2016, Society of
  Photo-Optical Instrumentation Engineers (SPIE) Conference Series, Vol. 9908,
  {MMP, the Multi Mini Prism device for ESPRESSO APSU: prototyping and
  integration}, 990872

\bibitem[{{Pariani} {et~al.}(2018){Pariani}, {Aliverti}, {Genoni}, {Oggioni},
  {Riva}, {Santana Tschudi}, {M{\'e}gevand}, {Cristiani}, \&
  {Pepe}}]{pariani:2018}
{Pariani}, G., {Aliverti}, M., {Genoni}, M., {et~al.} 2018, in Society of
  Photo-Optical Instrumentation Engineers (SPIE) Conference Series, Vol. 10706,
  \procspie, 107064H

\bibitem[{{Pariani} {et~al.}(2016){Pariani}, {Aliverti}, {Moschetti},
  {Landoni}, {Riva}, {Zerbi}, {M{\'e}gevand}, {Cristiani}, \&
  {Pepe}}]{pariani:2016}
{Pariani}, G., {Aliverti}, M., {Moschetti}, M., {et~al.} 2016, Society of
  Photo-Optical Instrumentation Engineers (SPIE) Conference Series, Vol. 9908,
  {Integration, alignment, and verification of the ESPRESSO Front-End}, 99087B

\bibitem[{{Pasquini} {et~al.}(2005){Pasquini}, {Cristiani}, {Dekker},
  {Haehnelt}, {Molaro}, {Pepe}, {Avila}, {Delabre}, {D'Odorico}, {Liske},
  {Shaver}, {Bonifacio}, {Borgani}, {D'Odorico}, {Vanzella}, {Bouchy},
  {Dessauges-Lavadsky}, {Lovis}, {Mayor}, {Queloz}, {Udry}, {Murphy}, {Viel},
  {Grazian}, {Levshakov}, {Moscardini}, {Wiklind}, \& {Zucker}}]{pasquini:2005}
{Pasquini}, L., {Cristiani}, S., {Dekker}, H., {et~al.} 2005, The Messenger,
  122, 10

\bibitem[{{Pasquini} \& {Hubin}(2018)}]{pasquini:2018}
{Pasquini}, L. \& {Hubin}, N. 2018, in Society of Photo-Optical Instrumentation
  Engineers (SPIE) Conference Series, Vol. 10702, \procspie, 1070204

\bibitem[{{Pasquini} {et~al.}(2009){Pasquini}, {Manescau}, {Avila}, {Delabre},
  {Dekker}, {Liske}, {D'Odorico}, {Pepe}, {Dessauges}, {Lovis}, {Megevand},
  {Queloz}, {Udry}, {Cristiani}, {Bonifacio}, {Dimarcantonio}, {D'Odorico},
  {Molaro}, {Vanzella}, {Viel}, {Haehnelt}, {Carswell}, {Murphy},
  {Garcia-Lopez}, {Herreros}, {Perez}, {Zapatero}, {Rebolo}, {Israelian},
  {Martin}, {Zerbi}, {Span{\`o}}, {Levshakov}, {Santos}, \&
  {Zucker}}]{pasquini:2009}
{Pasquini}, L., {Manescau}, A., {Avila}, G., {et~al.} 2009, Astrophysics and
  Space Science Proceedings, 9, 395

\bibitem[{{Pepe} {et~al.}(2011){Pepe}, {Lovis}, {S{\'e}gransan}, {Benz},
  {Bouchy}, {Dumusque}, {Mayor}, {Queloz}, {Santos}, \& {Udry}}]{pepe:2011}
{Pepe}, F., {Lovis}, C., {S{\'e}gransan}, D., {et~al.} 2011, \aap, 534, A58

\bibitem[{{Pepe} {et~al.}(2002){Pepe}, {Mayor}, {Rupprecht}, {Avila},
  {Ballester}, {Beckers}, {Benz}, {Bertaux}, {Bouchy}, {Buzzoni}, {Cavadore},
  {Deiries}, {Dekker}, {Delabre}, {D'Odorico}, {Eckert}, {Fischer}, {Fleury},
  {George}, {Gilliotte}, {Gojak}, {Guzman}, {Koch}, {Kohler}, {Kotzlowski},
  {Lacroix}, {Le Merrer}, {Lizon}, {Lo Curto}, {Longinotti}, {Megevand},
  {Pasquini}, {Petitpas}, {Pichard}, {Queloz}, {Reyes}, {Richaud}, {Sivan},
  {Sosnowska}, {Soto}, {Udry}, {Ureta}, {van Kesteren}, {Weber}, {Weilenmann},
  {Wicenec}, {Wieland}, {Christensen-Dalsgaard}, {Dravins}, {Hatzes},
  {K{\"u}rster}, {Paresce}, \& {Penny}}]{pepe:2002}
{Pepe}, F., {Mayor}, M., {Rupprecht}, G., {et~al.} 2002, The Messenger, 110, 9

\bibitem[{{Pepe} {et~al.}(2014){Pepe}, {Molaro}, {Cristiani}, {Rebolo},
  {Santos}, {Dekker}, {M{\'e}gevand}, {Zerbi}, {Cabral}, {Di Marcantonio},
  {Abreu}, {Affolter}, {Aliverti}, {Allende Prieto}, {Amate}, {Avila},
  {Baldini}, {Bristow}, {Broeg}, {Cirami}, {Coelho}, {Conconi}, {Coretti},
  {Cupani}, {D'Odorico}, {De Caprio}, {Delabre}, {Dorn}, {Figueira}, {Fragoso},
  {Galeotta}, {Genolet}, {Gomes}, {Gonz{\'a}lez Hern{\'a}ndez}, {Hughes},
  {Iwert}, {Kerber}, {Landoni}, {Lizon}, {Lovis}, {Maire}, {Mannetta},
  {Martins}, {Monteiro}, {Oliveira}, {Poretti}, {Rasilla}, {Riva}, {Santana
  Tschudi}, {Santos}, {Sosnowska}, {Sousa}, {Span{\'o}}, {Tenegi}, {Toso},
  {Vanzella}, {Viel}, \& {Zapatero Osorio}}]{pepe:2014}
{Pepe}, F., {Molaro}, P., {Cristiani}, S., {et~al.} 2014, Astronomische
  Nachrichten, 335, 8

\bibitem[{{Perryman} {et~al.}(2005){Perryman}, {Hainaut}, {Dravins}, {Leger},
  {Quirrenbach}, {Rauer}, {Kerber}, {Fosbury}, {Bouchy}, {Favata}, {Fridlund},
  {Gilmozzi}, {Lagrange}, {Mazeh}, {Rouan}, {Udry}, \&
  {Wambsganss}}]{Perryman2005}
{Perryman}, M., {Hainaut}, O., {Dravins}, D., {et~al.} 2005, {ESA-ESO Working
  Group on ``Extra-solar Planets''}, ESA-ESO Working Group on ``Extra-solar
  Planets'' Edited by M. Perryman et al. ESA

\bibitem[{{Queloz} {et~al.}(2000){Queloz}, {Mayor}, {Weber}, {Bl{\'e}cha},
  {Burnet}, {Confino}, {Naef}, {Pepe}, {Santos}, \& {Udry}}]{queloz:2000}
{Queloz}, D., {Mayor}, M., {Weber}, L., {et~al.} 2000, \aap, 354, 99

\bibitem[{{Rahmani} {et~al.}(2013){Rahmani}, {Wendt}, {Srianand}, {Noterdaeme},
  {Petitjean}, {Molaro}, {Whitmore}, {Murphy}, {Centurion}, {Fathivavsari},
  {D'Odorico}, {Evans}, {Levshakov}, {Lopez}, {Martins}, {Reimers}, \&
  {Vladilo}}]{rahmani:2013}
{Rahmani}, H., {Wendt}, M., {Srianand}, R., {et~al.} 2013, \mnras, 435, 861

\bibitem[{{Rauch} {et~al.}(2005){Rauch}, {Becker}, {Viel}, {Sargent}, {Smette},
  {Simcoe}, {Barlow}, \& {Haehnelt}}]{rauch:2005}
{Rauch}, M., {Becker}, G.~D., {Viel}, M., {et~al.} 2005, \apj, 632, 58

\bibitem[{{Redfield} {et~al.}(2008){Redfield}, {Endl}, {Cochran}, \&
  {Koesterke}}]{redfield:2008}
{Redfield}, S., {Endl}, M., {Cochran}, W.~D., \& {Koesterke}, L. 2008, \apjl,
  673, L87

\bibitem[{{Ricker} {et~al.}(2015){Ricker}, {Winn}, {Vanderspek}, {Latham},
  {Bakos}, {Bean}, {Berta-Thompson}, {Brown}, {Buchhave}, {Butler}, {Butler},
  {Chaplin}, {Charbonneau}, {Christensen-Dalsgaard}, {Clampin}, {Deming},
  {Doty}, {De Lee}, {Dressing}, {Dunham}, {Endl}, {Fressin}, {Ge}, {Henning},
  {Holman}, {Howard}, {Ida}, {Jenkins}, {Jernigan}, {Johnson}, {Kaltenegger},
  {Kawai}, {Kjeldsen}, {Laughlin}, {Levine}, {Lin}, {Lissauer}, {MacQueen},
  {Marcy}, {McCullough}, {Morton}, {Narita}, {Paegert}, {Palle}, {Pepe},
  {Pepper}, {Quirrenbach}, {Rinehart}, {Sasselov}, {Sato}, {Seager},
  {Sozzetti}, {Stassun}, {Sullivan}, {Szentgyorgyi}, {Torres}, {Udry}, \&
  {Villasenor}}]{ricker15}
{Ricker}, G.~R., {Winn}, J.~N., {Vanderspek}, R., {et~al.} 2015, Journal of
  Astronomical Telescopes, Instruments, and Systems, 1, 014003

\bibitem[{{Riemer-S{\o}rensen} {et~al.}(2017){Riemer-S{\o}rensen},
  {Kotu{\v{s}}}, {Webb}, {Ali}, {Dumont}, {Murphy}, \&
  {Carswell}}]{riemer:2017}
{Riemer-S{\o}rensen}, S., {Kotu{\v{s}}}, S., {Webb}, J.~K., {et~al.} 2017,
  \mnras, 468, 3239

\bibitem[{{Riemer-S{\o}rensen} {et~al.}(2015){Riemer-S{\o}rensen}, {Webb},
  {Crighton}, {Dumont}, {Ali}, {Kotu{\v{s}}}, {Bainbridge}, {Murphy}, \&
  {Carswell}}]{riemer:2015}
{Riemer-S{\o}rensen}, S., {Webb}, J.~K., {Crighton}, N., {et~al.} 2015, \mnras,
  447, 2925

\bibitem[{{Riva} {et~al.}(2014{\natexlab{a}}){Riva}, {Aliverti}, {Moschetti},
  {Land oni}, {Dell'Agostino}, {Pepe}, {M{\'e}gevand}, {Zerbi}, {Cristiani}, \&
  {Cabral}}]{riva:2014-2}
{Riva}, M., {Aliverti}, M., {Moschetti}, M., {et~al.} 2014{\natexlab{a}},
  Society of Photo-Optical Instrumentation Engineers (SPIE) Conference Series,
  Vol. 9147, {ESPRESSO front end: modular opto-mechanical integration for
  astronomical instrumentation}, 91477G

\bibitem[{{Riva} {et~al.}(2014{\natexlab{b}}){Riva}, {Conconi}, {Moschetti},
  {Dell'Agostino}, {Genoni}, {Aliverti}, {Pepe}, {M{\'e}gevand}, {Zerbi},
  {Cristiani}, {Cabral}, \& {Span{\`o}}}]{riva:2014-1}
{Riva}, M., {Conconi}, P., {Moschetti}, M., {et~al.} 2014{\natexlab{b}},
  Society of Photo-Optical Instrumentation Engineers (SPIE) Conference Series,
  Vol. 9147, {APSU @ ESPRESSO: final design towards the integration}, 91477D

\bibitem[{{Rosenband} {et~al.}(2008){Rosenband}, {Hume}, {Schmidt}, {Chou},
  {Brusch}, {Lorini}, {Oskay}, {Drullinger}, {Fortier}, {Stalnaker}, {Diddams},
  {Swann}, {Newbury}, {Itano}, {Wineland}, \& {Bergquist}}]{rosenband:2008}
{Rosenband}, T., {Hume}, D.~B., {Schmidt}, P.~O., {et~al.} 2008, Science, 319,
  1808

\bibitem[{{Roy} {et~al.}(2020){Roy}, {Halverson}, {Mahadevan}, {Stefansson},
  {Monson}, {Logsdon}, {Bender}, {Blake}, {Golub}, {Gupta}, {Jaehnig},
  {Kanodia}, {Kaplan}, {McElwain}, {Ninan}, {Rajagopal}, {Robertson}, {Schwab},
  {Terrien}, {Wang}, {Wolf}, \& {Wright}}]{arpita:2020}
{Roy}, A., {Halverson}, S., {Mahadevan}, S., {et~al.} 2020, \aj, 159, 161

\bibitem[{{Santana Tschudi} {et~al.}(2014){Santana Tschudi}, {Fragoso},
  {Amate}, {Rebolo}, {M{\'e}gevand}, {Zerbi}, \& {Pepe}}]{santana:2014}
{Santana Tschudi}, S., {Fragoso}, A., {Amate}, M., {et~al.} 2014, Society of
  Photo-Optical Instrumentation Engineers (SPIE) Conference Series, Vol. 9151,
  {Design of the opto-mechanical mounts of the ESPRESSO spectograph}, 915153

\bibitem[{{Santos} {et~al.}(2017){Santos}, {Adibekyan}, {Dorn}, {Mordasini},
  {Noack}, {Barros}, {Delgado-Mena}, {Demangeon}, {Faria}, {Israelian}, \&
  {Sousa}}]{santos:2017}
{Santos}, N.~C., {Adibekyan}, V., {Dorn}, C., {et~al.} 2017, \aap, 608, A94

\bibitem[{{Snellen} {et~al.}(2008){Snellen}, {Albrecht}, {de Mooij}, \& {Le
  Poole}}]{snellen:2008}
{Snellen}, I.~A.~G., {Albrecht}, S., {de Mooij}, E.~J.~W., \& {Le Poole}, R.~S.
  2008, \aap, 487, 357

\bibitem[{{Sotzen} {et~al.}(2020){Sotzen}, {Stevenson}, {Sing}, {Kilpatrick},
  {Wakeford}, {Filippazzo}, {Lewis}, {H{\"o}rst}, {L{\'o}pez-Morales}, {Henry},
  {Buchhave}, {Ehrenreich}, {Fraine}, {Garc{\'\i}a Mu{\~n}oz}, {Jayaraman},
  {Lavvas}, {Lecavelier des Etangs}, {Marley}, {Nikolov}, {Rathcke}, \&
  {Sanz-Forcada}}]{sotzen:2020}
{Sotzen}, K.~S., {Stevenson}, K.~B., {Sing}, D.~K., {et~al.} 2020, \aj, 159, 5

\bibitem[{{Sousa} {et~al.}(2015){Sousa}, {Santos}, {Adibekyan}, {Delgado-Mena},
  \& {Israelian}}]{sousa:2015}
{Sousa}, S.~G., {Santos}, N.~C., {Adibekyan}, V., {Delgado-Mena}, E., \&
  {Israelian}, G. 2015, \aap, 577, A67

\bibitem[{{Sousa}(2007)}]{sousa07}
{Sousa}, S. G. e.~a. 2007, A\&A, 469, 783

\bibitem[{{Tenegi} {et~al.}(2016){Tenegi}, {Santana}, {G{\'o}mez}, {Rodilla},
  {Hughes}, {M{\'e}gevand}, {Rebolo}, {Riva}, \& {Luis-Simoes}}]{tenegi:2016}
{Tenegi}, F., {Santana}, S., {G{\'o}mez}, J., {et~al.} 2016, Society of
  Photo-Optical Instrumentation Engineers (SPIE) Conference Series, Vol. 9912,
  {ESPRESSO optical bench: from mind to reality}, 99123K

\bibitem[{{Thompson}(1975)}]{thompson:1975}
{Thompson}, R.~I. 1975, \aplett, 16, 3

\bibitem[{{Udry} {et~al.}(2019){Udry}, {Dumusque}, {Lovis}, {S{\'e}gransan},
  {Diaz}, {Benz}, {Bouchy}, {Coffinet}, {Lo Curto}, {Mayor}, {Mordasini},
  {Motalebi}, {Pepe}, {Queloz}, {Santos}, {Wyttenbach}, {Alonso}, {Collier
  Cameron}, {Deleuil}, {Figueira}, {Gillon}, {Moutou}, {Pollacco}, \&
  {Pompei}}]{udry:2019}
{Udry}, S., {Dumusque}, X., {Lovis}, C., {et~al.} 2019, \aap, 622, A37

\bibitem[{{Uzan}(2011)}]{uzan:2011}
{Uzan}, J.-P. 2011, Living Reviews in Relativity, 14, 2

\bibitem[{{Van Eylen} {et~al.}(2018){Van Eylen}, {Agentoft}, {Lundkvist},
  {Kjeldsen}, {Owen}, {Fulton}, {Petigura}, \& {Snellen}}]{vaneylen:2018}
{Van Eylen}, V., {Agentoft}, C., {Lundkvist}, M.~S., {et~al.} 2018, \mnras,
  479, 4786

\bibitem[{{Vanzella} {et~al.}(2020){Vanzella}, {Meneghetti}, {Pastorello},
  {Calura}, {Sani}, {Cupani}, {Caminha}, {Castellano}, {Rosati}, {D'Odorico},
  {Cristiani}, {Grillo}, {Mercurio}, {Nonino}, {Brammer}, \&
  {Hartman}}]{vanzella:2020}
{Vanzella}, E., {Meneghetti}, M., {Pastorello}, A., {et~al.} 2020, \mnras
  [\eprint[arXiv]{2004.08400}]

\bibitem[{{Webb} {et~al.}(1999){Webb}, {Flambaum}, {Churchill}, {Drinkwater},
  \& {Barrow}}]{webb:1999}
{Webb}, J.~K., {Flambaum}, V.~V., {Churchill}, C.~W., {Drinkwater}, M.~J., \&
  {Barrow}, J.~D. 1999, \prl, 82, 884

\bibitem[{{Webb} {et~al.}(2011){Webb}, {King}, {Murphy}, {Flambaum},
  {Carswell}, \& {Bainbridge}}]{webb:2011}
{Webb}, J.~K., {King}, J.~A., {Murphy}, M.~T., {et~al.} 2011, \prl, 107, 191101

\bibitem[{{Wyttenbach} {et~al.}(2017){Wyttenbach}, {Lovis}, {Ehrenreich},
  {Bourrier}, {Pino}, {Allart}, {Astudillo-Defru}, {Cegla}, {Heng}, {Lavie},
  {Melo}, {Murgas}, {Santerne}, {S{\'e}gransan}, {Udry}, \&
  {Pepe}}]{wyttenbach:2017}
{Wyttenbach}, A., {Lovis}, C., {Ehrenreich}, D., {et~al.} 2017, \aap, 602, A36

\end{thebibliography}

\end{document}